\documentclass[twocolumn]{aastex631}

\usepackage{tablefootnote} 

\shorttitle{Mrk 501 during historically low activity}
\shortauthors{MAGIC Collaboration et al.}
\graphicspath{{./}{figures/}}

\begin{document}

\title{Multi-messenger characterization of Mrk\,501 during historically low X-ray and $\gamma$-ray activity}

\author{H.~Abe}
\affiliation{Japanese MAGIC Group: Institute for Cosmic Ray Research (ICRR), The University of Tokyo, Kashiwa, 277-8582 Chiba, Japan}
\author{S.~Abe}
\affiliation{Japanese MAGIC Group: Institute for Cosmic Ray Research (ICRR), The University of Tokyo, Kashiwa, 277-8582 Chiba, Japan}
\author{V.~A.~Acciari}
\affiliation{ Instituto de Astrof\'isica de Canarias and Dpto. de  Astrof\'isica, Universidad de La Laguna, E-38200, La Laguna, Tenerife, Spain}
\author{I.~Agudo}
\affiliation{ Instituto de Astrof\'isica de Andaluc\'ia-CSIC, Glorieta de la Astronom\'ia s/n, 18008, Granada, Spain}
\author{T.~Aniello}
\affiliation{ National Institute for Astrophysics (INAF), I-00136 Rome, Italy}
\author{S.~Ansoldi}
\affiliation{ Universit\`a di Udine and INFN Trieste, I-33100 Udine, Italy}
\affiliation{ also at International Center for Relativistic Astrophysics (ICRA), Rome, Italy}
\author{L.~A.~Antonelli}
\affiliation{ National Institute for Astrophysics (INAF), I-00136 Rome, Italy}
\author{A.~Arbet Engels}
\affiliation{ Max-Planck-Institut f\"ur Physik, D-80805 M\"unchen, Germany}
\author{C.~Arcaro}
\affiliation{ Universit\`a di Padova and INFN, I-35131 Padova, Italy}
\author{M.~Artero}
\affiliation{ Institut de F\'isica d'Altes Energies (IFAE), The Barcelona Institute of Science and Technology (BIST), E-08193 Bellaterra (Barcelona), Spain}
% ...
\collaboration{500}{(The MAGIC Collaboration)}
\author{M.~Cerruti$^\star$}
\affiliation{ Universitat de Barcelona, ICCUB, IEEC-UB, E-08028 Barcelona, Spain}
\affiliation{ Universit\'e Paris Cit\'e, CNRS, Astroparticule et Cosmologie, F-75013 Paris, France}
\author{J.~A.~Acosta-Pulido}
\affiliation{ Instituto de Astrof\'isica de Canarias and Dpto. de Astrof\'isica, Universidad de La Laguna, 38200 La Laguna, Tenerife, Spain }
\author{G.~Apolonio}
\affiliation{ Department of Physics and Astronomy, N283 ESC, Brigham Young University, Provo, UT 84602, USA}
\author{R.~Bachev}
\affiliation{ Institute of Astronomy and National Astronomical Observatory, Bulgarian Academy of Sciences, 72 Tsarigradsko shosse Blvd., 1784 Sofia, Bulgaria }
\author{M.~Balokovi\'{c}}
\affiliation{ Yale Center for Astronomy \& Astrophysics, 52 Hillhouse Avenue, New Haven, CT 06511, USA}
\affiliation{ Department of Physics, Yale University, P.O. Box 2018120, New Haven, CT 06520, USA}
\author{E.~Ben\'itez}
\affiliation{ Universidad Nacional Aut\'onoma de M\'exico, Instituto de Astronom\'ia, AP 70-264, CDMX 04510, Mexico}
\author{I.~Bj\"orklund}
\affiliation{ Aalto University Mets\"ahovi Radio Observatory, Mets\"ahovintie 114, 02540 Kylm\"al\"a, Finland}
\affiliation{ Aalto University Department of Electronics and Nanoengineering, P.O. BOX 15500, FI-00076 AALTO, Finland }
\author{V.~Bozhilov}
\affiliation{ Department of Astronomy, Faculty of Physics, University of Sofia, BG-1164 Sofia, Bulgaria}
\author{L.~F.~Brown}
\affiliation{ Connecticut College, Department of Physics, Astronomy and Geophysics, New London, CT 06320 }
\author{A.~Bugg}
\affiliation{ Department of Physics and Astronomy, N283 ESC, Brigham Young University, Provo, UT 84602, USA}
% ...



\begin{abstract}
We study the broadband emission of Mrk\,501 using multi-wavelength observations from 2017 to 2020 performed with a multitude of instruments, involving, among others, MAGIC, \textit{Fermi}-LAT, \textit{NuSTAR}, \textit{Swift}, GASP-WEBT, and OVRO. Mrk\,501 showed an extremely low broadband activity, which may help to unravel its baseline emission. Nonetheless, significant flux variations are detected at all wavebands, with the highest occurring at X-rays and very-high-energy (VHE) $\gamma$-rays. A significant correlation ($>$3$\sigma$) between X-rays and VHE $\gamma$-rays is measured, supporting leptonic scenarios to explain the variable parts of the emission, also during low activity. This is further supported when we extend our data from 2008 to 2020, and identify, for the first time, significant correlations between \textit{Swift}-XRT and \textit{Fermi}-LAT. We additionally find correlations between high-energy $\gamma$-rays and radio, with the radio lagging by more than 100 days, placing the $\gamma$-ray emission zone upstream of the radio-bright regions in the jet. Furthermore, Mrk\,501 showed a historically low activity in X-rays and VHE $\gamma$-rays from mid-2017 to mid-2019 with a stable VHE flux ($>$0.2\,TeV) of 5\% the emission of the Crab Nebula. The broadband spectral energy distribution (SED) of this 2-year-long low-state, the potential baseline emission of Mrk\,501, can be characterized with one-zone leptonic models, and with (lepto)-hadronic models fulfilling neutrino flux constraints from IceCube. We explore the time evolution of the SED towards the low-state, revealing that the stable baseline emission may be ascribed to a standing shock, and the variable emission to an additional expanding or traveling shock. 
\end{abstract}




\correspondingauthor{L.~Heckmann, D.~Paneque, S.~Gasparyan, M.~Cerruti, N.~ Sahakyan.}
\email{contact.magic@mpp.mpg.de}

\section{Introduction}
\label{sec:intro}
Blazars are some of the most energetic sources in our Universe and among the most prominent objects in the $\gamma$-ray sky. The new era of multi-messenger and multiwavelength (MWL) astronomy has widened our view into the Universe and possibilities to understand its riddles. For many blazars, very high energy (VHE; E$>$0.1~TeV) $\gamma$-rays provide us with a useful tool because, together with X-rays, they host most of the variable and rapidly evolving emission.

\begin{figure*}
   \centering
   \includegraphics[width=\textwidth]{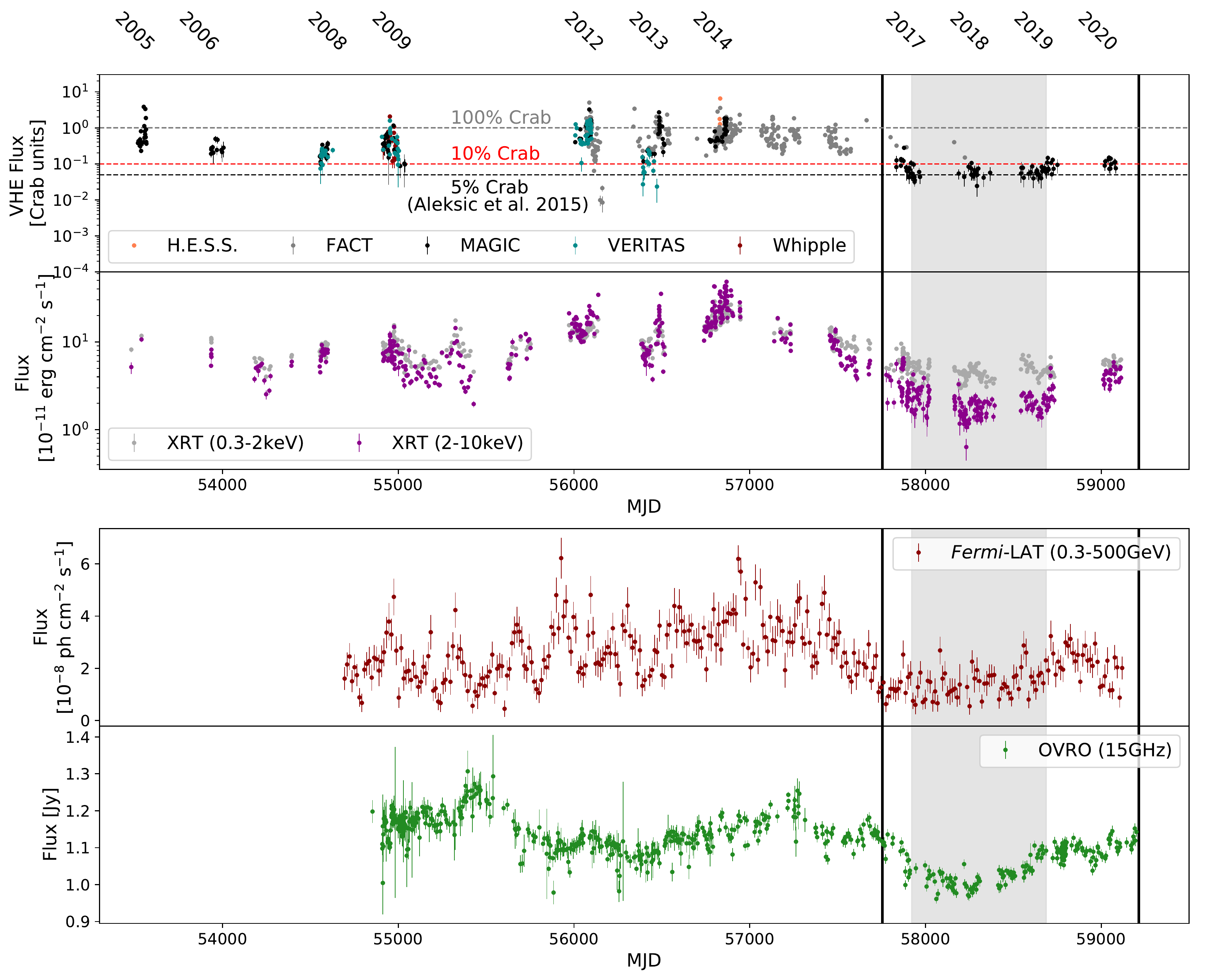}
   \caption{Long term Mrk\,501 light curve spanning from February 2005 until end of 2020 displaying (from top to bottom) all published VHE data (\citet{Mrk501_2005} with E~$>$~0.25~TeV, \citet{Mrk501_MAGIC_2006} with E~$>$~0.2 TeV, \citet{Mrk501_MAGIC_2008} with E~$>$~0.3~GeV, \citet{Mrk501_MAGIC_2009b} with E~$>$~0.3~TeV, \citet{Mrk501_MAGIC_2012} with E~$>$~0.2~TeV, \citet{Mrk501_MAGIC_2013} with E~$>$~0.2~TeV, \citet{Mrk501_MAGIC_2014} with E~$>$~0.15~TeV, \citet{Mrk501_HESS_2014}\textsuperscript{a} with E~$>$~2~TeV, \citet{FACT_Mrk501_2021} with E~$>$~0.75~TeV; only significant measurements $>$\,2$\sigma$ are displayed.); \textit{Swift}-XRT data; \textit{Fermi}-LAT  data in 14-day bins; OVRO data. The vertical black lines mark the 4-year period featured in this work, and the grey area marks the identified period with a very low activity (see Section~\ref{sect:MWL_lc} for details).}
   \small\textsuperscript{a} \footnotesize{For the data from \citet{Mrk501_HESS_2014} no original data set could be organized. Therefore the data points were extracted from the pdf and are displayed without error bars due to the lack of precision of this method.}
    \label{fig:long_LC_comb}%
\end{figure*}

Markarian 501 \citep[Mrk\,501; z=0.034,][]{Ulrich_1975} is a blazar that has been extensively studied during the last three decades. It belongs to the subclass of BL Lac objects, which are classified by their weak or missing broad emission lines in the optical spectrum \citep{UrryPad1995}. In the 1970s, Benjamin Markarian discovered Mrk\,501 while cataloging galaxies with excesses in the ultraviolet emission \citep{Markarian_1972}.
Two decades later, in 1996, Mrk\,501 was the second BL Lac object to be detected in very high energy (VHE) $\gamma$-rays with energies greater than 300 GeV by the Whipple Observatory Gamma Ray Collaboration \citep{Mrk501_1996}. As summarized in \citet{Mrk501_2005}, Mrk\,501 was regularly observed by the first generation of VHE instruments (HEGRA, Whipple, CAT) from 1996 to 2000 as shown by the light curve (LC) in Fig.~\ref{fig:long_LC_old}.  Subsequently, the new generation of  imaging atmospheric Cherenkov telescopes (IACTs), MAGIC, H.E.S.S., VERITAS and FACT, have been observing the source since 2005 together with regular observations in other wavebands \citep{Mrk501_2005,Mrk501_MAGIC_2006,Mrk501_MAGIC_2008,Mrk501_MAGIC_2009a, MRk501_Veritas_2009, Abdo_2011, Mrk501_MAGIC_2009b,Mrk501_MAGIC_2012, Mrk501_MAGIC_2013, Mrk501_MAGIC_2014, Mrk501_HESS_2014, FACT_Mrk501_2021}. A coarse overview of the broadband emission of Mrk\,501 from the year 2005 to the year 2020 is shown in Fig.~\ref{fig:long_LC_comb} including the time interval featured by this work, that spans from 2017 to 2020. The VHE fluxes are expressed using the flux of the Crab Nebula (unit C.U. = Crab units)\footnote{In this study we use the Crab nebula flux reported in \citet[][]{Crab_2015}} to ease comparison among the various IACTs.

This study focuses on the data collected during four years, from 2017 to 2020.  These data were collected within the framework of the coordinated multi-instrument observations of Mrk\,501 that started in the year 2008 \citep{Mrk501_MAGIC_2008}. They have been performed regularly every year since then with the goal of conducting detailed investigations of the broadband emission of Mrk\,501 during many distinct activity states. During the years 2017--2020 Mrk\,501 showed a quiescent broadband behavior with a remarkable feature, a very  low activity that lasted over two years, from mid-2017 to mid-2019 (indicated by the grey area in Fig.~\ref{fig:long_LC_comb}). For the VHE $\gamma$-rays and X-rays, this 2-year interval marks an historically low activity. For the first time since its discovery in VHE, the source remained for a long period of time at a VHE flux of about 0.05\,C.U., which is about five times lower than its typical activity. The proximity of Mrk\,501 and its intrinsic brightness, together with extensive observations with sensitive instruments, provide us with the unprecedented opportunity to investigate with accuracy the broadband emission of this archetypal TeV blazar during a period of extremely low activity.

The emission of blazars is expected to originate either purely from relativistic leptons or from a mix of relativistic leptons and hadrons and is known to produce a spectrum comprised by two distinctive components. The location of the low energy peak frequency can be used to distinguish blazars, including BL Lac type objects, into further subcategories: low synchrotron peaked blazars (LSPs) with a low-energy peak frequency of $\nu_{s}<10^{14}$\,Hz, intermediate synchrotron peaked blazars (ISPs)  with  $10^{14}\text{\,Hz}<\nu_{s}<10^{15}$\,Hz, and  high synchrotron peaked blazar (HSPs) with $\nu_{s}>10^{15}$\,Hz \citep{2010ApJ...716...30A}. Mrk\,501 is usually classified as an HSP, but the above-mentioned multi-year observations of Mrk\,501 have shown that it can also behave like an extreme HSP (EHSP, $\nu_{s}\geq10^{17}$\,Hz \citep{2001A&A...371..512C,2010ApJ...716...30A}) during extended periods of time (half year) and non-flaring activity \citep[][]{Mrk501_MAGIC_2012}.

In both leptonic and (lepto)-hadronic scenarios, synchrotron radiation from relativistic electrons inside the jet accounts for the radio to X-ray blazar emission for a typical HSP BL Lac type object. For purely leptonic scenarios, the observed $\gamma$-ray emission is produced when some of the synchrotron photons are inverse Compton scattered by relativistic electrons in the jet. This scenario is called synchrotron self-Compton (SSC) \citep[see for e.g.,][]{1992ApJ...397L...5M, 1996ASPC..110..436G, Tavecchio_Constraints, 1997MNRAS.292..646B} and is the most commonly applied blazar model to HSPs, and Mrk\,501 in particular. An alternative description provided by hadronic scenarios considers relativistic protons being responsible for the $\gamma$-ray production, either by synchrotron radiation from the hadrons or synchrotron radiation from secondary particles produced from hadron-photon interactions (see e.g. \citet{Mannheim93, Aharonian00, Anita_2001}; and \citet{Cerruti_review} for a recent review). The simplest scenarios consider a single emission zone inside the jet hosting these relativistic particles. The one-zone models have been successfully applied to explain the behavior of Mrk\,501 in the past for both the leptonic \citep[][]{Mrk501_MAGIC_2014, Mrk501_MAGIC_2013, Mrk501_MAGIC_2009b, Abdo_2011} as well as proton-induced models \citep[][]{Anita_2001,Anita_2003}. However, for some of the complex features seen among blazars, these simplified models fail, leading to many alternative explanations emerging, e.g. involving multiple emission zones or structured jets. A two-zone model was preferred to explain some multi-instrument data taken during flaring activities, as shown in \citet{Mrk501_MAGIC_2009b,Mrk501_MAGIC_2012}, and a structured jet was used to explain the evidences for a narrow spectral feature at 3\,TeV observed in the VHE emission of Mrk\,501, as shown in \citet{Mrk501_MAGIC_2014}. Additional claims for a stratified structure in the jet of Mrk\,501 are strongly supported by very-long-baseline interferometry (VLBI) images taken in the radio regime indicating a transverse structure \citep{Giroletti2004}. The emission mechanisms of Mrk\,501 are far from being understood, and there are already several observations that suggest the need for more complex scenarios than those related to a single emission region. 


Our detailed 4-year MWL data set and especially its detailed characterization of the historically low activity now allows us to investigate the capability of these different models to explain the emission during and around the low-state. On the one hand by investigating the nature of the low-state itself and on the other hand by evaluating its potential of being the baseline emission of Mrk\,501 that is typically hidden by  brighter and more variable components, that may be produced somewhere else along the jet. 

This paper is structured as follows: in Section~\ref{sec:instruments} we describe all instruments participating in the 2017-2020 campaign, as well as their data analyses. Section~\ref{sect:MWL_lc} summarizes the MWL behavior focusing on variability and correlation studies. The spectral studies and theoretical models applied in the campaign are described in Section~\ref{sec:spect}. The results are then put into a physical context in Section~\ref{sec:discuss} while Section~\ref{sec:summary} gives some concluding remarks and an outlook. 



\section{Instruments and analysis}
\label{sec:instruments}

This study focuses on the MWL data collected from Mrk\,501 during the 4-year period spanning from the beginning of the observational period in the year 2017 until the end of the observational period in the year 2020 (MJD 57754 to MJD 59214). The MWL fluxes during the above-mentioned 4-year time interval are depicted in Fig.~\ref{fig:MWL_LC}, and the sections below describe the details of the data collection and the strategies used to analyze the data of the instruments involved. For certain instruments, \textit{Fermi}-LAT, \textit{Swift}-XRT and OVRO, long term data since 2008 are available (Fig.~\ref{fig:long_LC_comb}) and added to the data set for part of the analysis.  

\subsection{MAGIC}
MAGIC (Major Atmospheric Gamma Imaging Cherenkov) consists of two IACTs separated by a distance of 85\,m. It is located at the Roque de los Muchachos Observatory, on the Canary island of La Palma at an altitude of 2243\,m above sea level. 
The telescopes work in an energy range between 50\,GeV and tens of TeVs, with a sensitivity above 100 GeV (300 GeV) of about 2\% (about 1\%) of
the Crab Nebula flux at low zenith angles ($<$~30$^\circ$) after 25~h of observations \citep[see Fig.~19 of][]{2016APh....72...76A}. With this performance, the MAGIC telescopes are very well suited to perform blazar observations in the VHE range.

During the 4-year period from 2017-2020, Mrk\,501 was observed by MAGIC for around 160\,hours,  yielding around 120\,hours after the data selection based on the atmospheric transmission and night sky background (NSB) levels. The data are analysed using the MARS (MAGIC Analysis and Reconstruction Software) package \citep{zanin2013, 2016APh....72...76A}. Owing to Mrk\,501 being one of the brightest sources in the MAGIC source catalog, it is possible to observe it even during moon conditions. The analysis is adjusted accordingly to the higher NSB levels as prescribed in \citet{2017APh....94...29A}. 

The VHE flux light curve is computed with a minimum energy of 0.2\,TeV in order to minimize the impact of the various observing conditions considered in this study, which can increase by a factor $\sim$2 the analysis energy threshold at zenith. Our selected energy threshold is compatible with the applied data selection (zenith$<$50$^\circ$ and exclusion of bright moon levels).  We bin the observations once night-wise and once on a weekly basis.  For bins with a significance of less than 2$\sigma$, the upper limits with 95\% confidence according to the Rolke method \citep{Rolke_2005} are computed. For the spectral reconstruction, we use a forward folding method assuming a simple power-law as the spectral model to obtain the relevant parameters which are summarized in Table~\ref{tab:spec_para_MAGIC} and the Tikhonov unfolding method to obtain the spectral data points \citep{2007unfold}. For each spectrum described in Section~\ref{sec:low-activity-SED}, we check if a log parabolic power-law model is preferred over a simple power-law model using a likelihood ratio test performed over the aggregated data in the corresponding time intervals. The log parabola describes slightly better the spectra, but the preference for this model is not statistically significant ($<$\,3$\sigma$).

\subsection{\textit{Fermi}-LAT}
\label{subsec:inst_fermi}

The Large Area Telescope (LAT) on board the \textit{Fermi} Gamma-ray Space Telescope (\textit{Fermi}) is constantly monitoring the high-energy sky since its launch in 2008. As a pair conversion instrument, it is sensitive to an energy range from 20\,MeV to beyond $300$\,GeV, and covers the whole sky every ${\sim}3$\,hours \citep{2009ApJ...697.1071A,2012ApJS..203....4A}. 

Owing to the low activity of Mrk\,501, as well as the need to characterize its $\gamma$-ray emission on relatively short time intervals,  we decided to use the unbinned-likelihood tools provided by the \texttt{FERMITOOLS} software\footnote{\url{https://fermi.gsfc.nasa.gov/ssc/data/analysis/}} (v1.0.10), which is more suitable than the binned analysis for situations with low-event statistics.  Table~ \ref{tab:Fermi_setting} summarizes the basic analysis settings that were used. The usage of 0.3\,GeV as minimum energy for the analysis (instead of the conventional 0.1\,GeV) reduces the detected number of photons from the source. However, this reduction is small for hard-spectrum sources (photon index $<$2) like Mrk\,501. On the other hand, the angular resolution (68\% containment) of photons above 0.1\,GeV is about 5\,deg, while it is about 2\,deg for photons above 0.3\,GeV. This means that a LAT analysis above 0.3\,GeV is less affected by the diffuse backgrounds (which are always softer than photon index 2), and hence will lead to an increase in the signal-to-noise ratio for hard sources. Additionally, the LAT analysis above 0.3\,GeV is less sensitive to possible contamination from non-accounted (transient) neighbouring sources, in comparison to a LAT analysis above 0.1\,GeV. The maximum energy range is chosen to overlap with the MAGIC energy range when reconstructing spectra. The fourth \textit{Fermi}-LAT source catalog \citep[4FGL;][]{2020ApJS..247...33A} is used to build a first model consisting of all sources within the region of interest (ROI) plus 5$^\circ$, which is a standard ROI to investigate. We fit the obtained model to our data set covering the time range from MJD 57754 (2017-01-01, 00:00:00) to MJD 59126 (2020-10-04, 00:00:00). The preliminary fit result is used to remove very weak components from the model (counts $<1$ or TS $<3$). 
Afterwards, the data set is divided in 98 bins each of 14\,day duration, and each bin is fit with the model. In the fitting procedure, only the normalization of bright sources (TS $>10$), sources close to the ROI center ($<3^{\circ}$) and the diffuse background are allowed to vary. Additionally, the spectral parameters of Mrk\,501 are allowed to vary. From this fit we obtain the flux values per time bin and produce the light curve for Mrk\,501. To check the impact of very variable sources in our ROI, we recalculated the light curve freeing the normalization and spectral indices of all sources with a variability index $>$ 100. The resulting flux values agree with the previous light curve within the statistical uncertainties. Furthermore, we perform spectral analyses for the 2-year period with very low activity, and for two-week bins centered around each of the three \textit{NuSTAR} observations conducted during the campaign (see Section~\ref{subsec:NuSTAR}). We use the same ROI model and approach as for the light curve. We checked the likelihood ratio between a power-law and log parabolic power-law model applied to the time intervals. Since for none of the spectra a preference of more than 3$\sigma$ is seen for the log parabolic fit, a power-law is chosen for Mrk\,501 for the spectral reconstruction. The parameters obtained are summarized in Table~\ref{tab:spec_para_Fermi}. For the spectral data points the number of spectral bins is chosen according to the time intervals and flux level. 

For part of the correlation analysis, a long-term \textit{Fermi}-LAT light curve is used including all data since its launch. We therefore apply the same procedure as described above to the data from MJD 54688 (2008-08-10, 00:00:00) to MJD 59126 (2020-10-04, 00:00:00), also using 14-day bins. The starting MJD is the earliest possible in 2008 that is in line with the time bins used for the 4-year epoch featured in this paper, and results in 317 time bins for the 2008-2020 period.

\begin{table}
\centering
\begin{tabular}{ l l }     
\hline\hline 
 Setting & Value  \\
\hline\hline   
 instrument response function & \texttt{P8R3\_SOURCE\_V2}  \\
 diffuse background model\footnote{\url{http://fermi.gsfc.nasa.gov/ssc/data/access/lat/\\BackgroundModels.html}} & \texttt{gll\_iem\_v07} \\ & \texttt{iso\_P8R3\_SOURCE\_V2\_v1} \\
 evtclass & 128 \\
 evttype & 3 \\
 ROI radius & 15$^{\circ}$  \\ 
 energy range & 0.3 to 500 GeV \\
 maximum zenith & 100$^{\circ}$  \\ 
\hline 
\end{tabular}
\caption{Settings used for the unbinned \textit{Fermi}-LAT analysis used in this work as described in Section~\ref{subsec:inst_fermi}}
\label{tab:Fermi_setting}
\end{table}

\subsection{\textit{NuSTAR}}
\label{subsec:NuSTAR}
\textit{NuSTAR} is one of NASA's Small Explorer satellites, sensitive in the hard X-ray band. It has two multilayer-coated telescopes, focusing the reflected X-rays on the pixillated CdZnTe focal plane modules, FPMA and FPMB.  The observatory provides a bandpass of 3-79 keV with spectral resolution of $\sim1$\,keV. The field of view of each telescope is $\sim13'$, and the half-power diameter of an image of a point source is $\sim1'$. This allows a reliable estimate and subtraction of  instrumental and cosmic backgrounds, resulting in an unprecedented sensitivity for measuring hard X-ray fluxes and spectra of celestial sources. For more details, see \citet{Harrison_2013}.

The data reported here (Table~\ref{tab:NuSTAR_data}) cover three observations in 2017 and 2018 that were planned as part of two dedicated {\it NuSTAR} proposals  (from PI Balokovic and PI Paneque). After screening for the South Atlantic Anomaly passages and Earth occultation, the observations yielded roughly 5 hours of on-target data per pointing.  The raw data products are processed separately for each pointing with the \textit{NuSTAR} Data Analysis Software (NuSTARDAS) package v.1.3.1 (via the script {\tt nupipeline}), producing calibrated and cleaned event files.  Source data are extracted from a region of $45''$ radius, centered on the centroid of X-ray emission, while the background is extracted from a $1.5'$ radius region roughly $5'$ North of the source location.  Spectra are binned in order to have at least 20 counts per rebinned channel.  We consider the spectral channels corresponding nominally to the 3-30\,keV energy range, where the source was detected. The mean net (background-subtracted) count rates in the modules FPMA and FPMB are consistent with each other. 

We perform spectral fits to the data, using the standard \textit{NuSTAR} response matrices and effective area files, via the \textit{NuSTAR} data analysis package {\tt nuproducts}.  We adopt a simple power-law model, absorbed by intervening material with Solar abundances and the Galactic column of $1.55 \times 10^{20}$\,cm$^{-2}$ \citep[][]{Kalberla_2005}. In all cases, the spectra are adequately described by such a simple model, but in all three cases, we see a modest improvement to the spectral fit when we attempt a more complex model, such as log-parabola. While this is not significant in any of the three pointings reported here, this spectral behavior is fully consistent with that reported in \citet{Mrk501_MAGIC_2013}. 

\subsection{\textit{Swift}}
The study reported in this paper makes use of two instruments on
board the \textit{Neil Gehrels Swift Gamma-ray Burst Observatory}
\citep{2004ApJ...611.1005G}; namely the X-ray
Telescope \citep[XRT,][]{2005SSRv..120..165B} and the Ultraviolet/Optical Telescope \citep[UVOT, ][]{Roming05}. The related observations were organized and performed within the framework of
planned extensive multi-instrument campaigns on Mrk\,501, that occur yearly since 2008 \citep{Mrk501_MAGIC_2008}.
In this study we consider all observations within the years 2017 and 2020 for both the UVOT and XRT instruments with an extension using the all data from 2005 to 2020 for the long-term studies on the XRT light curve. 

The \textit{Swift}-XRT observations are performed in the Windowed Timing (WT) and Photon Counting (PC) readout modes, and the data are processed using the XRTDAS software package (v.3.5.0) developed by the ASI Space Science Data Center (SSDC), and released by the NASA High Energy Astrophysics Archive Research Center (HEASARC) in the HEASoft package (v.6.26.1). The calibration files from \textit{Swift}-XRT CALDB (version 20190910) are used within the  \texttt{xrtpipeline} to calibrate and clean the events. The X-ray spectrum from each observation is extracted from the summed cleaned event file. 
For WT readout mode data, events for the spectral analysis are selected within a circle of 20-pixel ($\sim46''$) radius, which contains about 90\% of the point spread function (PSF), centered at the source position. For PC readout mode data, the source count rate is above $\sim$0.5\, counts\,s$^{-1}$ and data are significantly affected by pile-up in the inner part of the PSF. We remove pile-up effects by excluding events within a 4-6 pixel radius circles centered on the source position and use an outer radius of 30 pixels. The background is estimated from a nearby circular region with a radius of 20 and 40 pixels for WT and PC data, respectively. The ancillary response files (ARFs) are generated with the \texttt{xrtmkarf} task applying corrections for PSF losses and CCD defects using the cumulative exposure map.
The $0.3-10$\,keV source spectra are binned using the \texttt{grppha} task to ensure a minimum of 20 counts per bin, and then are modelled in XSPEC using power-law and log-parabola models (with a pivot energy fixed at 1\,keV) that include the photoelectric absorption due to a neutral-hydrogen column density fixed to the Galactic 21-cm value in the  direction of Mrk\,501, namely 1.55 $\times$ 10$^{20}$ cm$^{-2}$ \citep{Kalberla_2005}.

The {\it Swift}-UVOT data analysis reported here relates only to all the observations with the UV filters (namely W1, M2 and W2) 
performed during the {\it Swift} pointings to Mrk\,501, 259 exposures. 
Differently to the optical bands, the emission in the UV is not affected by the emission from the host galaxy, which is very low at these frequencies. We perform aperture photometry for all filters using the standard UVOT software within the HEAsoft package (v6.23) and the calibration included in the latest release of the CALDB (20201026).
The source photometry is evaluated following the recipe in \citet{Poole08}, extracting source counts from a circular aperture of 5\arcsec\,radius, and the background ones from an annular aperture of 26\arcsec\, and 34\arcsec\, for the inner and outer radii in all filters. The count rates are converted to fluxes using the standard zero points \citep{brev11} and finally de-reddened considering an $E(B-V)$ value of 0.017 \citep{schlegel1998,Schlafly_2011} for the UVOT filters effective wavelengths and the mean galactic interstellar extinction curve from \citet{Fitzpatrick1999}. 

\subsection{Optical}
\label{subsec:inst_opt}
We focus on the R-band for the optical waveband, as it is often done in previous studies of Mrk\,501, and HSPs, in general. The optical data are collected within the GASP program of the Whole Earth Blazar Telescope (WEBT) \citep[][]{villata2008, villata2009, 2017MNRAS.472.3789C, Raiteri_2017, Gazeas_2016} including the instruments: West Mountain (91 cm), Vidojevica (140 cm), Vidojevica (60 cm), University of Athens Observatory (UOAO), Tijarafe (40 cm),  Teide (STELLA-I), Teide (IAC80), St. Petersburg, Skinakas, San Pedro Martir  (84 cm),  Rozhen (200 cm), Rozhen (50/70 cm), Perkins (1.8m), New Mexico Skies (T21), New Mexico Skies (T11), Lulin (SLT), Hans Haffner, Crimean (70cm; ST-7; pol), Crimean (70cm; ST-7), Crimean (70 cm; AP7), Connecticut (51 cm), Burke-Gaffney, Belogradchik,  AstroCamp (T7), Abastumani (70 cm). Additional data were provided by AAVSO and by the Tuorla observatory using the KVA telescope.

The data analysis is performed using standard prescriptions. The host galaxy contribution is subtracted according to the \citet{nilsson2007} recipe for an aperture of 7.5\,\arcsec, which was adopted by the participating instruments. The R-band flux is then corrected for Galactic extinction assuming the values reported by \citet{Schlafly_2011}. In order to account for instrumental (systematic) differences among the analyses related to the various telescopes (i.e., due to different filter spectral responses and analysis procedures, combined with the strong host galaxy contribution), offsets of a few mJy have to be applied. To calculate the corresponding offsets, KVA is used as a reference due to its good time coverage taking into account simultaneous data within two days. For data sets containing majorly data collected in 2020, when KVA was not operational anymore, Hans Haffner is used as the reference. The corresponding offsets can be found in Table~\ref{tab:Offset_optical}. To further account for instrumental (systematic) uncertainties, a relative error of $2\%$ is added in quadrature to the statistical uncertainties of all the flux values, as done in previous works \citep[][]{Mrk501_MAGIC_2012}. Afterwards, the data sets from all the instruments are combined into a single R-band light curve, and binned in 1-day time intervals. 

\subsection{Radio}
We report here radio observations from the single-dish telescopes at the Owens Valley Radio Observatory (OVRO) operating at 15 GHz, the
Medicina observatory, operating at 8\,GHz and 24\,GHz, RATAN-600 at 4.7\,GHz, 11.2\,GHz and 22\,GHz, the Mets\"{a}hovi Radio Observatory at 37\,GHz, IRAM at 100 GHz and 230\,GHz, and also the interferometry observations from VLBA at 43\,GHz and SMA at 230\,GHz and 345\,GHz.

For Mets\"ahovi the detection limit of the telescope at 37\,GHz is on the order of 0.2\,Jy under optimal conditions. Data points with a signal-to-noise ratio below 4 are handled as non-detections. The flux density scale is set by observations of DR 21. Sources NGC\,7027, 3C\,274 and 3C\,84 are used as secondary calibrators. A detailed description of the data reduction and analysis is given in \citet{1998A&AS..132..305T}. The error estimate in the flux density includes the contribution from the measurement root mean square and the uncertainty of the absolute calibration.  
The data from  OVRO and Medicina were analysed following the prescription from \citet{Richards_2011} and \citet{2020MNRAS.492.2807G}, and provided by the instrument teams specifically for this study. The flux density measurements with the RATAN-600 radio telescope were obtained at three frequencies, 22~GHz, 11.2~GHz, and 4.7~GHz, over several minutes per object in transit mode \citep{1993IAPM...35....7P,2020gbar.conf...32S}. The data reduction procedures and main parameters of the antenna and radiometers are described in e.g., \cite{2016AstBu..71..496U} and \cite{2017AN....338..700M}. Flux density data of Mrk 501 at 230~GHz and 345~GHz were obtained with the Submillimeter Array (SMA) as part of a monitoring program of mm band gain calibrators \citep[][]{Gurwell_2007}.  Sources are periodically observed for 3 to 5 minutes, and calibrated against known standards (primarily Titan, Uranus, Neptune, or Callisto).  The light curve data are updated regularly at the SMA website
\footnote{\url{http://sma1.sma.hawaii.edu/callist/callist.html}}.

Additional radio data have been taken by the Very Long Baseline Array (VLBA). Since June 2014, Mrk\,501 has been observed monthly with the VLBA at 43\,GHz within the VLBA-BU-BLAZAR program, which includes 38 $\gamma$-ray AGNs. Fully calibrated data of Mrk\,501 are posted at the program website\footnote{\url{http://www.bu.edu/blazars/VLBA\_GLAST/1652.html}}. The data reduction is described in \citet{Jorstad_2017}. The total VLBA intensity individual measurements during this period are compatible with a constant emission with an average flux value of $0.34 \pm 0.03$\,Jy, and the typical flux uncertainty is about 0.2\,Jy. For the sake of clarity, these values are not reported in Fig.~\ref{fig:MWL_LC}.

For the single-dish instruments, Mrk\,501 is a point source, and therefore the
measurements represent an integration of the full source extension. The size of the radio emitting region is expected to be larger than that of the region of the jet that dominates the X-ray and $\gamma$-ray emission, known to vary on much shorter timescales than the radio emission. However, as reported by \citet{2011ApJ...741...30A}, there is a correlation between the 8 and 15\,GHz radio and the GeV  emission of blazars. In the case of Mrk\,501, \citet{Mrk501_MAGIC_2009b} showed that the radio core emission increased during a period of high $\gamma$-ray activity. 
Therefore, a significant fraction of the radio emission seems to be related to the $\gamma$-ray component, at least during some periods of time, and hence should be considered when studying and interpreting the overall broadband emission of Mrk\,501.

\subsection{Optical and radio polarization}
Linear polarization measurements have been taken in both the optical and radio band. Through the GASP-WEBT program, optical polarization data using the R-band was obtained with the 70\,cm telescope at the Crimean Observatory and with the 1.8\,m telescope at the Perkins Observatory with an aperture of 7.5\,\arcsec. Moreover, R-band polarimetry data was collected by the Nordic Optical Telescope (NOT) using the ALFOSC instrument\footnote{\url{http://www.not.iac.es/instruments/alfosc}} with an aperture of 1.5\,\arcsec. Publicly available data from the Steward Observatory\footnote{\url{http://james.as.arizona.edu/~psmith/Fermi/DATA/data.html}} collected in the 5000–7000\,Å band using an aperture of 3\,\arcsec completes our optical polarization data set. 

The GASP-WEBT optical polarization data are corrected for interstellar 
polarization (ISP) using field stars \citep[number 6, 1, and 4 in][]{Villata1998}.
The host galaxy contribution is taken into account assuming a seeing of 2\,\arcsec and using Table~B.1. in \citet{nilsson2007}. The degree of polarization
is corrected for the host galaxy contribution according to the prescription given in \citet{Weaver2020}. The analysis of the NOT data 
is done as described in \citet{Hovatte2016} and \citet{MAGIC2018b}. Due to the small aperture (1.5\,\arcsec) no ISP correction is applied, but for the host galaxy the correction is done in the same way as mentioned above. The same is the case for the Steward  observatory data. Since the Stokes parameters are ambiguous with respect to 180$^\circ$  shifts in the electric vector polarization angle (EVPA), the data are corrected using 180$^{\circ}$  shifts if differences between neighboring measurements are more than +90 (or less than $-$90) for observations within 15 days.

For polarization information in the radio regime, data obtained by the VLBA are used. A brief description of the polarization analysis can be found in \citet{Marscher_2021}. The absolute EVPA calibration is performed using either quasi-simultaneous observations with the Very Long Array of the several objects from the sample twice per year, or the D-term method provided by \citet{Gomez_2002}. The degree of polarization and EVPA at 43\,GHz are obtained from Stokes I, Q, and U parameters integrated over Stokes corresponding images at each epoch. 

\section{Multiwavelength light curves }
\label{sect:MWL_lc}
\begin{figure*}
   \centering
   \includegraphics[height=2.1\columnwidth]{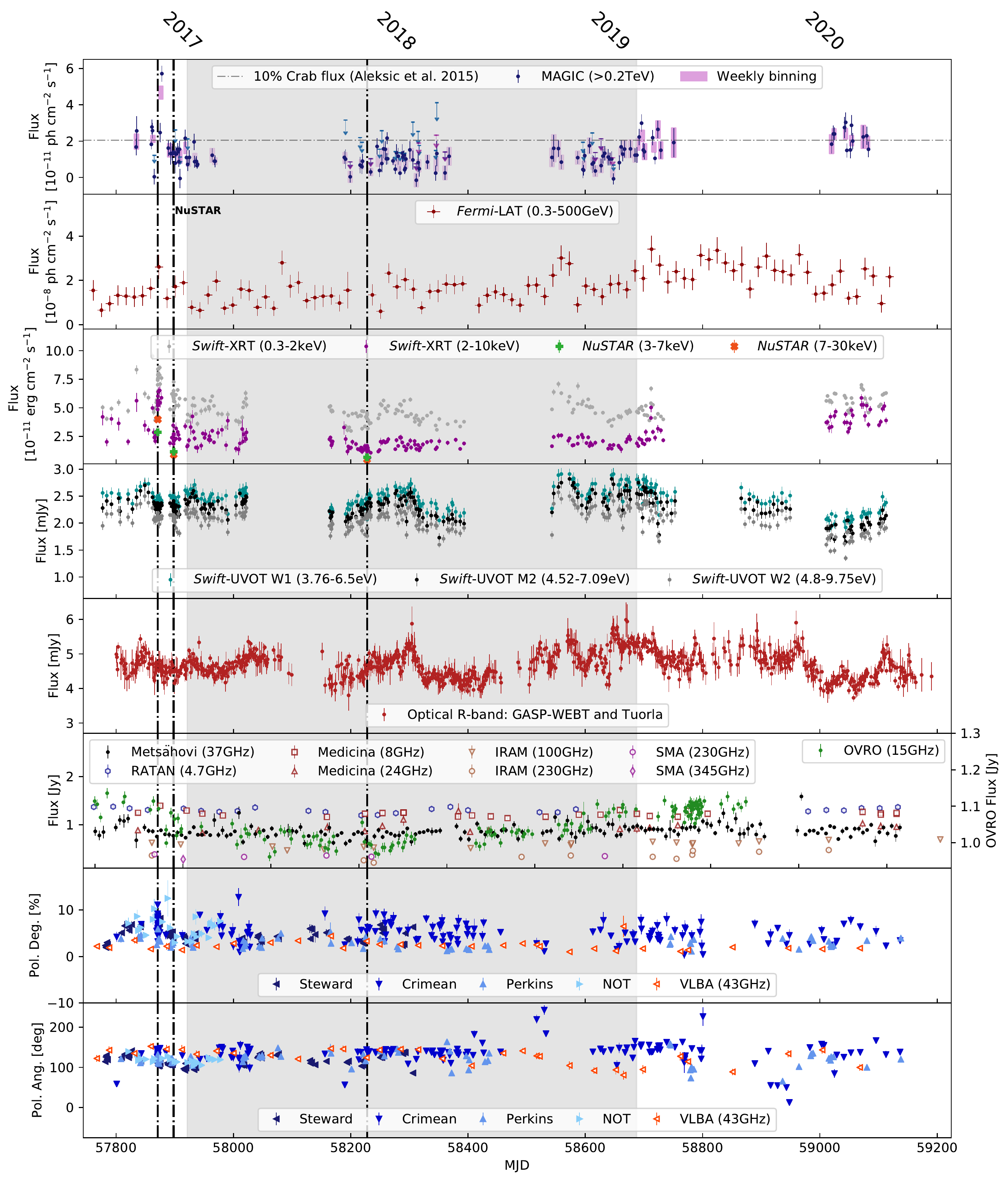}
   \caption{MWL light curve for the 4-year time interval, from MJD~57754 to MJD 59214. The grey area marks the identified very low-activity state spanning from 2017-06-17 to 2019-07-23 (MJD~57921 to MJD~58687), and the vertical dashed black lines depict the three long \textit{NuSTAR} observations conducted during the observing campaign on 2017-04-28, 2017-05-25 and 2018-04-20 (MJD~57871, MJD~57898, MJD~58228). Top to bottom: MAGIC fluxes in daily (blue markers) and weekly bins (violet markers) with the arrows additionally displaying the ULs for the non-significant bins ($<$2\,$\sigma$); \textit{Fermi}-LAT fluxes in 14\,day bins; X-ray fluxes in daily bins including \textit{Swift}-XRT and the three \textit{NuSTAR} observations; \textit{Swift}-UVOT; Optical R-band data from GASP-WEBT and Tuorla; Radio data including OVRO, Mets\"ahovi, Medicina, IRAM, RATAN-600 (for simplicity only the data points taken at 4.7\,GHz are shown, results taken at 22\,GHz and 11.2\,GHz are shown in Fig.~\ref{fig:lc_ratan}), SMA; polarization degree \& polarization angle observations in the optical R-band from Steward, Crimean, Perkins and NOT and the radio band from VLBA. See text in Section~\ref{sect:MWL_lc} for further details.}
    \label{fig:MWL_LC}%
\end{figure*}

Fig.~\ref{fig:MWL_LC} shows the collected MWL light curves from January 2017 until December 2020 (MJD 57754 to MJD 59214) complemented by the long term light curves starting in 2008 for certain wavebands shown in Fig.~\ref{fig:long_LC_comb}. 

In the VHE regime, the flux above 0.2~TeV varies around 10\% C.U. in the beginning of 2017, followed by a decline in the middle of 2017. It then stays below 10\% C.U. for two years until it rises again in the middle of 2019. We use a Bayesian block algorithm \citep{bb_2013ApJ...764..167S} on all significant measurements ($>$\,2$\sigma$) of the weekly binned MAGIC light curve to determine the start/stop of this time period of extremely low activity, yielding a single block with the lowest flux that spans from 2017-06-17 until 2019-07-23 (from MJD 57921 until MJD 58687). Using also the measurements below 2$\sigma$ with their bigger uncertainties, the same time interval is identified. During this 2-year low-state period, the VHE fluxes are consistent with a constant flux hypothesis ($\chi^2$/dof: $36/30$) with an average VHE flux of $0.99 \pm 0.05\times10^{-11}$\,ph\,cm$^{-2}$s$^{-1}$. This VHE flux corresponds to about 5\% C.U., and is less than 5 times lower than the typical (non-flaring) VHE flux of Mrk\,501, as shown in the top panel of Fig.~\ref{fig:long_LC_comb}.

A similar behavior is seen in the X-rays, where the flux decreases substantially during the years 2017--2019. For both VHE and X-rays this is an historically low activity, as clearly shown in the multi-year light curves from Fig.~\ref{fig:long_LC_comb}.

Additionally, three long \textit{NuSTAR} observations have been performed in the X-ray band. The first two on 2017-04-28 (MJD~57871) and 2017-05-25  (MJD~57898), and the last one on 2018-04-20 (MJD~58228), during the very low activity. For all three observations we investigated their intra-night behavior as shown in Fig.~\ref{fig:nustar_intranight}. The observations were performed during different flux states. In the 3--7~keV band, the flux ranges from $\sim 0.6$ to $3.0 \times 10^{-11}$ erg cm$^{-2}$ s$^{-1}$, which is about 5 times lower than the X-ray fluxes from the previous \textit{NuSTAR} observations between April and July 2013, which ranged from $\sim 3.7$ to $12.1 \times 10^{-11}$ erg cm$^{-2}$ s$^{-1}$ in the 3--7~keV band \citep{Mrk501_MAGIC_2013}.
Within each of the three above-mentioned pointings, we did not detect any source variability from one orbit to another (orbit-to-orbit variability $<~\sim$5\%) and we find no change in the hardness ratio of the source as a function of time. This indicates that there was no
 significant flux or spectral variability on the time scales from hours to a half-day during any of the three observations considered here. The spectra from each of the three \textit{NuSTAR} observations are reported in Table~\ref{tab:NuSTAR_data}. We further note that the shape of the X-ray spectra from the four \textit{NuSTAR} observations of Mrk\,501 in 2013 were generally harder than the ones reported here.  This is consistent with the general behavior of HSP-type blazars (such as Mrk\,501) of 'harder when brighter' X-ray spectral correlation. 

For the other wavebands shown in Fig.~\ref{fig:MWL_LC}, no remarkable behavior can be distinguished comparing the 2-year low-activity period to the full 4-year time interval. However, from the long-term light curves in Fig.~\ref{fig:long_LC_comb}, one can see that the $\gamma$-rays measured with \textit{Fermi}-LAT and the radio emission measured with OVRO also report a flux level that is substantially lower than the typical ones, during the previous years. 

Considering the previously observed 'harder when brighter' behavior in Mrk\,501 \citep[][]{Mrk501_2005,Mrk501_MAGIC_2009b,Mrk501_MAGIC_2012}, we had a look at the spectral parameters in the different wavebands. For both MAGIC and \textit{Fermi}-LAT, no significant variations in the spectral behavior can be distinguished, the spectral shape appears to remain constant throughout the 4-year period using the 14-day/7-day binning for the LAT/MAGIC light curve. This, however, might be due to the relatively low flux and therefore limited sensitivity of these instruments to determine small changes in the spectral shape. On the other hand, in the X-ray and the UV range, where the sensitivity of the instruments is better than for $\gamma$-rays, a 'harder when brighter' behavior is measured, as shown in Fig.~\ref{fig:hardness}.

\subsection{Fractional variability}
\label{subsec:var_fvar}
The fractional variability $F_{var}$ is used, as given by Eq. 10 in \citet{2003MNRAS_frac_var}, to estimate the degree of variability in each waveband.  Its uncertainty is defined using the estimates and descriptions in \citet{2008MNRAS_deltaf_var} and
\citet{Mrk501_MAGIC_2008}, resulting in Eq.~2 in \citet{Mrk501_MAGIC_2008}.

\begin{figure}
  \resizebox{\hsize}{!}{\includegraphics{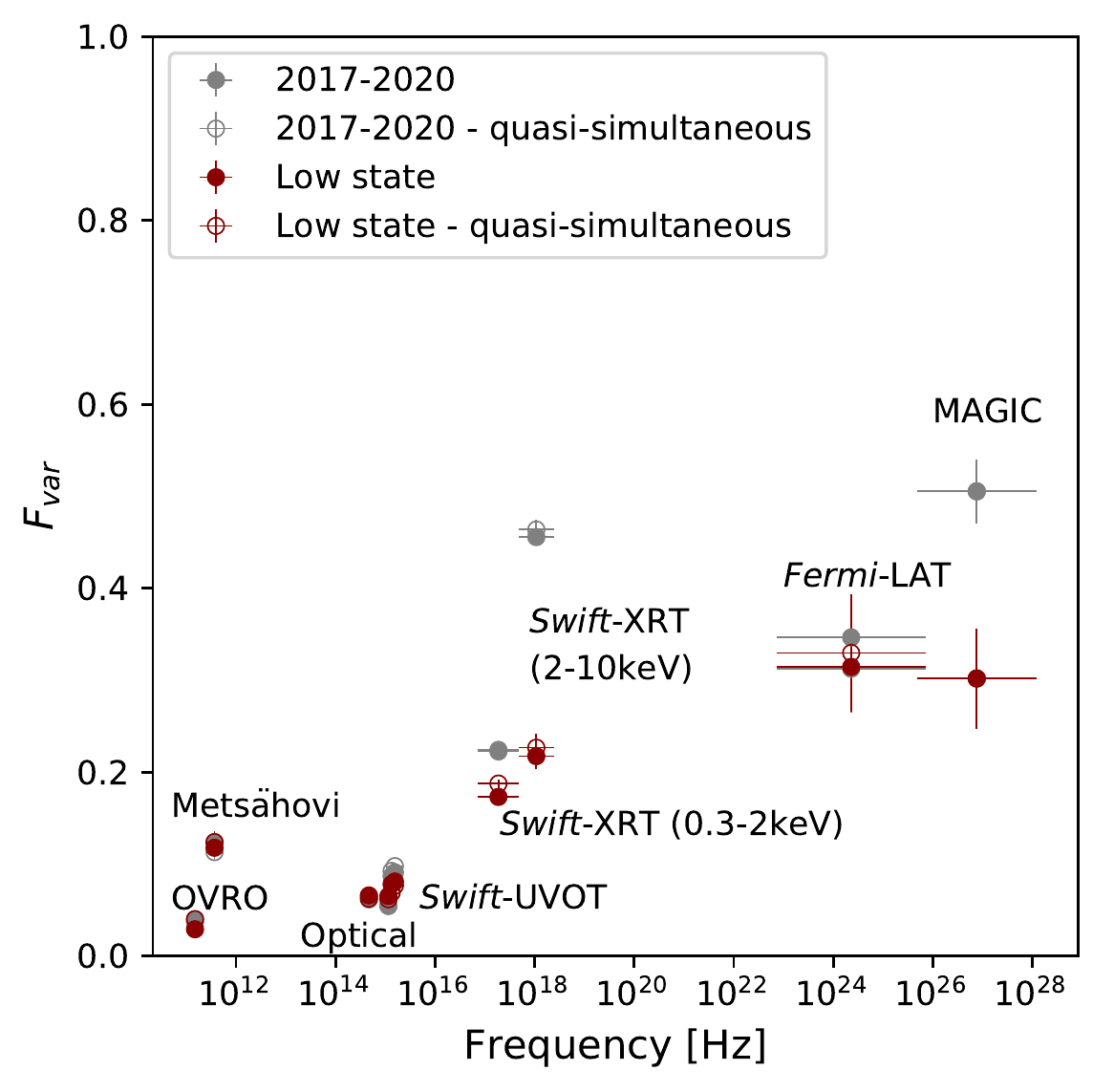}}
  \caption{Fractional variability $F_{var}$ for the light curves displayed in Fig.\,\ref{fig:MWL_LC} as described in Section~\ref{subsec:var_fvar}. Daily flux bins are used for all instruments, apart from MAGIC and  \textit{Fermi}-LAT, where 7-day and 14-day flux bins are used, respectively. $F_{var}$ is computed for the 4-year time interval (grey) and the 2-year low-activity period (red). Open markers depict the results for only quasi-simultaneous data to the weekly binned MAGIC observations while filled markers show the whole data set.}
  \label{fig:Fvar}%
\end{figure}

The fractional variability is computed for the light curves shown in Fig.~\ref{fig:MWL_LC} for both the whole 4-year period, as well as the 2-year low-activity period. The results are displayed in Fig.~\ref{fig:Fvar} for both the whole data set as well as only quasi-simultaneous data to the weekly binned MAGIC observations. Table~\ref{tab:Fvar} summarizes the results for the whole data set. Most light curve measurements used for this study are retrieved from daily bins, except \textit{Fermi}-LAT, with its 14-day binning, and MAGIC, with its 7-day binning. The 7-day binning is chosen for this analysis to mitigate the impact of the low flux levels, and hence the low number of significant single-night measurements (especially during the above-mentioned 2-year low-state period). However, when using a 1-day binning for the MAGIC fluxes, and discarding the non-significant fraction of flux measurements, the obtained fractional variability values are $F_{var}^{\text{2017 to 2020}} = 0.52\pm 0.03$, and $F_{var}^{\text{low-state}} = 0.25\pm 0.05$. These are very similar to values reported in Table~\ref{tab:Fvar} obtained with the 7-day binning. We note that the differences in the temporal bins used to characterize the variability may affect the comparability of $F_{var}$ between different wavebands. 

\begin{table}
\centering
\begin{tabular}{ c | c  c }     
 & Full data set (4-y) & Low-state (2-y) \\
 & 2017 to 2020 & 2017.5 to 2019.5 \\
\hline   
MAGIC &  0.505 $\pm$ 0.035 & 0.302 $\pm$ 0.055 \\
LAT & 0.347 $\pm$ 0.031 & 0.314 $\pm$ 0.047 \\
XRT (2-10keV) &  0.456 $\pm$ 0.009 & 0.217 $\pm$ 0.014 \\
XRT (0.3-2keV)  & 0.222 $\pm$ 0.003 & 0.173 $\pm$ 0.004 \\
UVOT W2 & 0.091 $\pm$  0.002  & 0.081 $\pm$ 0.003 \\
UVOT M2 &  0.087 $\pm$ 0.003 &  0.078 $\pm$ 0.003 \\
UVOT W1 & 0.054 $\pm$  0.003 & 0.065 $\pm$ 0.003 \\
Optical R-band & 0.062  $\pm$ 0.002 & 0.066$\pm$ 0.002 \\
Mets\"ahovi &  0.124  $\pm$  0.005 & 0.118 $\pm$ 0.007 \\
OVRO  &  0.040  $\pm$ 0.001  & 0.029 $\pm$ 0.002 \\
\end{tabular}
\caption{Fractional variability $F_{var}$ for the light curves displayed in Fig.~\ref{fig:MWL_LC}. 
Daily flux bins are used for all instruments, apart from MAGIC and  \textit{Fermi}-LAT, where 7-day and 14-day flux bins are used, respectively. $F_{var}$ is computed separately for the 4-year time interval and the 2-year low-activity period.} 
\label{tab:Fvar}
\end{table}

For the two time intervals considered, the radio, optical and UV frequencies show only mild flux variability. As reported in previous works \citep{Mrk501_MAGIC_2013,Mrk501_MAGIC_2012}, Mets\"ahovi shows a larger variability compared to the other radio data. In the case of the 2-year low-state, the fractional variability increases with energy after the eV regime, but then it reaches a plateau for the GeV and TeV regime. On the other hand, the 4-year epoch shows a two peak structure with the 2-10\,keV data showing a similar variability level as the VHE data. 

\subsection{Correlations}
\label{sec:correlations}
\begin{table*}
\centering
\begin{tabular}{ c c c c c c c c c }  
\hline\hline   
 MAGIC & \textit{Fermi}-LAT & \textit{Swift}-XRT & \textit{Swift}-UVOT & Optical R-band & Mets\"ahovi & OVRO  & Pol. Deg. & Pol Ang. \\
 \hline\hline   
 7\,days & 14\,days & 3.5\,days &  3.5\,days & 3.5\,days & 3.5\,days & 7\,days & 1\,days & 3.5\,days  \\ 
\hline 
\end{tabular}
\caption{Binning used for the correlation analysis described in Section~\ref{subsec:var_corr} for the different wavebands} 
\label{tab:DCF_binning}
\end{table*}
\label{subsec:var_corr}


\begin{figure*}
\gridline{\fig{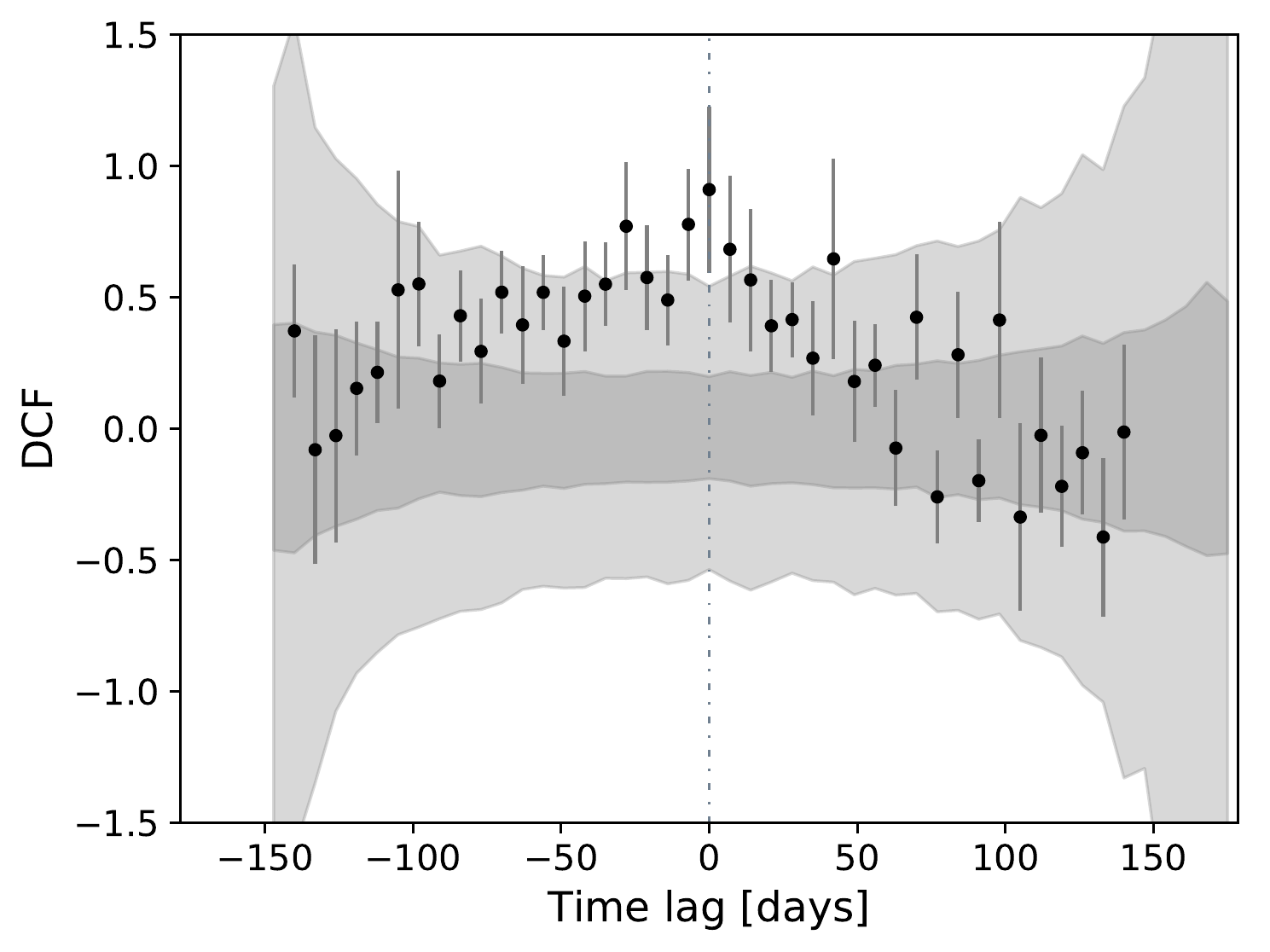}{0.45\textwidth}{(a) \textit{ MAGIC ($>$0.2\,TeV) vs. Swift}-XRT (2-10\,keV)} 
          \fig{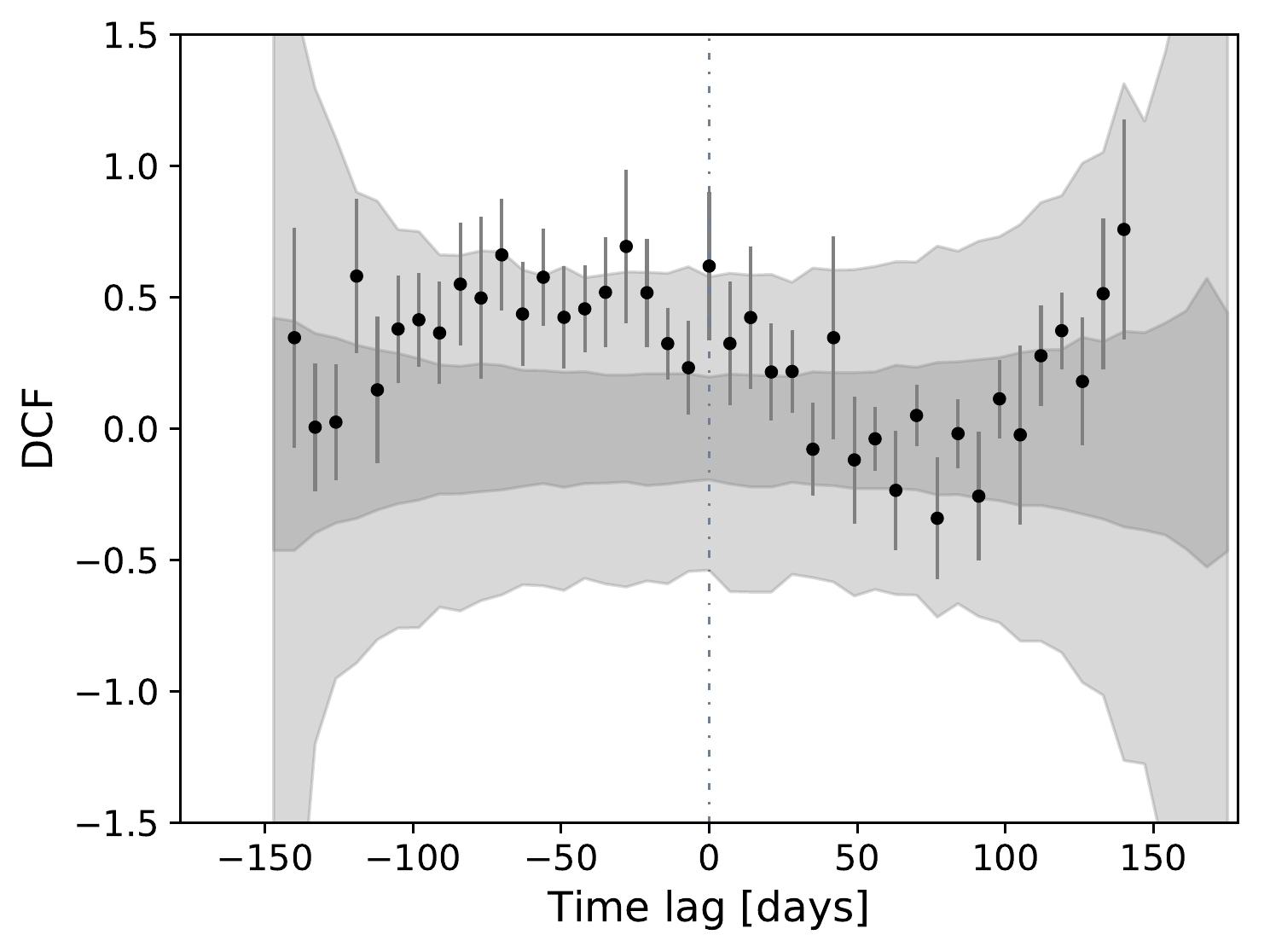}{0.45\textwidth}{(b) \textit{ MAGIC ($>$0.2\,TeV) vs. Swift}-XRT (0.3-2 \,keV) }
}
\caption{Discrete correlation function DCF computed between MAGIC and different energy ranges of \textit{Swift}-XRT light curves shown in Fig.~\ref{fig:MWL_LC} using a binning of 7\,days. It is computed for different time shifts, time lags, applied to the LCs. The 1$\sigma$ and 3$\sigma$ confidence levels obtained by simulations  as described in Section~\ref{subsec:var_corr} are shown by the dark and light grey bands, respectively.}
\label{fig:dcf_1}
\end{figure*}

Conventional methods to compute correlations between data sets are often hard to apply to astrophysical light curves due to the differences in sampling, varying uncertainties and the often long gaps between observations. The discrete correlation function (DCF) provides an assumption-free representation of the correlation without interpolating in time \citep{1988ApJ_DCF}. \par

To compute the DCF between two light curves, an appropriate time binning is chosen and applied to both light curves in the same manner, yielding pairs of flux data points. We choose the appropriate binning for each light curve by considering the typical decay times in the auto-correlation distributions of each waveband. Additionally, the compatibility of the chosen bins with the original binning of the different light curves is taken into account. The time bins used for the correlation analysis for all instruments can be found in Table~\ref{tab:DCF_binning}. We then compute the DCF using Eq. 4 in \citet{1988ApJ_DCF} if the number of flux pairs is bigger than 10. In order to identify delayed correlations, we shift the light curves with respect to each other for certain time intervals, and then repeat the correlation computations. For the UVOT data, we choose to select only the W1 filter for the correlation analyses because the light curves of all three UV filters show very similar behavior. An easy way of estimating  the significance of the DCF values may be exploited by using the 1$\sigma$ errors as defined in Eq. 5 in \citet{1988ApJ_DCF}. However, when the light curves include correlated red-noise data, which is the case for blazars, these 1$\sigma$ may not be suitable to determine the significance of the correlation correctly. In these situations, the calculated significance could be overestimated, as discussed in  \citet{2003ApJ...584L..53U}. Therefore, in order to better assess the reliability of the significance in our correlations,  we use dedicated Monte Carlo simulations that take into account the actual sampling and flux measurement uncertainties. This strategy, described further below, allows us to determine  1$\sigma$ and 3$\sigma$ confidence levels that apply for the specific light curves probed, and hence provide a robust evaluation of whether a correlation is significant or not. For the purpose of comprehensibility and completeness, both the 1$\sigma$ errors from Eq. 5 in \citet{1988ApJ_DCF} and the dedicated MC-derived 1$\sigma$ and 3$\sigma$ contours are displayed in our results.
\par

\begin{figure}
    \resizebox{\hsize}{!}{   \includegraphics{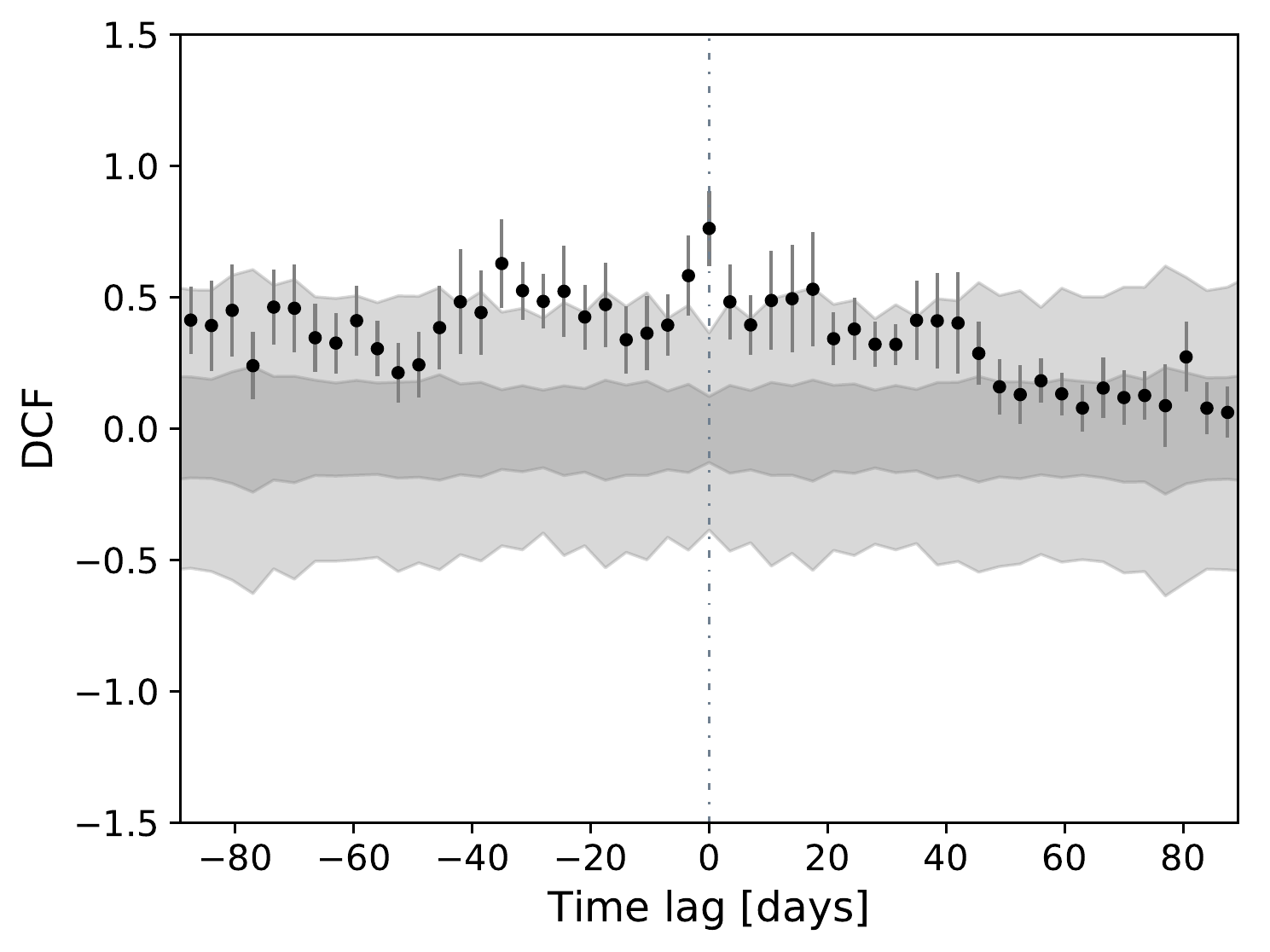}}
    \caption{Discrete correlation function DCF computed for the Swift-XRT (2-10 keV) and Swift-XRT (0.3-2 keV) light curves shown in Fig.~\ref{fig:MWL_LC} using a binning of 3.5\,days. It is computed for different time shifts, time lags, applied to the LCs. The 1$\sigma$ and 3$\sigma$ confidence levels obtained by simulations  as described in Section~\ref{subsec:var_corr} are shown by the dark and light grey bands, respectively.}  
    \label{fig:dcf_xrt-le_xrt-he}
\end{figure}


\begin{figure*}
\gridline{\fig{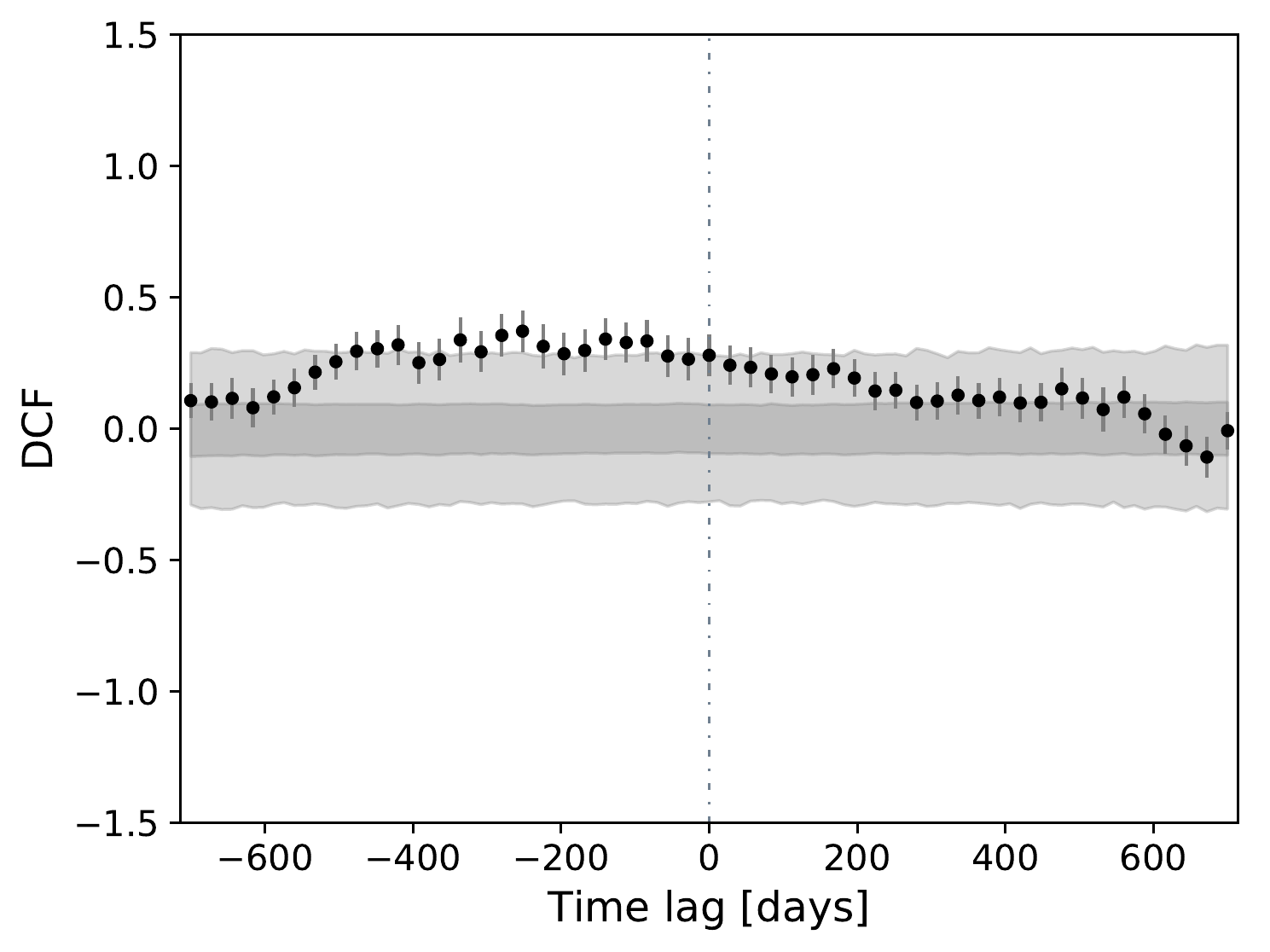}{0.45\textwidth}{(a) DCF computed using the untreated light curves from Fig.~\ref{fig:long_LC_comb}.} 
          \fig{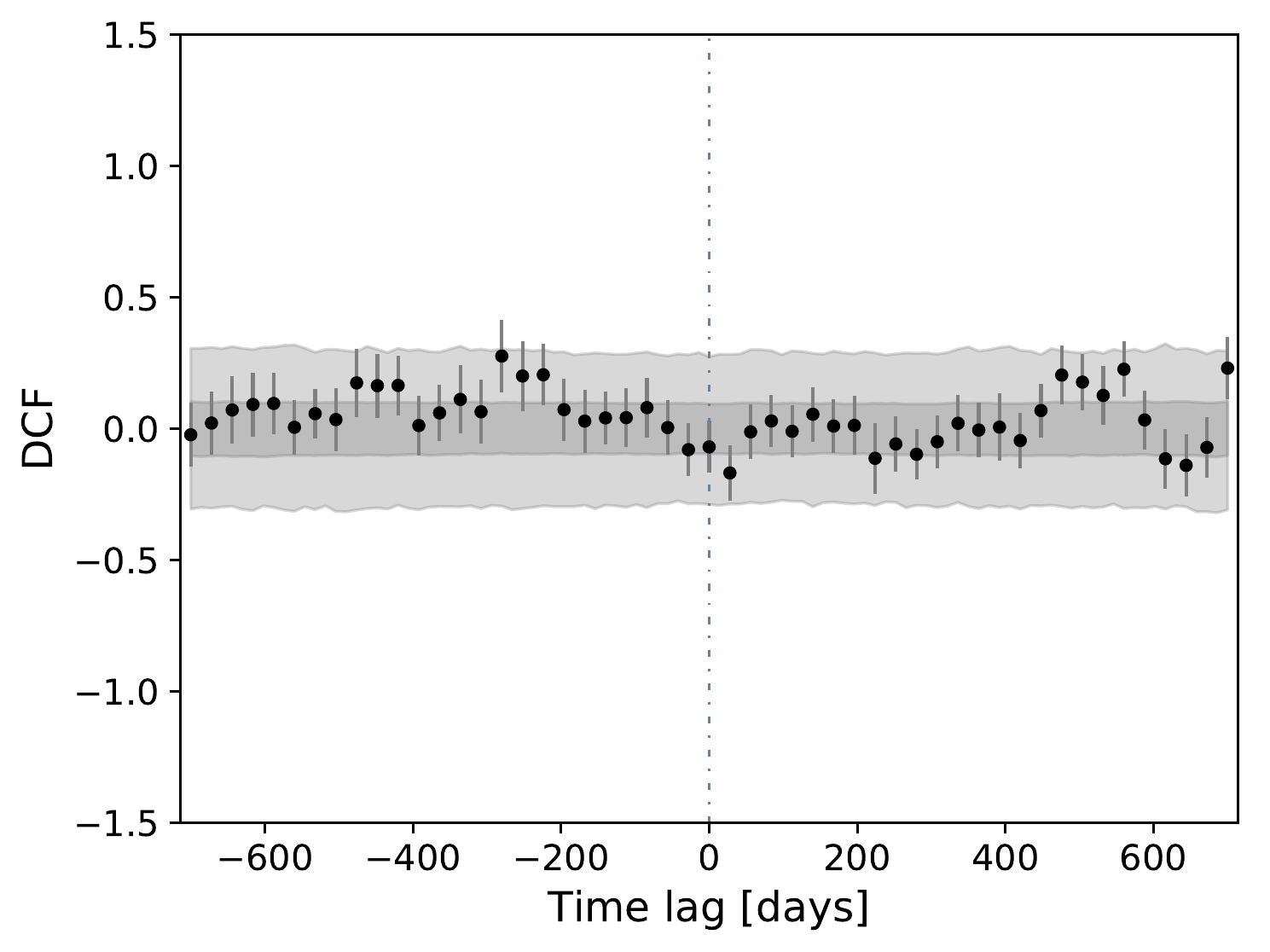}{0.45\textwidth}{(b) Same as panel (a), but with the light curves detrended, as described in Section~\ref{subsec:var_corr}, before computing the correlations.}
}
\caption{Discrete correlation function DCF computed for the \textit{Fermi}-LAT (0.3-500\,GeV) and OVRO (15\,GHz) light curves from the 12-year data set (2008–2020) shown in Fig.~\ref{fig:long_LC_comb} using a binning of 14\,days. It is computed for different time shifts, time lags, applied to the LCs. The 1$\sigma$ and 3$\sigma$ confidence levels obtained by simulations are shown by the dark and light grey bands, respectively. }
\label{fig:dcfs_1}
\end{figure*}

\begin{figure*}
\gridline{\fig{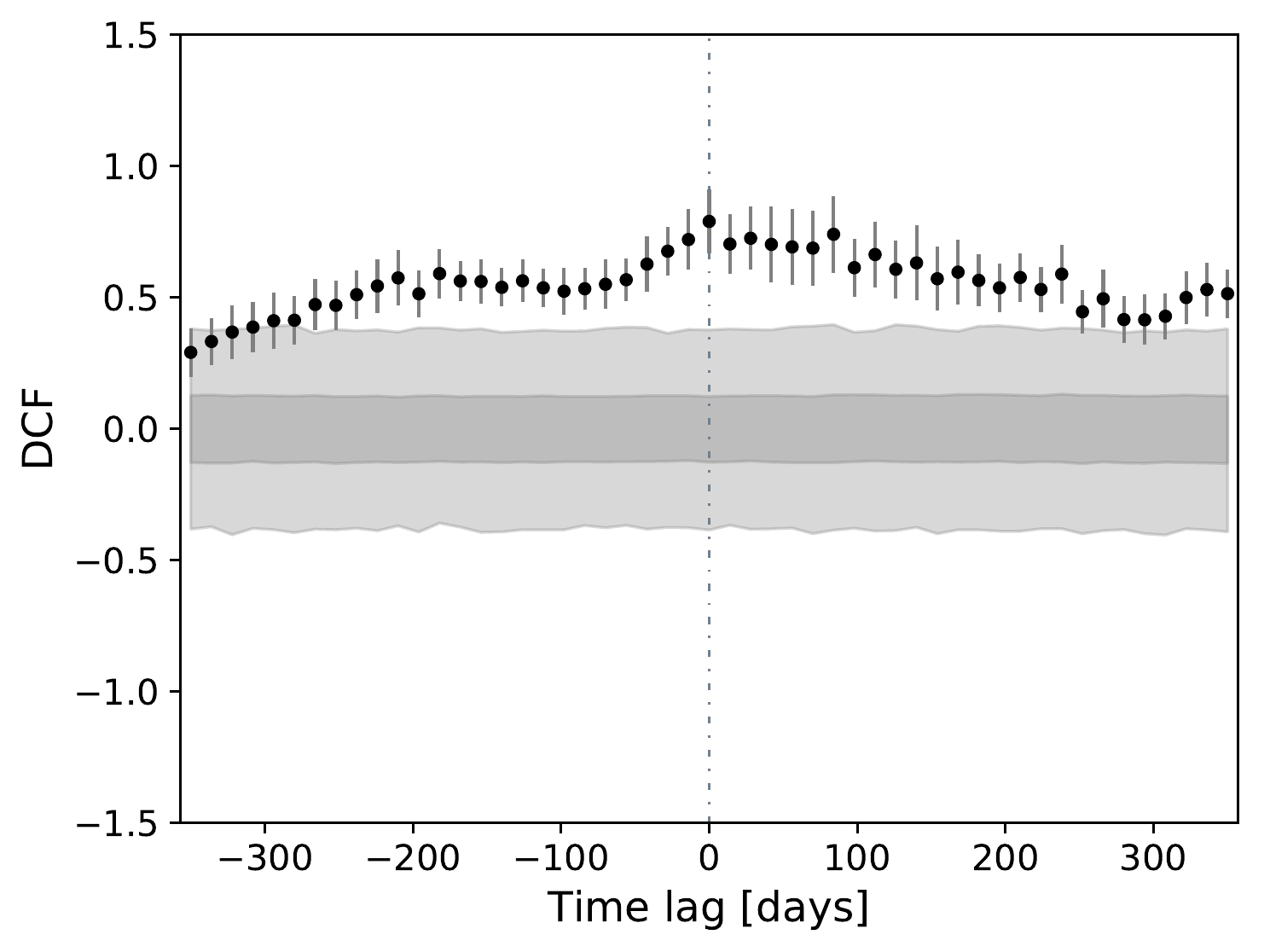}{0.45\textwidth}{(a) DCF computed for the untreated \textit{Fermi}-LAT (0.3-500\,GeV) vs. \textit{Swift}-XRT (2-10\,keV) light curves from Fig.~\ref{fig:long_LC_comb}.}
          \fig{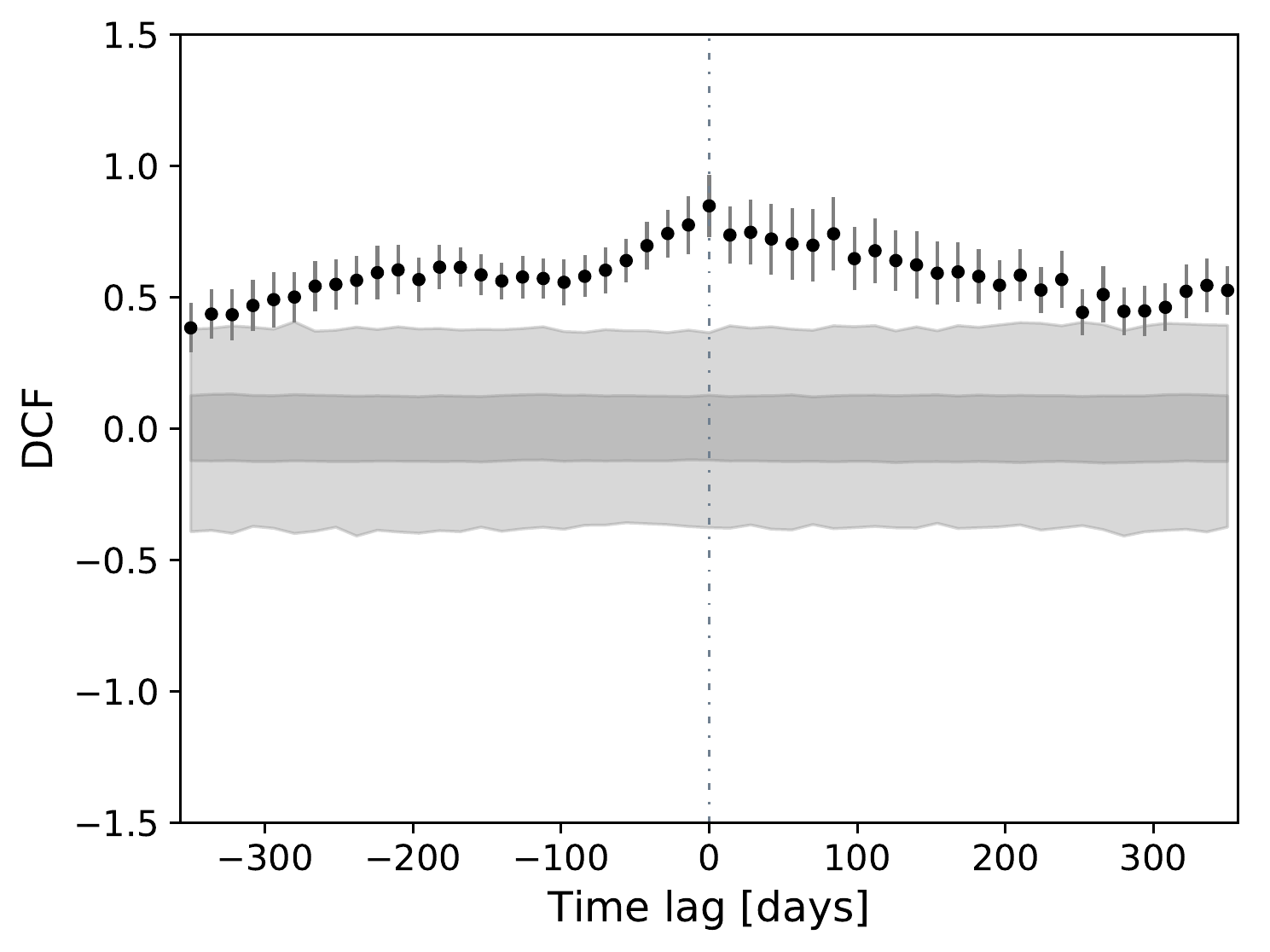}{0.45\textwidth}{(b) DCF computed for the untreated \textit{Fermi}-LAT (0.3-500\,GeV) vs. \textit{Swift}-XRT (0.3-2\,keV) light curves from Fig.~\ref{fig:long_LC_comb}.}
}
\gridline{\fig{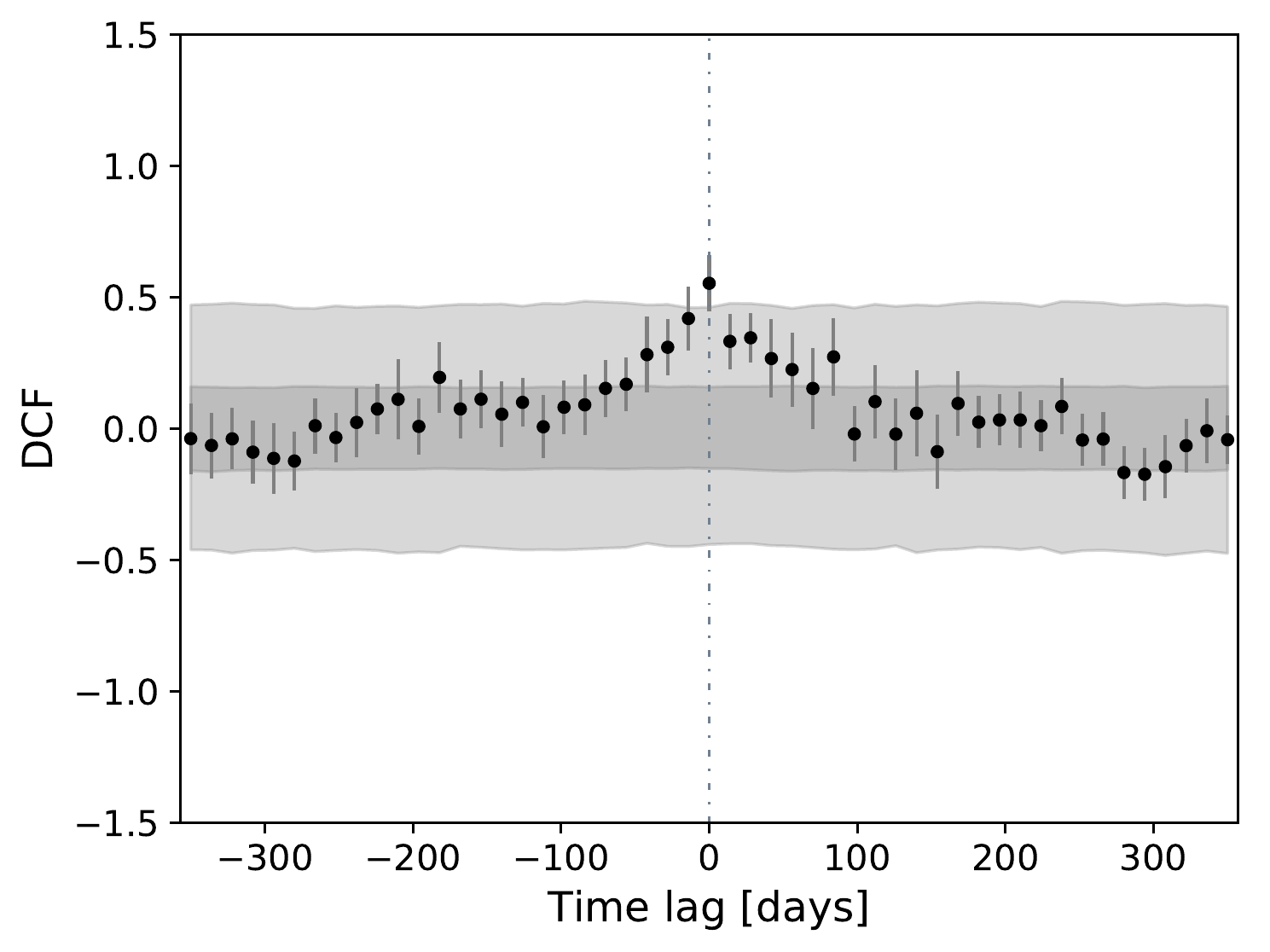}{0.45\textwidth}{(c) Same as panel (a), but with the light curves detrended, as described in Section~\ref{subsec:var_corr}, before computing the correlations.} 
          \fig{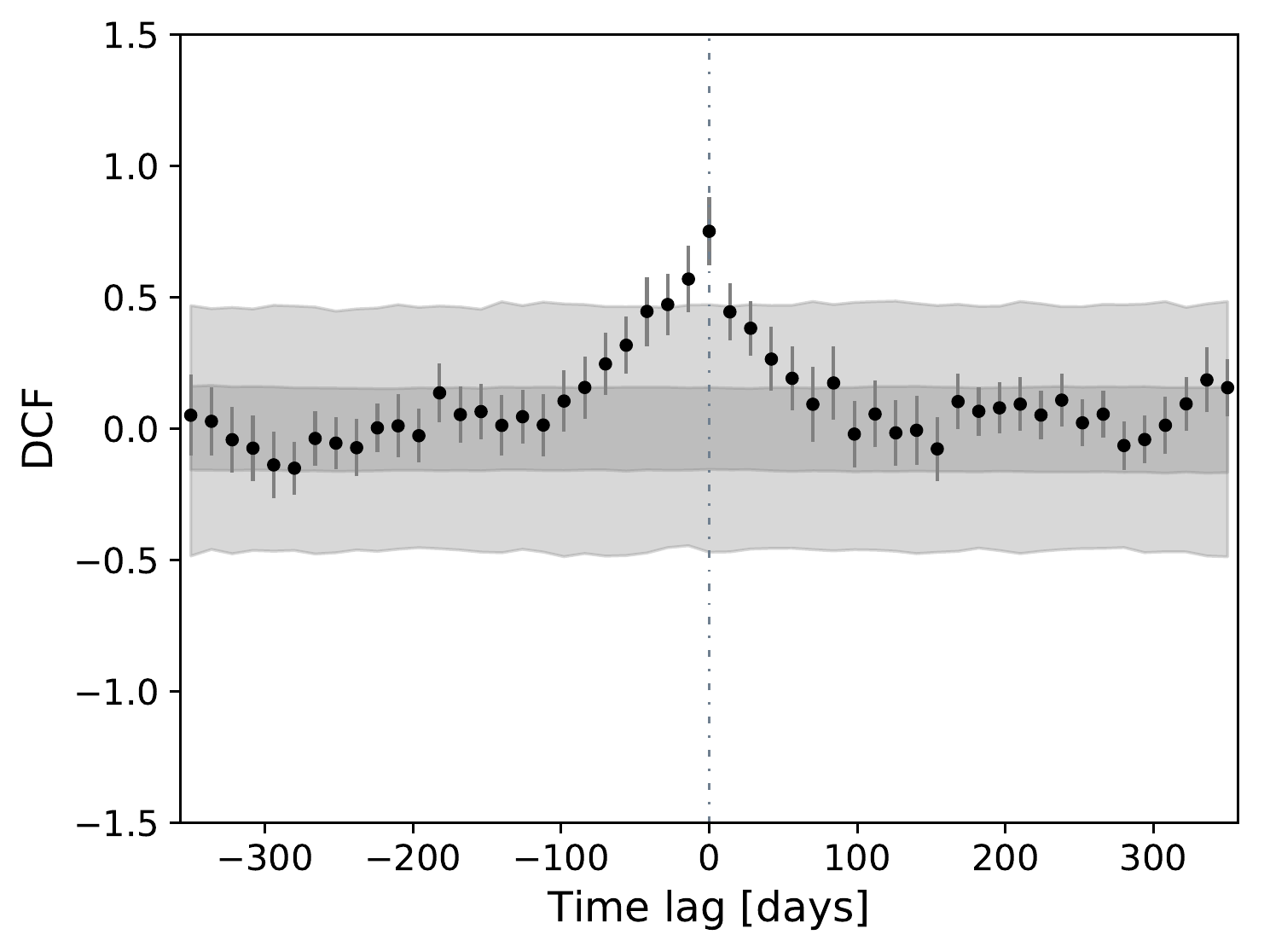}{0.45\textwidth}{(d) Same as panel (b), but with the light curves detrended, as described in Section~\ref{subsec:var_corr}, before computing the correlations.}
}
\caption{Discrete correlation function DCF computed for the \textit{Fermi}-LAT (0.3-500\,GeV) and different energy ranges of \textit{Swift}-XRT light curves of the 12-year data set (2008–2020) shown in Fig.~\ref{fig:long_LC_comb} using a binning of 14\,days. It is computed for different time shifts, time lags, applied to the LCs. The 1$\sigma$ and 3$\sigma$ confidence levels obtained by simulations are shown by the dark and light grey bands, respectively.}
    \label{fig:dcfs_lt}
\end{figure*}

In order to estimate the statistical significance of the DCF measurements, we simulate 10,000 uncorrelated light curves for each waveband using the DELCgen package\footnote{Connolly, S. D., 2016, Astrophysics Source Code Library, record ascl:1602.012} \citep{2013MNRAS_DELCgen}. The underlying power density spectrum (PDS) of the original light curve is used to reproduce the level of variability in the simulations and the same time sampling as for the original data sets is applied. We choose a power-law as the function to estimate the PDS after no significant preferences for more complicated functions are seen in our evaluations. Since our flux distributions are mostly compatible with Gaussian distributions, we select the Timmer\&Koenig method \citep{1995A&A_TimmerKoenig} to simulate the light curves. Uncertainties on the simulated flux points are estimated by using the minimum error of the original light curve as a minimum error, and drawing random additional errors from a distribution constructed from the original relative error distributions. This for example results in a minimum error of $\mathbf{1.1 \times 10^{-12}}$\,ph\,cm$^{-2}$\,s$^{-1}$ being added to all simulated MAGIC data points complemented by additional errors drawn from a distribution with a distribution peaked around $\mathbf{2.0\times10^{-12}}$\,ph\,cm$^{-2}$\,s$^{-1}$. For the optical data, a minimum error of 0.06\,mJy is added, with additional errors being drawn from a distribution peaking around 0.1\,mJy with a tail towards higher values.
For the \textit{Fermi}-LAT light curve, the original relative error distribution shows a different distribution at higher than at lower flux level. Therefore the original relative error distribution is divided into two samples using the flux level of 1.3$\times 10^{-8}$\,ph\,cm$^{-2}$s$^{-1}$ as a limit where the differences in error distributions appear. For each simulated flux point, the relative error is then drawn using the sample corresponding to its flux level as the weight. For all other wavebands no dependency on the flux level is found. Afterwards, the same binning, time shifting and correlation calculation methods as for the real light curves are applied to the simulated ones. The resulting DCF distributions allow us to derive confidence levels at which random correlation from originally uncorrelated light curves can be excluded. The computed significance contours derived in this manner consider the specific sampling and flux measurement errors of the two light curves being used, and hence are a reliable evaluation of the significance of our correlations. However, when the same computation is performed a large number, N, of times, as we do when computing the DCF for different time lags, one should also consider the "look elsewhere effect". The effect could artificially increase the chance probability of obtaining, for a random time lag $\Delta T$, a DCF value above the 3$\sigma$ contour. One could potentially correct for this effect by considering the trial factors, as described in \citet{Gross_2010,Algeri_2016}. However, the "look elsewhere effect" does not affect the DCF obtained for physically meaningful time lags, such as $\Delta T=0$, or general trends, as e.g. the broader peaks spanning over many consecutive time lags.

A 3$\sigma$-correlation (DCF value $>$ 3$\sigma$ confidence level) at zero time lag is found between the high energy (HE) X-ray (2-10\,keV) range, measured with \textit{Swift}-XRT, and the VHE $\gamma$-ray ($>$0.2~TeV) range, measured with MAGIC, as shown in Fig.~\ref{fig:dcf_1}a). Additionally, a 3$\sigma$-correlation in the form of a structure can be seen with XRT lagging behind MAGIC peaking at a time lag of 28\,days. For the low energy (LE) X-ray range (0.3-2\,keV), the same two features can be seen in Fig.~\ref{fig:dcf_1}b). We note here that, while significant VHE-X-ray correlations have always been observed in the emission of Mrk\,501 during flaring activities, such a correlation has always been elusive during periods of very low activity \citep[e.g.][]{Aleksi2015a,Mrk501_MAGIC_2009b}. The improved sensitivity to characterize the VHE emission, in comparison to past measurements, and the extensive data set collected during these 4 years, made possible the measurement of a correlated behaviour with a significance above 3$\sigma$ during a period of historically low VHE and X-ray activity. 

Similarly, when comparing the two energy ranges in X-rays with each other, as shown in Fig.~\ref{fig:dcf_xrt-le_xrt-he}, 3$\sigma$-correlations are seen with a clear peak at time lag 0 and a broader structure around time lag -35\,days, suggesting that at least a fraction of the low energy photons are lagging behind the more energetic ones. 


\begin{figure*}
\gridline{\fig{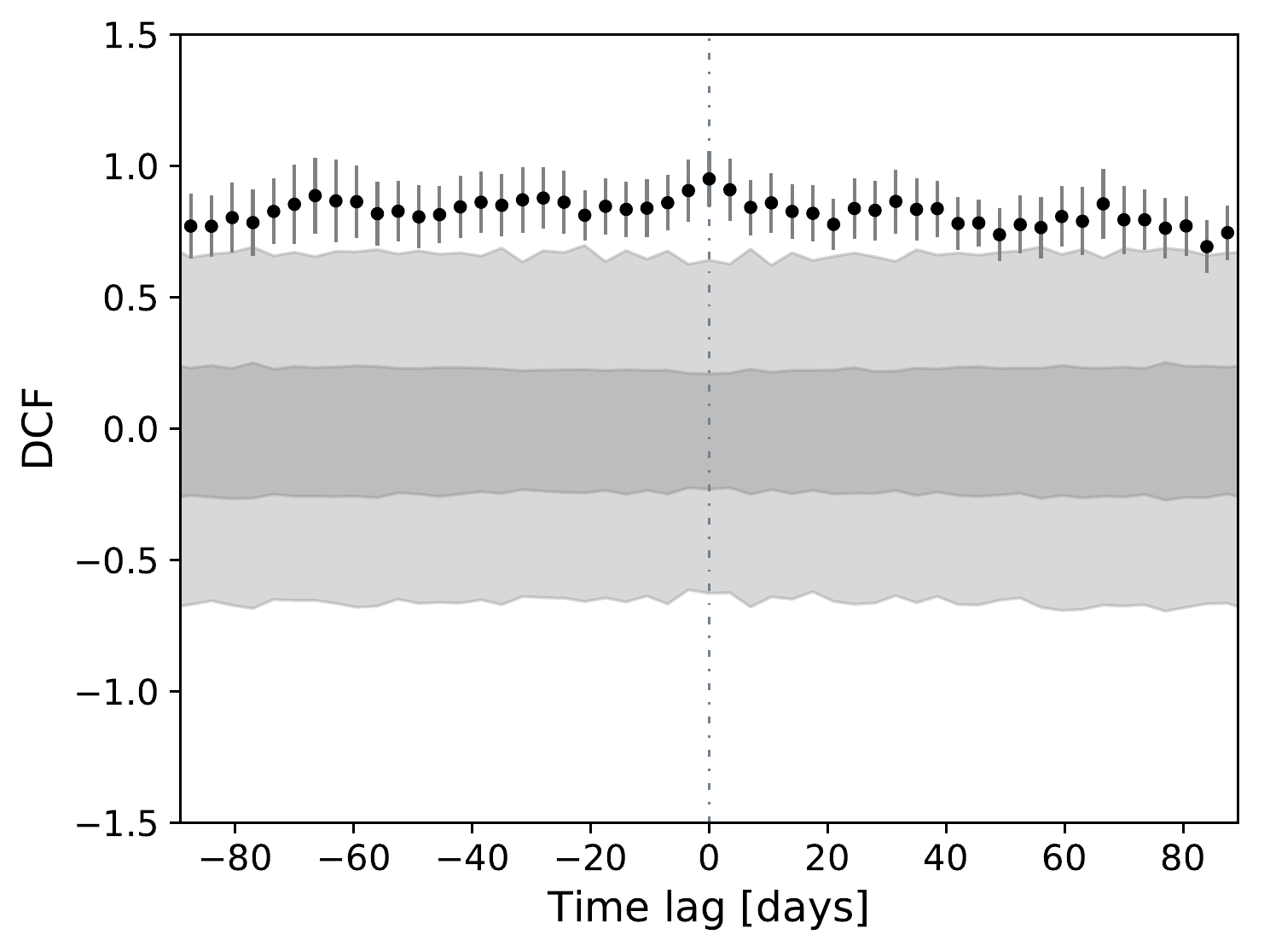}{0.45\textwidth}{(a) DCF computed using the untreated light curves from Fig.~\ref{fig:long_LC_comb}.}
          \fig{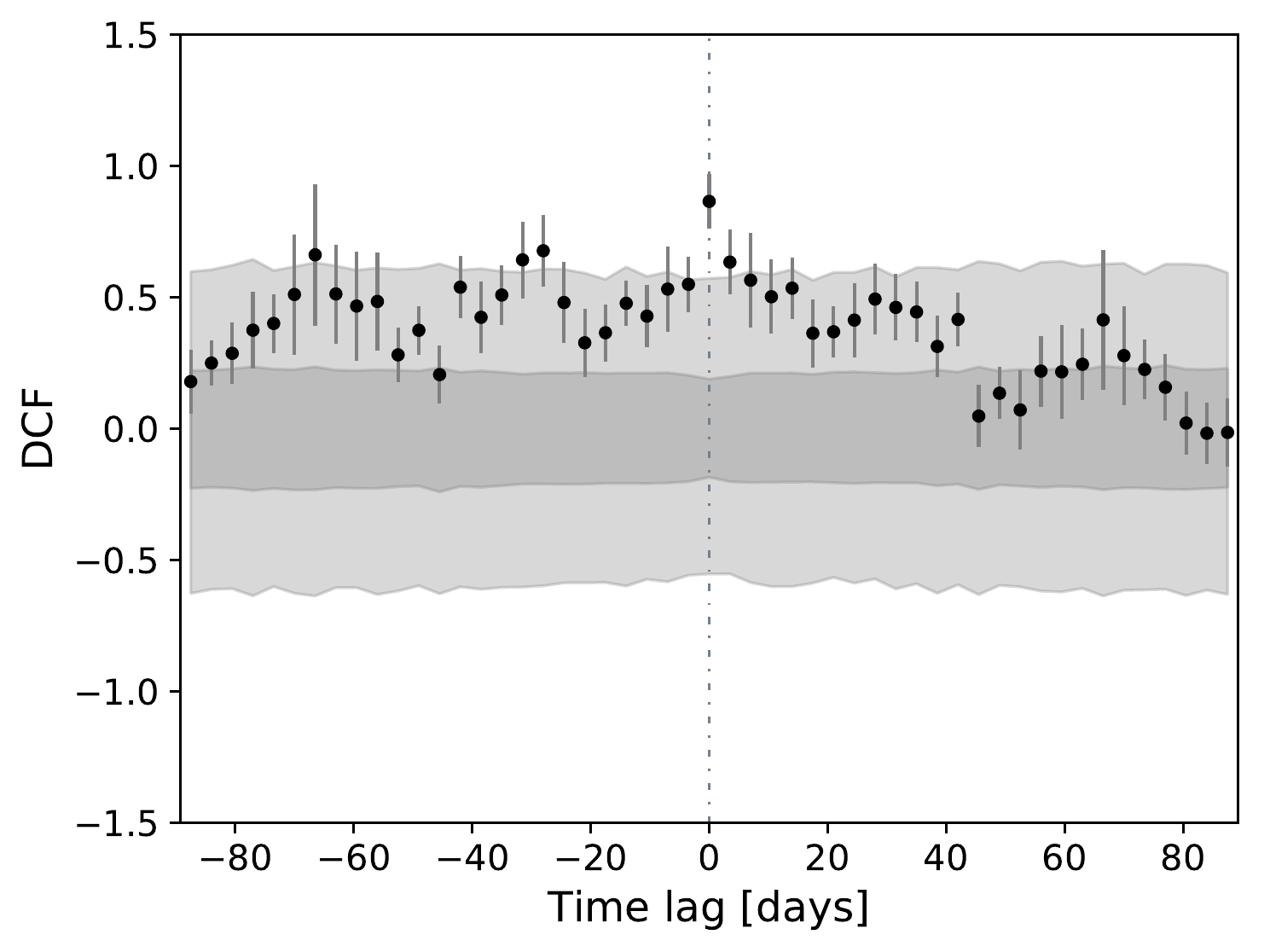}{0.45\textwidth}{(b) Same as panel (a), but with the light curves detrended, as described in Section~\ref{subsec:var_corr}, before computing the correlations.}
}

\caption{Discrete correlation function DCF computed for the \textit{Swift}-XRT (2-10\,keV) vs. \textit{Swift}-XRT (0.3-2\,keV) light curves from the 12-year data set (2008–2020) shown in Fig.~\ref{fig:long_LC_comb} using a binning of 3.5\,days. It is computed for different time shifts, time lags, applied to the LCs. The 1$\sigma$ and 3$\sigma$ confidence levels obtained by simulations are shown by the dark and light grey bands, respectively.}
    \label{fig:dcfs_detrend_xrt}
\end{figure*}

Between OVRO and MAGIC, a 3$\sigma$-correlation is found for most positive time lags, as shown in Fig.~\ref{fig:dcfs_2}c). Since these correlations vanish when detrending the light curves as described further below in more detail, we can characterize them as long term features. Other 3$\sigma$-correlations can be seen for a few negative time lags. But they occur simultaneous with an increase of the confidence levels (due to the smaller number of data pairs used in the computation), and hence we cannot say if they are physically meaningful, or due to potentially non-accounted uncertainties due to the limitations in the time range used. We also observe a marginally significant ($<$$\sim$$3\sigma$) correlation when comparing the HE X-rays with OVRO where the DCF values rise for positive time lags with two values reaching the 3$\sigma$ level at 42 days and 84 days, as shown in Fig.~\ref{fig:dcfs_2}a) (see also Table \ref{tab:DCF}). This behavior is not seen when using the LE band of the X-ray data set, as shown in Fig.~\ref{fig:dcfs_2}b). These correlations disappear when de-trending the light curves, and hence they can be ascribed to the long-term behaviour in the light curves at HE X-rays and radio.

\begin{table*}
\centering
\begin{tabular}{ c c c  c c }     
\multicolumn{2}{l}{\textbf{4-year data set (2017--2020)}} & & &\\
\hline 
Instrument 1 & Instrument 2 & Time lag [days]  & DCF  & 3$\sigma$ confidence level DCF \\
\hline \hline 
MAGIC ($>$0.2\,TeV) & \textit{Swift}-XRT (2-10\,keV)  &  0 & 0.91
 $\pm$ 0.32  & 0.54  \\
MAGIC ($>$0.2\,TeV) & \textit{Swift}-XRT (2-10\,keV)  &  -28 & 0.77  $\pm$ 0.24  & 0.59 \\
\hline   
MAGIC ($>$0.2\,TeV) & \textit{Swift}-XRT (0.3-2\,keV) &  0 & 0.62 $\pm$ 0.28  &  0.58 \\
MAGIC ($>$0.2\,TeV) & \textit{Swift}-XRT (0.3-2\,keV)  &  -28 & 0.70 $\pm$ 0.29 & 0.60 \\
\hline   
\textit{Swift}-XRT (2-10\,keV)  &  \textit{Swift}-XRT (0.3-2\,keV)  &  0 & 0.76  $\pm$ 0.14   &  0.37 \\
\textit{Swift}-XRT (2-10\,keV)  &  \textit{Swift}-XRT (0.3-2\,keV)  &  -35 & 0.63  $\pm$ 0.17   & 0.45 \\
\hline  
MAGIC ($>$0.2\,TeV) &  OVRO (15\,GHz)  &  28	& 1.08  $\pm$ 0.26  &  0.72 \\
MAGIC ($>$0.2\,TeV) &  OVRO (15\,GHz)  &   49 &	0.97 $\pm$  0.30 & 0.75 \\
\hline  
\textit{Fermi}-LAT (0.3-500\,GeV)  & OVRO (15\,GHz)  &   -112 & 0.92  $\pm$ 0.24  & 0.67  \\
\hline 
\textit{Swift}-XRT (2-10\,keV) &  OVRO (15\,GHz)  &  42	& 0.64 $\pm$  0.21  &  0.62 \\
\textit{Swift}-XRT (2-10\,keV) &  OVRO (15\,GHz) &  84	& 0.69 $\pm$ 0.21   & 0.69 \\
\hline  
UVOT W1 (3.76-6.5\,eV) & Optical (R-band) &  0 & 0.93  $\pm$ 0.11   &  0.7 \\
\hline 
\textit{Fermi}-LAT (0.3-500\,GeV) & Optical (R-band) &  0  &  0.47 $\pm$  0.14  &  0.6 \\
\hline  
OVRO (15\,GHz)  & Mets\"ahovi (37\,GHz)  &   -28 & 0.67  $\pm$ 0.16  & 0.63  \\
\multicolumn{5}{c}{} \\
\multicolumn{2}{l}{\textbf{12-year data set (2008--2020)}} & & &\\
\hline
Instrument 1 & Instrument 2 & Time lag [days]  & DCF  & 3$\sigma$ confidence level DCF \\
\hline   \hline
\textit{Fermi}-LAT (0.3-500\,GeV)  &  \textit{Swift}-XRT (2-10\,keV)  &  0	 & 0.79	$\pm$ 0.12 & 0.38  \\
\textit{Fermi}-LAT (0.3-500\,GeV)  &  \textit{Swift}-XRT (0.3-2\,keV)  &  0	 & 0.85	$\pm$ 0.12 & 0.36  \\
\hline
\textit{Swift}-XRT (2-10\,keV)   &  \textit{Swift}-XRT (0.3-2\,keV)  &  0	 & 0.95	$\pm$ 0.11 & 0.65  \\
\textit{Swift}-XRT (2-10\,keV)   &  \textit{Swift}-XRT (0.3-2\,keV)  &  -28	 & 0.88	$\pm$ 0.12 & 0.68 \\
\hline
\textit{Fermi}-LAT (0.3-500\,GeV)  & OVRO (15\,GHz)  &  -126	& 0.39 $\pm$ 0.08  & 0.28 \\
\textit{Fermi}-LAT (0.3-500\,GeV)  & OVRO (15\,GHz)  &  -238	 & 0.4	$\pm$ 0.09 & 0.30  \\
\multicolumn{5}{c}{} \\
\multicolumn{2}{l}{\textbf{Detrended 4-year data set (2017--2020)}} & & &\\
\hline
Instrument 1 & Instrument 2 & Time lag [days]  & DCF  & 3$\sigma$ confidence level DCF \\
\hline   \hline
MAGIC ($>$0.2\,TeV) & \textit{Fermi}-LAT (0.3-500\,GeV)  &  0	 & 0.67	 	$\pm$ 0.42  & 0.63  \\
\multicolumn{5}{c}{} \\
\multicolumn{2}{l}{\textbf{Detrended 12-year data set (2008--2020)}} & & &\\
\hline
Instrument 1 & Instrument 2 & Time lag [days]  & DCF  & 3$\sigma$ confidence level DCF \\
\hline   \hline
\textit{Fermi}-LAT (0.3-500\,GeV)  &  \textit{Swift}-XRT (2-10\,keV)  &  0	 & 0.55	$\pm$ 0.11 & 0.46 \\
\textit{Fermi}-LAT (0.3-500\,GeV)  &  \textit{Swift}-XRT (0.3-2\,keV)  &  0	 & 0.75	$\pm$ 0.13 &  0.47 \\
\hline
\textit{Swift}-XRT (2-10\,keV)   &  \textit{Swift}-XRT (0.3-2\,keV)  &  0	 & 0.87	$\pm$ 0.10 &  0.57 \\
\textit{Swift}-XRT (2-10\,keV)   &  \textit{Swift}-XRT (0.3-2\,keV)  &  -28	 & 0.68	$\pm$ 0.14 & 0.61  \\
\end{tabular}
\caption{Values for the discrete correlation function DCF computed as described in Section~\ref{subsec:var_corr} for different pairs of the light curves shown in Fig.~\ref{fig:long_LC_comb} and Fig.~\ref{fig:MWL_LC}, and with the binning described in Table~\ref{tab:DCF_binning}. The DCF values and the respective 3$\sigma$ contour boundary are reported for the relevant time
lags that are discussed in the text. } 
\label{tab:DCF}
\end{table*}

Other 3$\sigma$-correlations are found between \textit{Fermi}-LAT and OVRO with a time delay of 112\,days by OVRO, as displayed in Fig.~\ref{fig:dcfs_3}a). Since for both OVRO and \textit{Fermi}-LAT data have been taken continuously since 2008, we expanded the analysis to their long-term light curves displayed in Fig.~\ref{fig:long_LC_comb} to refine our picture of the correlation. A similar DCF distribution can be seen in Fig.~\ref{fig:dcfs_1}a), but this time with smaller statistical uncertainties, due to the much larger data set spanning over a time window that is three times larger. The highest degrees of correlation occur at time lags of 126 and 238\,days, but there is a 3$\sigma$-correlation that persistently occurs for negative time lags larger than 3 months.

We additionally took advantage of the multi-year long observations of Mrk\,501 with \textit{Swift}-XRT, and quantified the correlation between X-rays and $\gamma$-rays using the XRT and LAT data from the period 2008--2020.  Interestingly, while the 2017-2020 data set does not show any correlation between X-ray and $\gamma$-ray fluxes, the 12-year data set reveals a very clear correlation between these two bands, as shown in   Fig.~\ref{fig:dcfs_lt}. We note that the correlation exists for both XRT energy bands, 0.3--2~keV and 2--10~keV. The main difference between the 4-year and the 12-year data sets is easily explained. In the longer data set, Mrk\,501 did show several periods of extended high activity, and hence it becomes easier to measure a correlated behaviour between the fluxes. We note that the highest DCF values occur at a time lag zero, but with a very broad and somewhat asymmetric bump with a 3$\sigma$-correlation, that extends over a time lag of $\pm$1 year. This indicates that the measured correlation is dominated by the flux variations with timescales of about 1 year, such as the substantial decrease in the activity during the period 2017-2020, in comparison with the activity during the period 2008--2016. A very similar and expected behavior can be seen for the long-term correlation between the LE X-rays and the HE X-rays (Fig. \ref{fig:dcfs_detrend_xrt}a)).

Since the correlations in our 12-year data set are dominated by long-term (month to years) time scales, we additionally apply a detrending as described in Section~5.1. in \citet{Lindfors_2016} to the light curves. We slightly adapt step 2 of the procedure by not using the variance of the original high-energy light curve to scale the variance of the polynomial fit to the low-energy data, but rather use the variance of a polynomial fit to the high-energy data for the scaling. In this way we are matching the variability of the long term behaviors of both light curves. For our data set this is especially relevant, since the large flares also include very variable behavior on short time scales in the X-ray range (Fig.~\ref{fig:long_LC_comb}), we would otherwise over-correct the light curves leading to negative dips at these variable time epochs. The procedure allows us to subtract the long term behavior from our light curves and provides us with a view on the shorter time scales and correlations.

When the long-term behaviour is removed from the light curves, one sees that the correlation between the $\gamma$-ray and radio light curves disappears, as shown in Fig.~\ref{fig:dcfs_1}b). However, this is not the case for the correlation between the $\gamma$-ray and X-ray fluxes, as shown in Fig.~\ref{fig:dcfs_lt}c) and  Fig.~\ref{fig:dcfs_lt}d), where prominent bumps centered at time lag zero appear. When considering the correlation between the two X-ray energy bands, namely 0.3--2~keV and 2--10~keV (see Fig,~\ref{fig:dcfs_detrend_xrt}b)), the detrended light curves yield a DCF plot with a clear correlation peak centered at time lag zero, but also some indications of peaks that repeat with a period of about 30 days. The potential periodicity in the X-ray emission of Mrk\,501 is discussed further in Section~\ref{sec:periodicity}.


For the rest of our 4-year data set, we see a clear correlation between the UV and the R-band (see Fig.~\ref{fig:dcfs_3}b)). The highest DCF values occur at a time lag zero, but the peak extends over $\pm$1 week, due to the relatively slow flux variations in these two bands. The correlation analysis between the R-band and the HE $\gamma$-ray emission measured with \textit{Fermi}-LAT does show a small DCF peak at about time lag of zero, but the DCF values do not reach the 3$\sigma$ significance level (see Fig.~\ref{fig:dcfs_3}c)).

We did not find a 3$\sigma$-correlation for any other combination of untreated (not detrended) light curves, except for a marginally significant correlation with a delay of -28\,days between the two long-term radio data sets from OVRO and Mets\"ahovi (see  Fig.~\ref{fig:dcfs_3}d)). However, if we apply the detrending described above to our 4-year data set, we can identify a 3$\sigma$-correlation between the VHE and HE $\gamma$-rays as depicted in Fig.~\ref{fig:dcf_magic_fermi_detrend}.

Table~\ref{tab:DCF} reports the DCF values and the 3$\sigma$ contour limits for the various energy band combinations and the relevant time lags that were discussed above in the text. 
%

As for the polarization data, we do not find any significant correlation between the polarization degree and the flux levels, neither for optical nor for the 43\,GHz radio (see Fig.~\ref{fig:pol_deg_corr}). Concerning the EVPA, in the optical R-band and the radio frequency probed, 43\,GHz, the preferred values fluctuate around 130 degrees, and they are independent (not correlated) of the polarization degree (see Fig.~\ref{fig:pol_ang_optical}a) and Fig.~\ref{fig:pol_ang_optical}b)). Due to 180$^\circ$ ambiguity and the corrections applied, some of the EVPA values are above 180$^\circ$ for the optical data. Here the Von Mises distribution can be used to check the accurate description of the data, which results in a mean value of $133.7\pm 17.1~^\circ$. This matches the EVPA angle of $134\pm 5~^\circ$ reported from the first polarization measurements in the X-rays of Mrk\,501 \citep{IXPE_Mrk501} as well as the measured jet direction of 119.7 $\pm$ 11.8 $^\circ$ from \citet[][]{Weaver_2022}. When comparing the polarization properties in the optical and radio regime with each other, no correlation can be seen between the polarization degree and angle (see Fig.~\ref{fig:pol_corr_opt_rad}).

\subsection{Periodicity}
\label{sec:periodicity}
Both the auto-correlation of the LE and HE X-ray ranges as well as their correlation with each other (Fig. \ref{fig:dcfs_detrend_xrt}) indicate the possibility of a periodic behavior of $\sim$30\,days. Previously for the large 1997 flare a periodicity both in the VHE and the X-ray data was claimed at a time scale of 23\,days \citep{Osone_2006,Kranich_1999} and explained with a binary black hole system as the center of Mrk\,501 \citep[][]{Rieger_2000}. Additionally, a periodicity of 332\,days was reported in \citet[][]{Bhatta_2019} for the \textit{Fermi}-LAT data set with a significance of 99.4\,\%, slightly below the 3$\sigma$ level, as well as at 195\,days with a significance slightly above 90\%.

Exploiting the availability of our 12-year data set, we apply the Lomb–Scargle periodogram (LSP) \citep[][]{Lomb_1976,Scargle_1982}. Details on the implementation and computation can be found in Appendix Section \ref{sec:app_period} as well as on the significance estimation which is based on the same simulations as described in Section~\ref{sec:correlations}.

No significant ($>$\,3$\sigma$) periodicity can be detected for the two X-ray energy ranges (see  Fig.~\ref{fig:LSP_xrt}). All peaks in the signal LSP including the one at $f\sim1/30$\,days$^{-1}$ are accompanied by corresponding rises in the confidences limits and the signal does not go above the 3$\sigma$ confidence limits. The peaks can be attributed to the time sampling of the X-ray light curve which is coordinated together with the Earth based telescopes (see details in Appendix Section \ref{sec:app_period}). Similarly, for the VHE light curve no periodicity can be identified.

Furthermore, no significant periodicity is found for the \textit{Fermi}-LAT (Fig.~\ref{fig:LSP2}a)) or OVRO (Fig.~\ref{fig:LSP2}b)) data. However, we can reproduce the reported weak hint for a '$330$-day' periodicity in the \textit{Fermi}-LAT light curve at a significance of 2.3\,$\sigma$ as well as at $\sim 200$\,days at 2.2\,$\sigma$.

\citet{Oneill_2022} laid out a standard in which they proposed a 3$\sigma$ global significance cutoff, which clearly, Mrk\,501 does not meet. For this reason, we interpret all claimed periodicities in Mrk\,501 as red noise for now, although we will continue to monitor the source for periodicities that reach our cutoff threshold.


\section{Characterization and theoretical modeling of the broadband  SED}
\label{sec:spect}
As described in Section~\ref{sect:MWL_lc}, Mrk\,501 was found at an historically low activity level lasting for more than two years. The detailed multi-instrument coverage of Mrk\,501 during this 2-year-long period of low activity, which includes three \textit{NuSTAR} observations, enables us to investigate the nature of the low-state as well as its prior evolution.

\begin{figure*}
\gridline{\fig{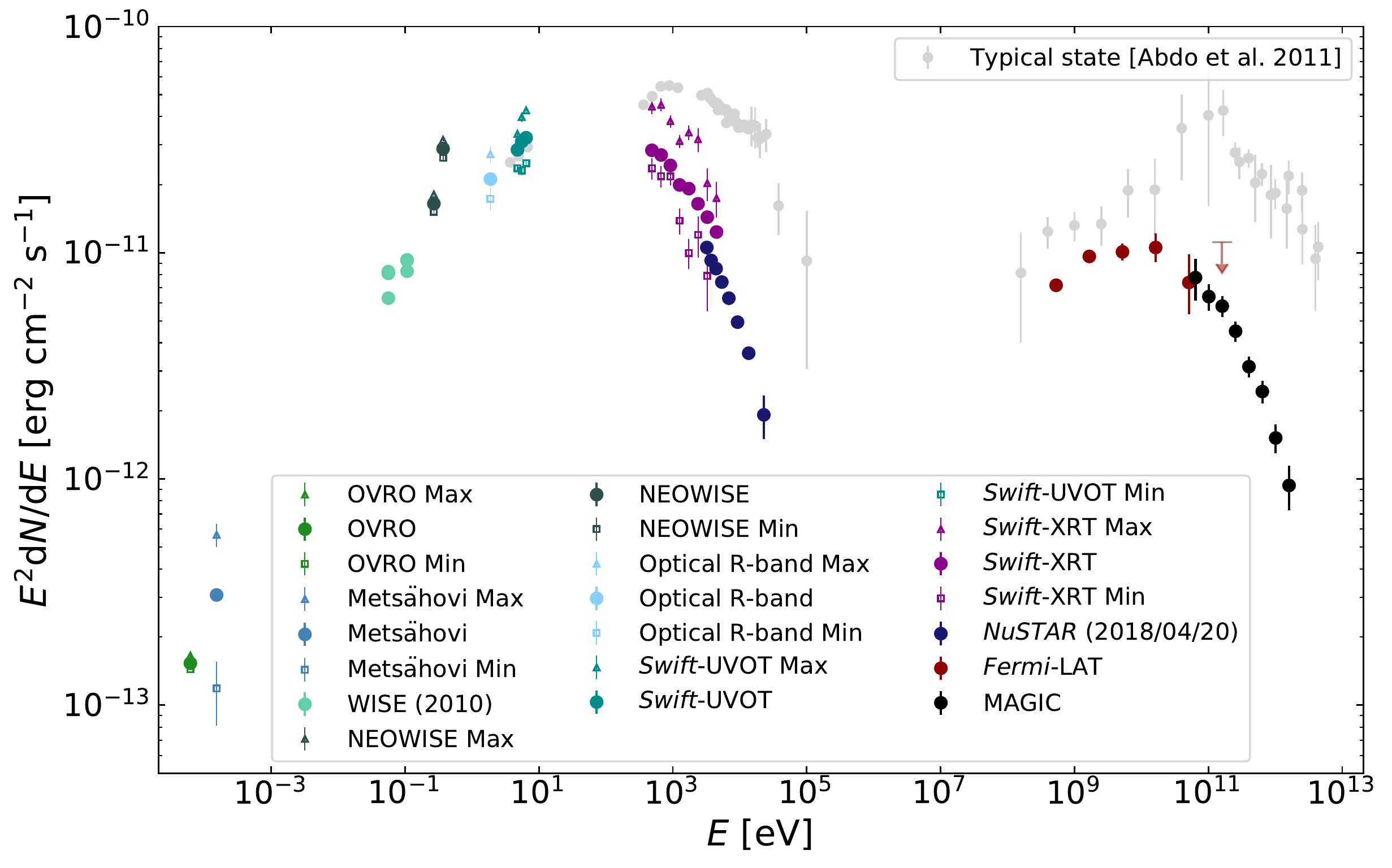}{0.6\textwidth}{(a) Broadband SED of the identified low-state period spanning from 2017-06-17 to 2019-07-23 (MJD~57921 to MJD~58687), which includes the \textit{NuSTAR} observation on 2018-04-20  (MJD~58228). Archival WISE data points from 2010 are also depicted. For comparison purposes, the typical (non-flaring) SED of Mrk\,501 from \citet{Abdo_2011} is shown with light-grey markers.}
}
\gridline{\fig{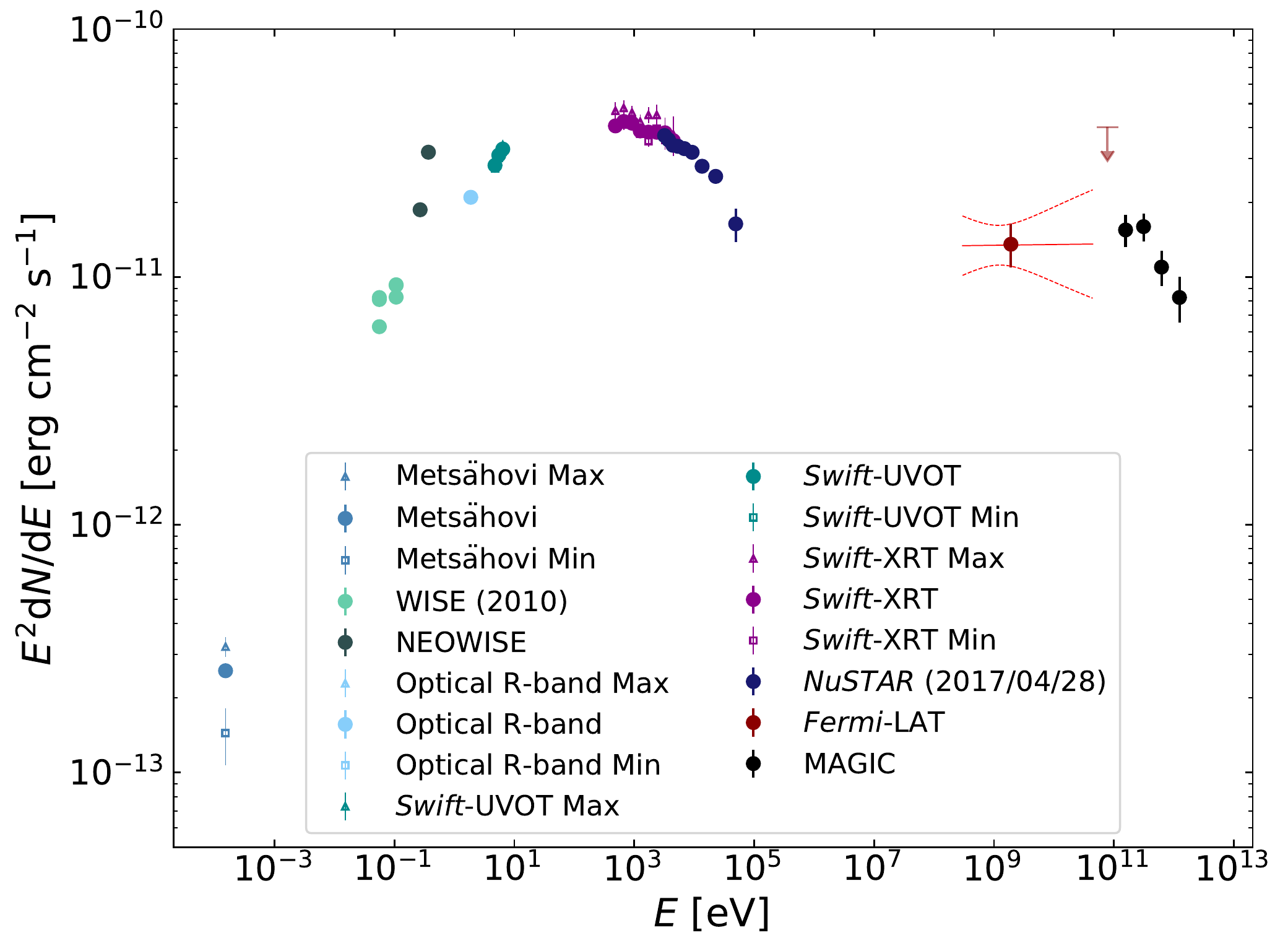}{0.49\textwidth}{(b) Broadband SED around the first \textit{NuSTAR} observation (\textit{NuSTAR}-1) on 2017-04-28 (MJD~57871). For MAGIC and \textit{Fermi}-LAT, the $\gamma$-ray spectra is derived using data from a 2-week interval centered at \textit{NuSTAR}-1. 
         The spectral fit result for \textit{Fermi}-LAT is shown (red dashed lines), as well as archival WISE points from 2010.}
          \fig{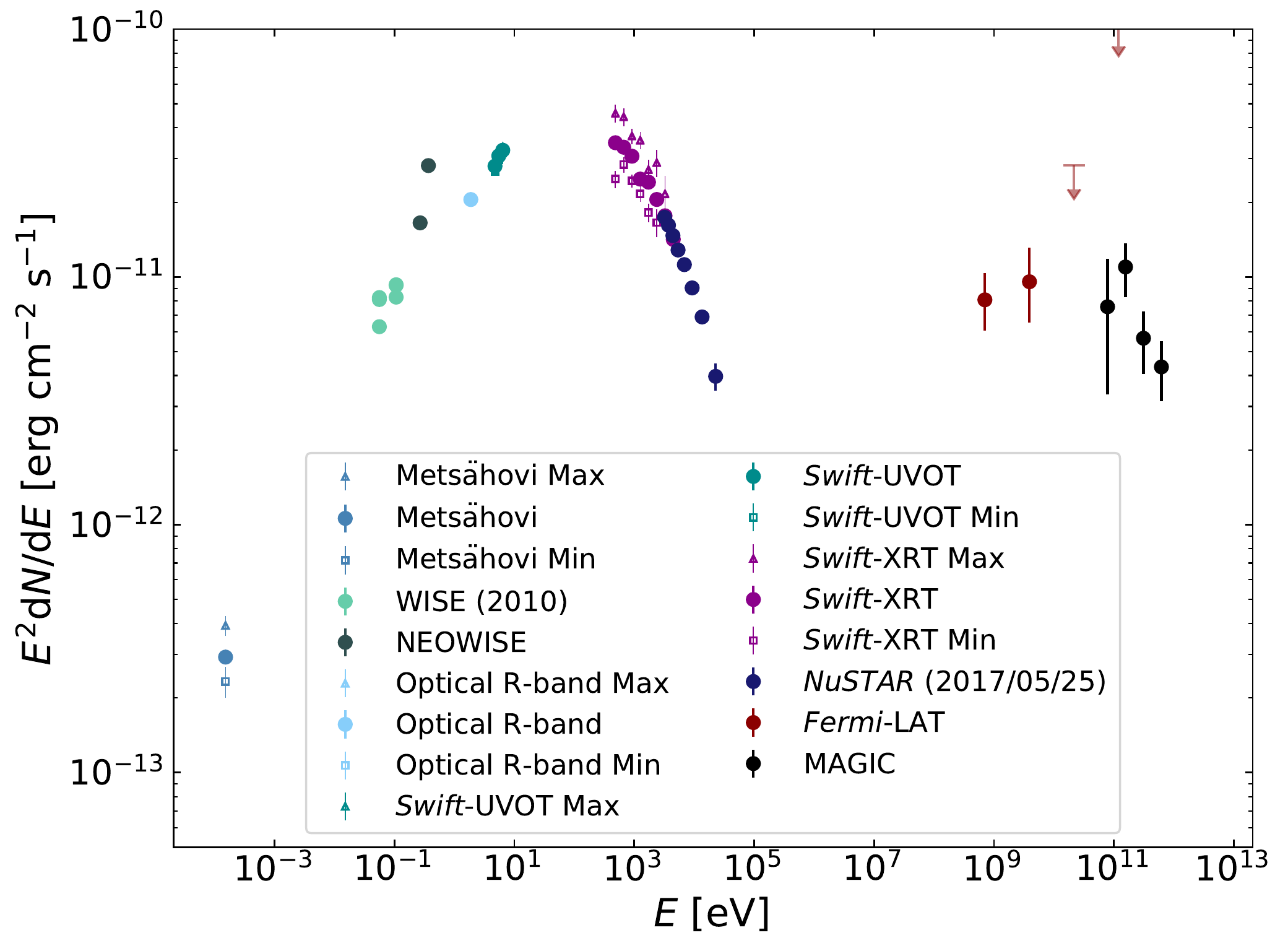}{0.49\textwidth}{(c) Broadband SED around the second \textit{NuSTAR} observation (\textit{NuSTAR}-2) on 2017-05-25 (MJD~57898).
          For \textit{Fermi}-LAT, the $\gamma$-ray spectrum is derived using data from a 2-week interval centered at \textit{NuSTAR}-2. Archival WISE data from 2010 are also shown.}
}

\caption{Broadband spectral energy distributions (SEDs) for different observation states during the 2017--2020 campaign, as described in Section~\ref{sec:spect}.}
    \label{fig:seds}
\end{figure*}

\subsection{Broadband SEDs during the 2017--2020 campaign}

The 2-year-long period without substantial flux variations allows one to average a large amount of data to compute a broadband SED with small statistical uncertainties, despite the historically low activity of the source. This is particularly important at $\gamma$-ray energies, where the sensitivity of the instruments is more limited, in comparison with optical or X-ray instruments, and both \textit{Fermi}-LAT and MAGIC would have limitations to deliver accurate spectra for timescales of days or even weeks (for such low source activity). Therefore, the entire 2-year data set from 2017-06-17 to 2019-07-23 (MJD~57921 to MJD~58687) was used to derive the \textit{Fermi}-LAT and MAGIC spectra shown in  Fig.~\ref{fig:seds}a).  For the other wavebands, we determine the average of the spectral points weighted with their uncertainties from the individual observations, which is shown together with the minimum and maximum spectral flux values within the 2-year-long epoch considered. Owing to the low flux variability in the radio, optical and X-ray bands, the weighted average spectral points are used as a good proxy of the emission of Mrk\,501 during this 2-year-long low-state. 
The hard X-ray spectrum derived with \textit{NuSTAR} data from 2018-04-20 (MJD~58228) is also depicted, showing a good agreement (within 30\%) with the weighted average \textit{Swift}-XRT spectrum in the overlapping energy range. This agreement is consistent with the relatively low flux variability at X-rays mentioned above, and suggests that the \textit{NuSTAR} spectrum from 2018-04-20 is a good approximation of the hard X-ray emission of Mrk\,501 during this 2-year-long epoch. As a consistency check, Fig.~\ref{fig:sed_nustar3} shows the broadband SED during the above-mentioned 2-year-long period of low activity, together with the SED around the \textit{NuSTAR} observation from 2018-04-20, using MAGIC and \textit{Fermi}-LAT spectra derived with data within $\pm$ 1\,week of the \textit{NuSTAR} observation. In radio, optical and soft X-ray data the sensitivity is good enough to use the individual observations (typically less than 1\,hour long). The good agreement among the spectra further supports the use of the \textit{NuSTAR} observation from 2018-04-20 as a good proxy of the hard X-ray emission of the 2-year-long integrated (weighted average) spectrum of Mrk\,501.

For comparison purposes, Fig.~\ref{fig:seds}a) also depicts the typical (non-flaring) state of Mrk\,501 from \citet{Abdo_2011}. This helps visualize the historically low-activity state of Mrk\,501 during the 2-year time interval that goes from  2017-06-17 to 2019-07-23 (MJD 57921 to MJD 58687).

For the first \textit{NuSTAR} observation (\textit{NuSTAR}-1), 2017-04-28 (MJD~57871)  no simultaneous VHE observations are available. Therefore we choose to use all nights inside $\pm$ 1\,week for the computation of both the MAGIC as well as the \textit{Fermi}-LAT spectra. For the radio, optical and X-ray frequencies, we use simultaneous data (within $\pm$ 5\,hours of the \textit{NuSTAR} observation) for all instruments.
The resulting SED is shown in Fig.~\ref{fig:seds}b). Since the \textit{Fermi}-LAT spectral analysis yields only one flux point, the figure additionally depicts the spectral shape (with statistical uncertainties) derived with our analysis.

The second \textit{NuSTAR} observation (\textit{NuSTAR}-2), on 2017-05-25 (MJD~57898), was performed simultaneous to the observations carried out with the MAGIC telescopes, which allows us to derive a SED with simultaneous (within $\pm$ 4\,hours) data from radio to VHE $\gamma$-rays.  The only exception is the spectrum from \textit{Fermi}-LAT, which, in order to reduce the statistical uncertainties (and owing to the lack of significant flux variability), is derived with data integrated  within $\pm$1 week of \textit{NuSTAR}-2. The corresponding multi-instrument SED is shown in 
Fig.~\ref{fig:seds}c).

Additionally, we added infra-red (IR) points taken by the NEOWISE mission\footnote{\url{https://irsa.ipac.caltech.edu/Missions/wise.html}} \citep{NEOWISE}. For the low state all simultaneous data were averaged as done for the other low-energy wavebands. No simultaneous data were available for the two \textit{NuSTAR} time intervals. We chose the closest NEOWISE observation, which took place on MJD~57807. However, NEOWISE only makes use of two (W1, W2) of the four filters of the WISE spacecraft. In order to include also information about the other two filters, we added archival WISE data from 2010  \citep[][]{WISE_general, WISE_datarealease} to all three SEDs from Fig.~\ref{fig:seds}, and used them for our spectral studies based on the observation of very low variability in the IR band. 

Thereafter, all three SEDs (low-state, \textit{NuSTAR}-1 \& \textit{NuSTAR}-2) are characterized within different theoretical scenarios, as described in the following sections. Owing to the very low variability at radio and optical frequencies and the consideration that this low-energy emission is dominated by the contributions from different, more extended and outer regions of the jet \citep[see e.g.,][]{Mrk501_MAGIC_2014}, the radio and optical flux points are treated as upper limits in the adjustment of the theoretical models to the data.  For all theoretical models, we consider a $\gamma$-ray absorption according to the EBL (extra-galactic background light) model of Franceschini \citep[][]{Franceschini}, which, for the distance of Mrk\,501 and the energies considered here, is perfectly compatible with that of many other EBL models \citep[see e.g.,][]{2011MNRAS.410.2556D}. 

It is worth noting that, even for accurately determined SEDs such as the ones presented here, there is an ample degeneracy in the model parameters from the various theoretical scenarios. Therefore, the following results should not be interpreted as unique solutions, but rather as plausible theoretical scenarios that are able to explain the broadband data in line with our physical understanding of the underlying mechanisms.

\subsection{Theoretical modeling of the historically low-activity state of Mrk\,501}
\label{sec:low-activity-SED}

\begin{table*}
    {\centering
    \caption{Parameter values from the leptonic one-zone SSC models used to describe the low-state SED of Mrk\,501 derived with data from 2017-06-17 to 2019-07-23 (MJD~57921 to MJD~58687), as described in Section~\ref{sec:baseline_lep}, and shown in Fig.~\ref{fig:fit_lep_bb}. In the two model realizations, the radius of the emission region $R'$ is fixed to $1.14\times10^{17}$\,cm, the Doppler factor $\delta$ to 11 and the magnetic field $B'$ to 0.025\,G. For the radiating electron distribution, a broken power-law is used with the spectral indices fulfilling $\alpha_{2}=\alpha_{1} +1$. The minimum energy of the electrons $\gamma_{\text{min}}'$ is set to 1000.  The table reports, for the broken power-law describing the shape of the electron distribution, the first spectral index $\alpha_{1}$, the break energy $\gamma_{br}'$ and the maximum energy $\gamma_{max}'$ as well as the electron luminosity $L_{\text{e}}$, the energy densities held by the electron population $U_{\text{e}}'$ and the magnetic field $U_{\text{B}}'$, their ratio, and the total jet luminosity $L_{\text{jet}}$. For the EBL $\gamma$-ray absorption at a redshift $z$=0.034, the model from Franceschini \citep{Franceschini} is used.}
    \label{tab:fits_lep_bb}
    \begin{minipage}{.49\textwidth}
        \centering
        \begin{tabular}{ c | c c c c  }     
             & $L_{\text{e}}$ & $\alpha_{1}$ & $\gamma_{\text{br}}'$ &$\gamma_{\text{max}}'$  \\
             & [erg/s] & & & \\
            \hline\hline   
            Modified Naima & $7.7\times10^{43}$ & 2.6 & $2.0\times10^{5}$*  &  $1.2\times10^{6}$ \\
            \hline  
            Jetset & $8.4\times10^{43}$ & 2.6 & $2.0\times10^{5}$* & $1.2\times10^{6}$  \\
        \end{tabular}
    \end{minipage}
    \hfill
    \begin{minipage}{0.45\textwidth}
        \centering
        \begin{tabular}{c c c c c}
            & $U_{\text{e}}'$ &  $U_{\text{B}}'$ & $U_{\text{e}}'$/$U_{\text{B}}'$ &  $L_{\text{jet}}$  \\
            &   [erg/cm$^3$] & [erg/cm$^3$] &   & [erg/s]\\
             \hline\hline   
            &$5.2\times10^{-4}$ & $2.5\times10^{-5}$ &  21  & $8.8\times10^{43}$ \\
            \hline
            &$5.7\times10^{-4}$ & $2.5\times10^{-5}$ & 23 & $9.6\times10^{43}$ \\
        \end{tabular}
    \end{minipage} 
}
\par\vspace{0.5cm}
*\text{Fixed to the cooling break}\par
\end{table*}

To exploit the historically low-activity broadband SED of Mrk\,501 and explore different blazar scenarios to describe this sort of "baseline broadband emission", we employ various widely used theoretical frameworks that consider leptonic, hadronic, and lepto-hadronic scenarios. For all models the low-energy component is mainly produced by synchrotron radiation of relativistic electrons. As leptonic scenarios we consider models where the high-energy component is also produced purely by the emission of relativistic electrons. For the scenarios where relativistic protons are contributing to the emission, we are following the conventions defined in \citet{Cerruti_review}. 
We define hadronic models as the ones where purely hadronic-initiated emission processes, i.e. proton synchrotron, dominate the high-energy component. For the lepto-hadronic models, SSC processes are responsible for a significant part of the high-energy component but proton initiated processes are present as well leading to an expected emission of neutrinos while staying in a similar parameter space as for the leptonic models. 

\subsubsection{Leptonic}
\label{sec:baseline_lep}
To evaluate a leptonic origin of the low-state SED, we choose a stationary one-zone SSC model \citep[see for e.g.,][]{1996ASPC..110..436G,Tavecchio_Constraints}.
For the model fitting, we compare two independent frameworks. 

Within the first framework, in order to constrain the model parameters more efficiently, the naima package \citep{2015ICRC...34..922Z}, is modified to derive the best-fit and uncertainty distributions of spectral model parameters through MCMC (Markov chain Monte Carlo) sampling of their likelihood distributions. Our prior constraints of the model parameter space obtained via "fit by eye" strategy, and the data likelihood function is passed on to \textit{emcee} \citep{2013PASP..125..306F}, which is a python implementation of the Goodman \& Weare’s Affine Invariant Markov chain Monte Carlo Ensemble sampler \citep[][]{EMCEE_2010}. 

As a second framework, we use the public open source C/Python framework jetset version 1.2.2\footnote{\url{https://github.com/andreatramacere/jetset/tree/1.2.2}} \citep{Tramacere_2009,2011ApJ_jetset,2020ascl_jetset}. We first approximate the spectral shape and use it together with basic information about the electron distribution and jet properties as input for a pre-fit to constrain the parameter space. Afterwards, a full spectral fit is carried out using the minuit minimizer. The resulting best fit is then used as a prior for a MCMC chain using as recommended by the instructions on the algorithm, 128 walkers, 10 steps of burn-in, and 50 run steps, varying all free parameters from the minuit fit. This enables us to both improve the best fit as well as obtain a confidence interval on the resulting model.

We use the same reference parameters for both frameworks.
Our emission zone is assumed to be a spherical blob with radius $R'$ filled with relativistic electrons. The radiating electron distribution is described by a broken power-law where the high-energy power-law index $\alpha_2$ is connected to the low-energy index $\alpha_1$ via the relation $\alpha_2=\alpha_1+1$. For all models the parameters referred to in the blob frame are marked as primed while unmarked ones are in the observer frame.

Since no short time scale variability is observed in our data, a value of $R'=1.14\times 10^{17}$\,cm is chosen for the emission region size. This is consistent with the values used in a previous work reporting the typical (non-flaring) broadband emission of Mrk\,501 \citep{Abdo_2011} based on the observed low variability taking into account our choice for the Doppler factor.  We assume a viewing angle of $\approx$ $1/\Gamma_b$, with $\Gamma_b$ being the bulk Lorentz factor. Therefore, $\Gamma_b$ and the Doppler factor $\delta$ can be considered equal. 
For the Doppler factor, we fixed $\delta=11$, which showed a good agreement between the models and our data in our preliminary fits where we tried a grid of values around $\delta=12$ from \citet{Abdo_2011}. For this, we first used a step size of 2 and then optimized further using a step size of 1 for the region between 10 and 12. The literature shows that, because of short-time flux variability and VHE $\gamma$-ray opacity arguments, values of $\delta>20$ are often necessary to explain adequately the data from BL Lacs \citep[][]{Tavecchio_Constraints}. On the other hand, high Doppler factors imply a small angle between our line of sight and the blazar, which cause tensions with the idea of a parent population of inefficiently accreting AGN including both BL Lacs and  Fanaroff–Riley Type I (FR-I) radio galaxies \citep{2000A&A...358..104C, Tavecchio_doppler}. The Doppler factor $\delta=11$ used in this study is in good agreement with blazar unification schemes, as well as being consistent with our low flux variability levels and the required transparency for the measured VHE $\gamma$-rays. However, it is still considerably higher than the Doppler factors derived in the radio regime \citep{Finke_2019} and in strong tension with the slow jet component speeds observed in the radio \citep[][]{Piner_2010, Giroletti2004}. This supports the scenario where part of radio emission is assumed to be produced in a different, more extended region of the jet than the X-ray to TeV emission. 

For the magnetic field $B'$, we obtain $B' \sim $ 0.025 G in our first fits, and fix it at this value. For both frameworks we fix the break energy $\gamma_{\text{br}}'$ of the electron distribution according to the cooling break defined by Eq.~30 in \cite{Tavecchio_Constraints} with the cooling time being equal to the escape time of the radiating particles $t_{cool}' = t_{esc}' = \nu \frac{R'}{c}$ with our choice of $\nu=1$. Longer escape times with $\nu=2,3,...$ are possible as well, but prior investigations do not show any major changes for our physical conclusions. A $\nu$ of two would, for example, still allow the ambient parameters to stay at the same level as for the results presented below and just the electron parameters would change to slightly lower values for the energies and the spectral indices. Additionally, we fixed the minimum energy of the electrons to $\gamma_{\text{min}}'=1000$ to further constrain the fitting procedure and following the discussion in \citet{Mrk501_MAGIC_2008}. We note that our ability to constrain the model parameter $\gamma_{\text{min}}'$ is 
 strongly limited by the lack of data at MeV energies, as well as the contributions from other (more extended) regions to the emission at radio and optical frequencies.

\begin{figure}
    \resizebox{\hsize}{!}{\includegraphics{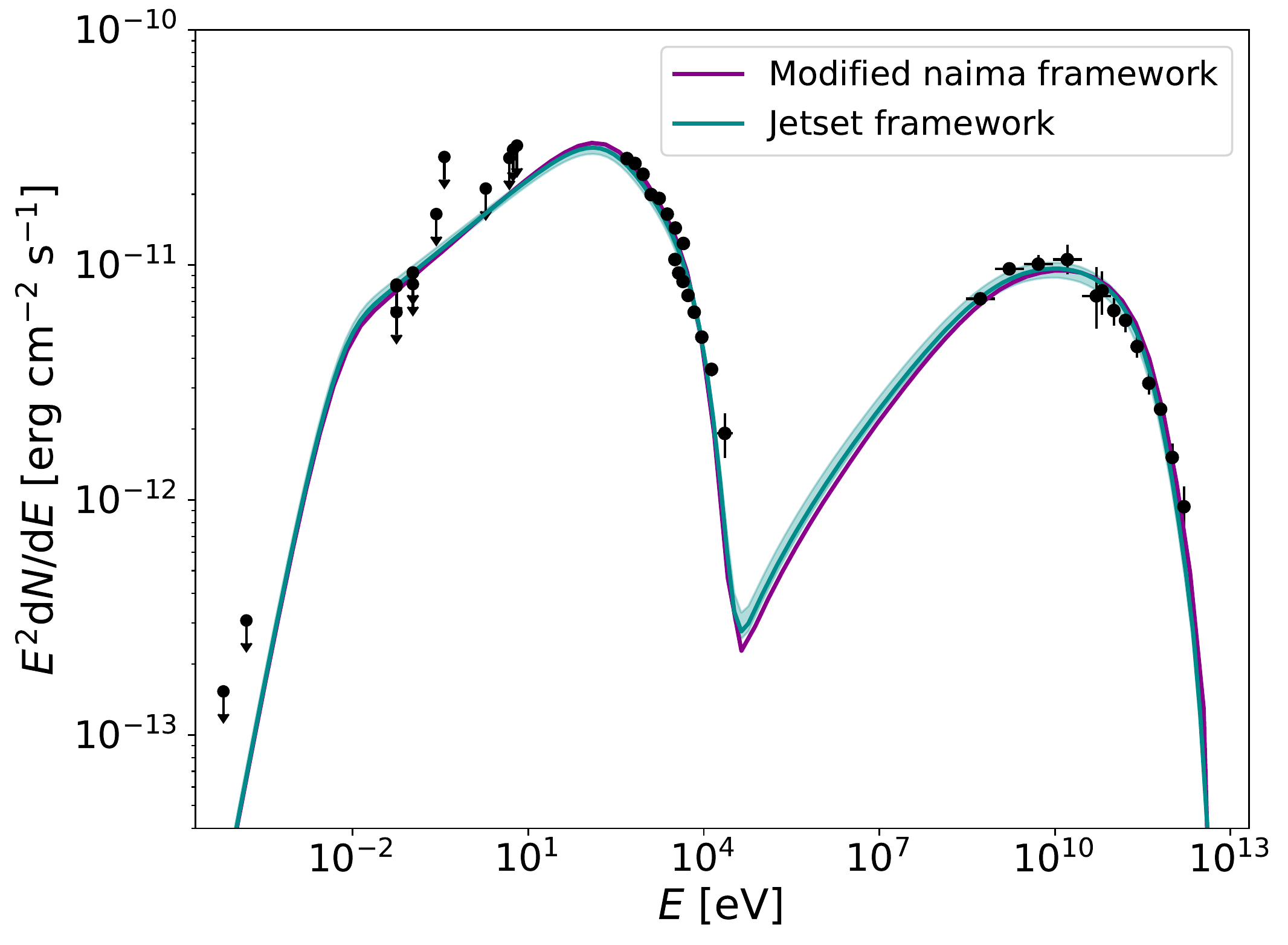}}
    \caption{Leptonic one-zone models that describe the broadband SED of the low-state of Mrk\,501 derived with data from 2017-06-17 to 2019-07-23 (MJD~57921 to MJD~58687), as described in Section~\ref{sec:baseline_lep}. The corresponding model parameters are reported in Table~\ref{tab:fits_lep_bb}. Data with frequencies in the UV or lower are considered as upper limits for the modeling of the blazar emission, and are therefore depicted with arrows. For the jetset model (cyan) the confidence interval of the MCMC fit is displayed since the framework allows for it.}
    \label{fig:fit_lep_bb}
\end{figure}

The resulting low-state SED data-model adjustment with the theoretical frameworks mentioned above is shown in Fig.~\ref{fig:fit_lep_bb}, and the corresponding model and energy parameters are reported in Table~\ref{tab:fits_lep_bb}. The two theoretical frameworks used yield very similar results. 

\subsubsection{Hadronic}
\label{subsec:hadronic}

From a purely electromagnetic perspective, leptonic radiative models and hadronic radiative models can fit blazar SEDs equally well \citep[see e.g.][]{Cerruti_review}. While for luminous blazars hadronic models face the difficulty of requiring often super-Eddington powers, and can thus be disfavoured from an energetic point of view, this is not true for HSP BL Lac type objects, and in particular when low flux states are studied. In this case, leptonic and hadronic emission scenarios are truly degenerate in their photon emission and can only be distinguished via the detection of neutrinos, naturally produced in hadronic interactions while absent in leptonic ones. With this in mind, we investigate here a hadronic modeling of the SED of Mrk\,501 in its historically low-activity state described above (see Fig.~\ref{fig:seds}a)). 

In order to apply this modeling, we again employ two different numerical frameworks. The first numerical code, the LeHa code, used for the hadronic modeling is described in \citet{Cerruti15}. The code simulates photon and neutrino emission from a spherical plasmoid (with radius $R'$) in the jet, moving with Doppler factor $\delta$ and filled with a homogeneous and entangled magnetic field $B'$. The plasmoid contains a population of primary electrons and protons. Proton-photon interactions via photo-meson and Bethe-Heitler channels (of protons on synchrotron photons produced by primary electrons) inject secondary particles in the emitting region: these secondary particles trigger synchrotron-supported pair-cascades for which the synchrotron emission at equilibrium is computed. \\

The second code is SOPRANO\footnote{\url{https://www.amsdc.am/soprano/}} \citep[see e.g.][]{2022MNRAS.509.2102G}, a time-dependent code including leptonic and hadronic processes designed to study the particle interaction mechanisms in different astrophysical objects. The code follows the temporal evolution of the isotropic distribution functions of primary injected particles and the secondaries produced in photo-pair and photo-pion interactions, alongside the evolution of photon and electron/positron distribution functions. In this case, the final spectrum is computed by evolving the kinetic equations for several dynamical time scales to guarantee that the steady-state condition is achieved.

Important to note is that when we are comparing the model parameter values between the different frameworks, the first gives the parameters of the steady-state solution of the particle distributions while the second gives the particle properties at injection. Nonetheless, the results are compatible between the different codes.

\begin{figure}
    \resizebox{\hsize}{!}{\includegraphics{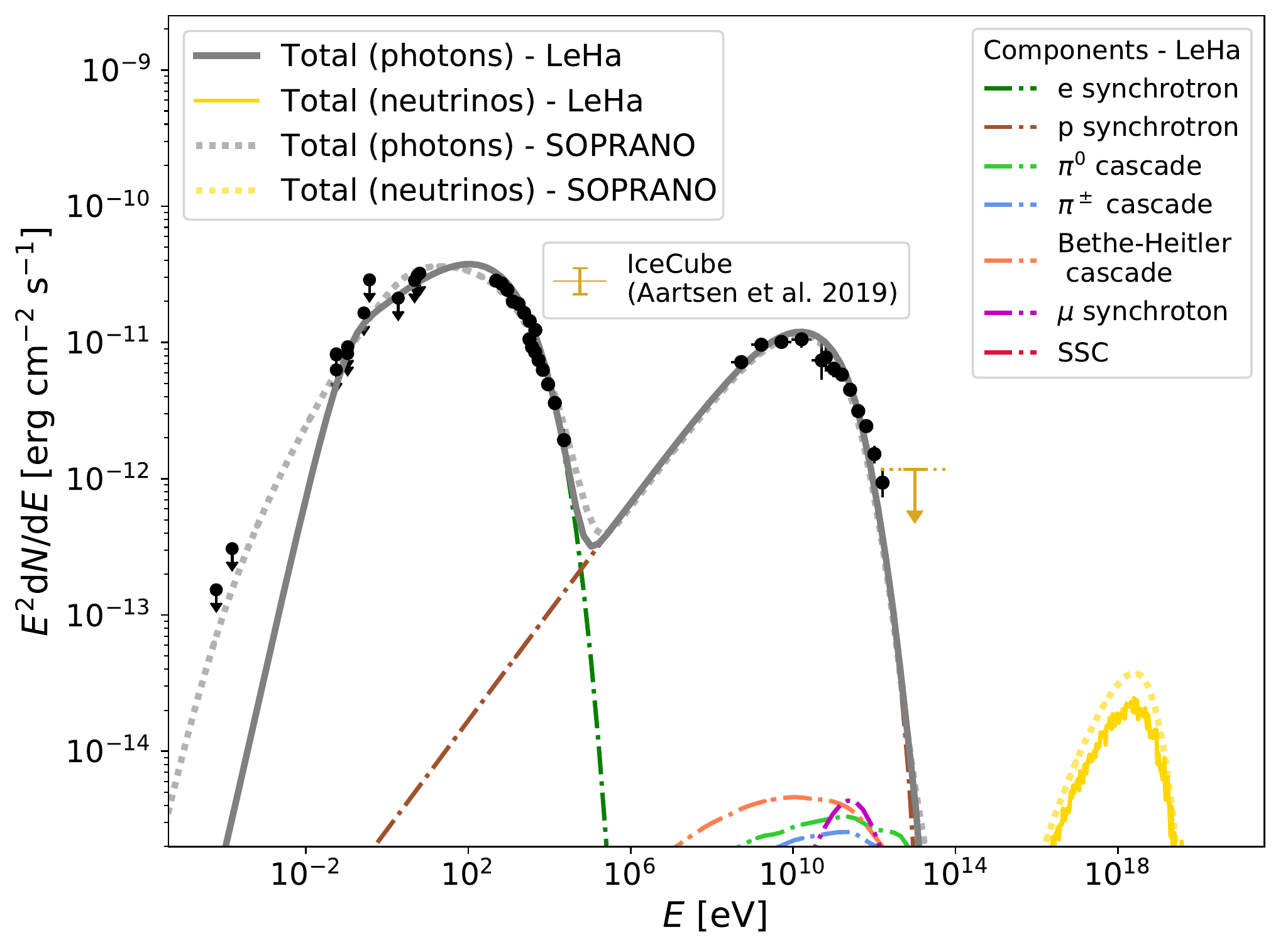}}
   \caption{Hadronic one-zone models that describe the broadband SED of the low-state of Mrk\,501 derived with data from 2017-06-17 to 2019-07-23 (MJD~57921 to MJD~58687), as described in Section~\ref{sec:baseline_lep}. The grey solid line describes the results of the LeHa code with the individual components shown by the color-dashed lines. The model obtained with the SOPRANO code is shown by the grey dashed line. The corresponding model parameters are reported in Table~\ref{tab:fits_had} and Table~\ref{tab:fits_had_Sargis}. Data with frequencies in the UV or lower are considered as upper limits for the modeling of the blazar emission,  and therefore depicted with arrows. Additionally, the neutrino flux estimate is shown by the yellow curve (solid for the LeHA results, dashed for the SOPRANO result), together with the upper limit from Icecube \citep{IceCube_UL2020} depicted by the golden upper limit.}
   \label{fig:fit_hadronic}
\end{figure}

We first make the same assumptions 
as for the leptonic case and choose the radius $R'$ to be $1.14\times10^{17}$\,cm and $\delta=11$. Simple power-law distributions are chosen for the radiating electron and proton distributions. The maximum proton energy is determined by equating the acceleration timescale \citep[see e.g.][]{Rieger07} and the escape timescale ($t_{\text{esc}}' \sim R'/c$). The minimum proton energy is not constrained by the data, we therefore fix it to $\gamma_{\text{min,p}}'=1$ following the discussion in \citet{Cerruti15}. To be in line with a slow variability, the magnetic field is decreased as much as possible while looking for a fitting scenario (a higher magnetic field being associated with a faster synchrotron cooling). The resulting models from the two frameworks are shown in Fig.~\ref{fig:fit_hadronic}. For simplicity, all individual components of the emission are only shown for the LeHa framework while for the SOPRANO result only the total emission in photons and neutrinos is indicated. A more detailed comparison including the distinguishing components of our hadronic and lepto-hadronic scenarios is shown in Fig.~\ref{fig:fit_hadronic_comp}. The parameter values are reported in Table~\ref{tab:fits_had} for the LeHa framework stating the steady-state particle properties as is done for the leptonic case, while Table~\ref{tab:fits_had_Sargis} states the parameters of the SOPRANO framework displaying the injected particle properties. For the injected spectrum a simple power-law is used to describe the distribution while it evolves into a broken power-law for the steady-state spectrum. For some cases, the break energy $\gamma_{\text{br}}'$ obtained in the optimization of the model is higher than the maximum particle energy $\gamma_{\text{max}}'$ constrained by the acceleration time scale. For these cases the results are shown for a simple power-law. In these hadronic scenarios the $\gamma$-ray emission is ascribed to proton synchrotron radiation. Emission from pair-cascades (triggered by pion decay and Bethe-Heitler pair production) and from muons (synchrotron radiation) is subdominant and largely absorbed by the EBL. The total jet powers required to produce the observed photon flux are $3.1\times10^{46}$ erg/s and  $3.0\times10^{46}$ erg/s for the two frameworks, which is sub-Eddington for typical super-massive black hole masses.\\

The hadronic scenario enables the prediction of an estimated neutrino spectrum which can then be converted into an IceCube detection rate per year using the instrument effective area by \citet[][]{Icecube_data_2021}. For our model, the expected neutrino rate is $1.1\times10^{-5}$ events per year for the LeHa model and $ 1.5\times10^{-5}$ events per year for the SOPRANO model. Since this result does not come from a fit, and is not a unique solution, hadronic models exploiting other parts of the parameter space could lead to different explanations for the SED, and different neutrino spectra can be produced. A brighter neutrino emission can be achieved if the emitting region is more compact (smaller size and higher particle density), resulting in a higher proton-photon interaction rate, but this is limited by our choice of $R'$ based on the observed low variability.

\begin{table}
    \centering
    {\caption{Parameter values obtained by the LeHa code for the hadronic and lepto-hadronic one-zone models used to describe the low-state SED of Mrk\,501 derived with data from 2017-06-17 to 2019-07-23 (MJD~57921 to MJD~58687), as described in Section~\ref{subsec:hadronic} and shown in Fig.~\ref{fig:fit_hadronic} (hadronic), and in Section~\ref{subsec:lepto-hadronic} and Fig.~\ref{fig:fit_leptohadronic} (lepto-hadronic).
    In these model realizations, the radius of the emission region $R'$ is fixed to $1.14\times10^{17}$\,cm, and the Doppler factor to $\delta=$11. 
    For the radiating particle distributions simple and broken power-laws are used. The table reports the magnetic field $B'$, the radius $R'$ of the emitting region and, for the assumed power-law distribution for the electrons and protons, the density ratio $n_{\text{e}}'/n_{\text{p}}'$ and the total density of radiating particle N$_{\text{total}}'$, the slopes $\alpha_{\text{e,1}}$, $\alpha_{\text{e,2}}$, $\alpha_{\text{p}}$ and the minimum, maximum and break energies  $\gamma_{\text{min,e}}'$,  $\gamma_{\text{min,p}}'$,  $\gamma_{\text{max,e}}'$,  $\gamma_{\text{max,p}}'$, $\gamma_{\text{br,e}}'$. We also show the  energy densities held by the electron population $U_{\text{e}}'$, the proton population $U_{\text{p}}'$ and the magnetic field $U_{\text{B}}'$, their ratios and the total jet luminosity $L_{\text{jet}}$. For the EBL $\gamma$-ray absorption at a redshift $z$=0.034, the model from Franceschini \citep{Franceschini} is used.}
    \label{tab:fits_had}
        \begin{tabular}{ c | c  | c}     
        
            & Hadronic & Lepto-hadronic \\
            \hline \hline
            $B'$ [G] & 3 & 0.025 \\
             $R'$ [cm] & $1.14 \times 10^{17}$ & $1.14 \times 10^{17}$ \\
             N$_{\text{total}}'$ [1/cm$^{3}$] & 3 & $2.3 \times 10^{4}$\\
             $n_{\text{e}}'$/$n_{\text{p}}'$ & 2.2 & 0.01 \\
             $\alpha_{\text{e,1}}$ & 2.5 & 2.6\\
             $\alpha_{\text{e,2}}$ & - & 3.6 \\

             $\gamma_{\text{min,e}}'$ & 400  & $1 \times 10^{3}$ \\
             $\gamma_{\text{br,e}}'$ & - & $2.0 \times 10^{5}$ \\
             $\gamma_{\text{max,e}}'$  & $3.5 \times 10^{4}$ & $1.2 \times 10^{6}$ \\
             $ \alpha_{\text{p}}$ & 2.2 & 2.0\\
             $\gamma_{\text{min, p}}'$ & 1& 1\\
             $\gamma_{\text{max, p}}'$ & $1.1 \times 10^{10}$ & $2 \times 10^{7}$ \\
            &&\\
            $U_{\text{e}}'$ [erg/cm$^{3}$]  & $2.2 \times 10^{-7}$ & $5.3 \times 10^{-4}$\\
            $U_{\text{B}}'$ [erg/cm$^{3}$] & 0.36 & $2.5 \times 10^{-5}$\\
            $U_{\text{p}}'$ [erg/cm$^{3}$] & 0.05 & $5.6$\\
            $U_{\text{e}}'$ / $U_{\text{B}}'$ & $6.1 \times 10^{-7}$ & $21.2$\\
            $U_{\text{p}}'$ / $U_{\text{B}}'$ & 0.14 & $2.3 \times 10^{5}$\\
            $L_{\text{jet}}$ [erg/s] & $3.1 \times 10^{46}$ & $1.7 \times 10^{48}$\\
        \end{tabular}
        }
\end{table}

\subsubsection{Lepto-hadronic}
\label{subsec:lepto-hadronic}

\begin{figure}
   \centering
   \includegraphics[width=0.49\textwidth]{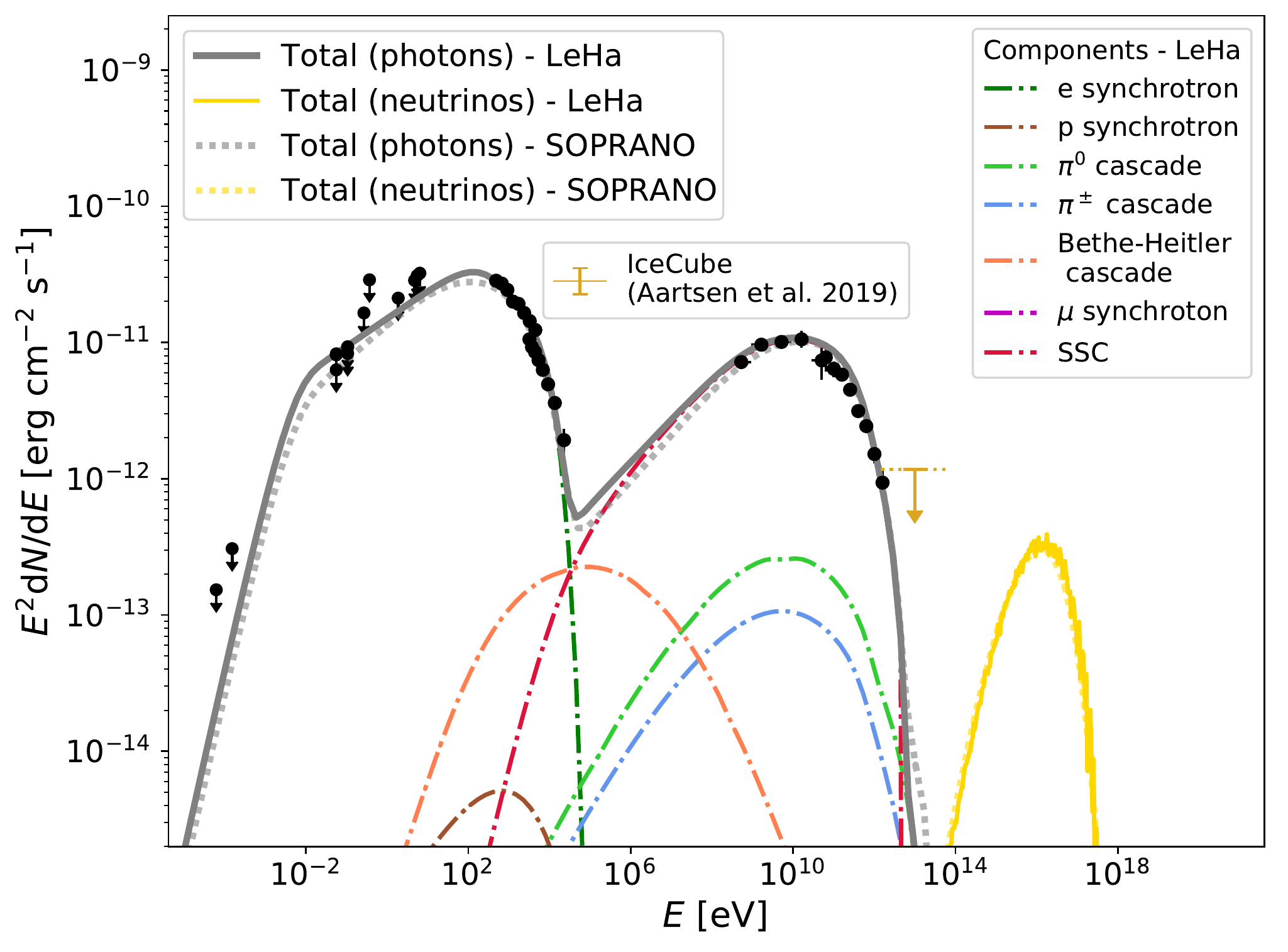}
   \caption{Lepto-hadronic one-zone models that describe the broadband SED of the low-state of Mrk\,501 derived with data from 2017-06-17 to 2019-07-23 (MJD~57921 to MJD~58687), as described in Section~\ref{subsec:lepto-hadronic}. The grey solid line describes the results of the LeHa code with the individual components shown by the color-dashed lines. The model obtained with the SOPRANO code is shown by the grey dashed line. The corresponding model parameters are reported in Table~\ref{tab:fits_had} and Table~\ref{tab:fits_had_Sargis}.  Data with frequencies in the UV or lower are considered as upper limits for the modeling of the blazar emission, and therefore depicted with arrows. The different components of the models are shown by the colorful dashed lines, while the total model is displayed by the grey line. Additionally, the neutrino flux estimate is shown by the yellow curve (solid for the LeHA results, dashed for the SOPRANO result), together with the upper limit from Icecube \citep{IceCube_UL2020} shown by the golden upper limit.}
   \label{fig:fit_leptohadronic}
\end{figure}

In recent years special attention has been given to mixed lepto-hadronic models, in which the high-energy SED component is associated to a combination of both leptonic (inverse-Compton) and hadronic (emission by cascades triggered by hadronic interactions) processes. It is of particular interest due to the fact that the first evidence for the detection of joint photon and neutrino emission from the direction of a blazar, the 2017 flare of TXS 0506+056 \citep[][]{Icecube_TXS}, supported this kind of emission scenario, disfavouring a proton synchrotron one \citep[see e.g.][]{Gao18, Keivani18, Cerruti_2019}. In the lepto-hadronic solutions considered for this work, the bulk of the high-energy SED component is due to SSC, while the hadronic components are subdominant and can emerge (and dominate the SED) in hard-X-rays, filling in the SED dip, and in the VHE band. In this scenario, the proton synchrotron emission is very suppressed, mainly due to the lower magnetization of the emitting region with respect to proton synchrotron solutions. We explore this lepto-hadronic model starting from the SSC solution described in Section~\ref{sec:baseline_lep}, and adding a proton distribution with index $\alpha_{p} = 2.0$ and $\gamma_{max,p}' = 2 \times 10^7$. Again both the LeHa code as well as the SOPRANO code as described in Section~\ref{subsec:hadronic} are utilized. The models are shown in Fig.~\ref{fig:fit_leptohadronic} with the more detailed comparison given in Fig.~\ref{fig:fit_leptohadronic_comp}.The parameter values are reported in Table~\ref{tab:fits_had} for the LeHa framework stating the steady-state particle properties as is done for the leptonic case, while Table~\ref{tab:fits_had_Sargis} states the parameters from the SOPRANO framework displaying the injected particle properties. Compared to the proton-synchrotron solution, this scenario is much more demanding in terms of energetics, with a total jet power of $1.7\times10^{48}$ erg/s and $4.4\times10^{47}$ erg/s for the two frameworks, which are already super-Eddington for a black hole mass of $10^9 M_{\odot}$ ($L_{\text{Edd}} \sim 10^{47}$\,erg/s). However, this is not a strong constraint on the model: the solution shown here is not the result of the fit and not achieved by minimizing the jet power; the proton distribution is also conservative with respect to the total power, and harder distributions, or the introduction of a low-energy cut-off ($\gamma_{min,p}' > 1$) would significantly lower the jet power. As an example, a similar electromagnetic emission can be achieved by hardening the proton distribution to $\alpha_p = 1.8$ and lowering the proton power to $6.6\times10^{47}$ erg/s. The expected IceCube neutrino rates for the solutions shown in Figure~\ref{fig:fit_leptohadronic} are  $5.8 \times 10^{-3}$ events per year for the LeHa model and $6.3\times10^{-3}$ events per year for the SOPRANO model. This value is larger than the proton synchrotron solution, closer to the IceCube detection energy range. It is a rate that remains however low (less than one event in two decades) and consistent with the non-detection of Mrk~501 as a point like source in the IceCube data. \\

\subsection{Theoretical modeling of the temporal evolution of the broadband SEDs}
The two \textit{NuSTAR} observations performed in April and May 2017, right before the 2-year-long period of historically low activity, provide us with the opportunity to characterize the overall decrease in the broadband emission of Mrk\,501 down to this sort of "baseline emission". The variable behavior of Mrk\,501 is often ascribed to variations in a radiatively efficient electron population \citep[see e.g.,][]{Mrk501_MAGIC_2009b,Mrk501_MAGIC_2012,Mrk501_MAGIC_2014}, and hence this temporal evolution study concentrates on SSC scenarios. In these theoretical frameworks, hard X-rays contain information from the dynamics of the highest energy electrons. Because of the very low activity of Mrk\,501 during the time interval considered here, hard X-ray instruments such as \textit{Swift}-BAT or INTEGRAL do not have sufficient sensitivity to detect the X-ray emission, and thus only \textit{NuSTAR} has the capability to provide an accurate characterization of the hard X-ray emission of Mrk\,501.

First, we study the temporal evolution of the broadband SED using the same one-zone SSC scenario employed to describe the historically low activity state reported in Section~\ref{sec:baseline_lep}. We take the model parameters reported in Table~\ref{tab:fits_lep_bb} as a starting point, and evaluate what parameters need to vary to explain the data from the previous months. As a second approach, we consider the 2-year-long low-state SED reported in Section~\ref{sec:baseline_lep} as sort of steady (or very slowly variable) "baseline broadband emission", and evaluate the presence of an additional region (located somewhere else along the jet of Mrk\,501) whose emission is variable on timescales of weeks and months, and is responsible for the blazar activity during the \textit{NuSTAR}-1 and \textit{NuSTAR}-2 observations. 

\subsubsection{One-zone}
\label{subsec:1zone}
\begin{figure*}
\gridline{\fig{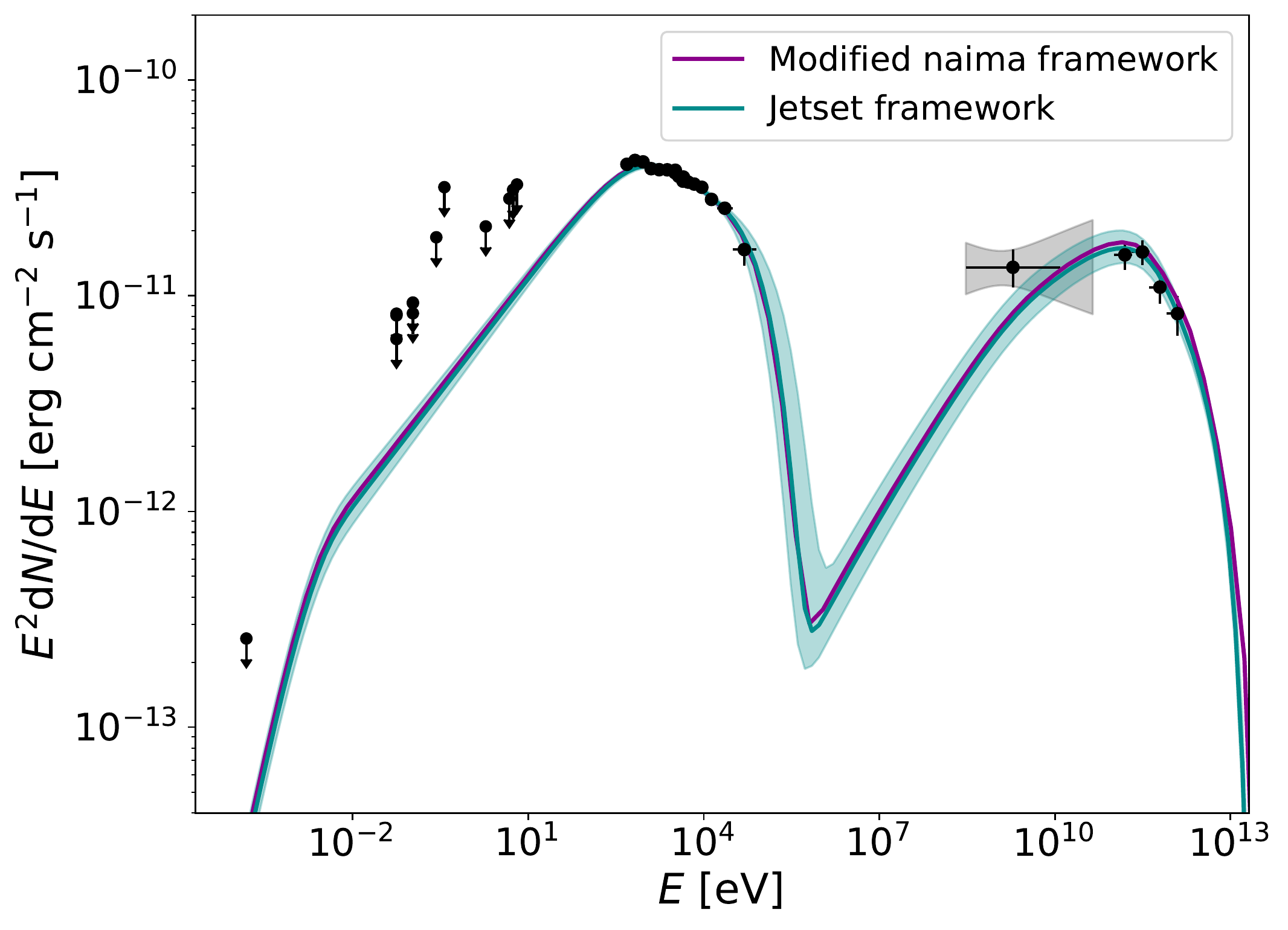}{0.48\textwidth}{(a) SED and related SSC models for the \textit{NuSTAR}-1 observation.}
          \fig{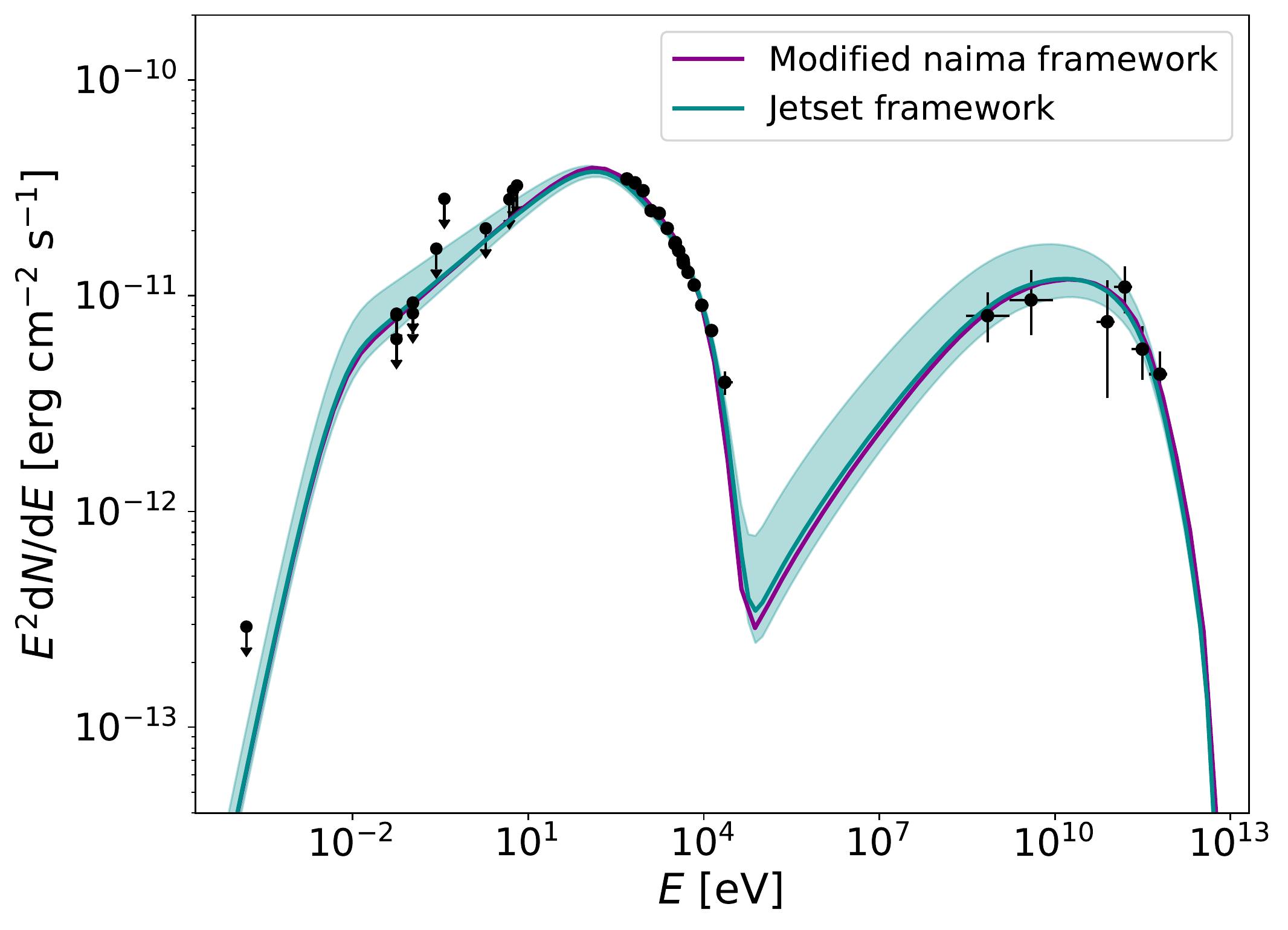}{0.48\textwidth}{(b) SED and related SSC models for the \textit{NuSTAR}-2 observation.}
}

\caption{Leptonic one-zone SSC models that describe the broadband SED of Mrk\,501 during the \textit{NuSTAR}-1 and \textit{NuSTAR}-2 observations on 2017-04-28  (MJD 57898) and 2017-05-25 (MJD~57871), respectively. See Section~\ref{subsec:1zone} for further details. The corresponding model parameters are reported in Table~\ref{tab:fits_lep_evolution}. Data with frequencies in the UV or lower are considered as upper limits for the modeling of the blazar emission,  and therefore depicted with arrows. For the jetset model (cyan) the confidence interval of the MCMC fit is displayed since the framework allows for it. }
    \label{fig:fits_lep_evolution}
\end{figure*}

As a starting point, we adopt the SSC scenario shown in Fig.~\ref{fig:fit_lep_bb}, with the model parameters reported in Table~\ref{tab:fits_lep_bb}. Subsequently, we change the model parameter values to describe the SEDs from the months before the 2-year-long low-state. The environmental parameters, including the magnetic field, are kept as close as possible to the values from the low-state model, while most of the model parameters describing the electron distribution are allowed to vary. The model parameters that describe well the SEDs during the \textit{NuSTAR}-1 and \textit{NuSTAR}-2 observations  are reported in Table~\ref{tab:fits_lep_evolution}, and the model curves (together with the SED data) are displayed in Fig.~\ref{fig:fits_lep_evolution}. 


\begin{table*}
    {\centering
    \caption{Parameter values from the leptonic one-zone SSC models used to describe the broadband SED of Mrk\,501 during the \textit{NuSTAR}-1 and \textit{NuSTAR}-2 observations on 2017-04-28  (MJD 57898) and 2017-05-25 (MJD~57871), that are shown in Fig.~\ref{fig:fits_lep_evolution}. See Section~\ref{subsec:1zone} for further details.
    In the various model realizations,  the radius of the emission region $R'$ is fixed to $1.14\times10^{17}$\,cm, and the Doppler factor $\delta$ to 11. For the radiating electron distribution, a broken power-law is used with the spectral indices fulfilling $\alpha_{2}=\alpha_{1} +1$. The minimum energy of the electrons $\gamma_{\text{min}}'$ is set to 1000.  The table reports, for the broken power-law describing the shape of the electron distribution, the first spectral index $\alpha_1$, the break energy $\gamma_{br}'$ and the maximum energy $\gamma_{max}'$ as well as the electron luminosity $L_{\text{e}}$, the energy densities held by the electron population $U_{\text{e}}'$ and the magnetic field $U_{\text{B}}'$, their ratio, and the total jet luminosity $L_{\text{jet}}$. For the EBL $\gamma$-ray absorption at a redshift $z$=0.034, the model from Franceschini \citep{Franceschini} is used.}
    \label{tab:fits_lep_evolution}
    \begin{minipage}{.49\textwidth}
        \centering
        \begin{tabular}{ c | c c c c  }     
            \multicolumn{5}{l}{a) Model for \textit{NuSTAR}-1 with a magnetic field of $B'=$0.01\,G} \\    
             & $L_{\text{e}}$ & $\alpha_1$ & $\gamma_{\text{br}}'$ &$\gamma_{\text{max}}'$  \\
             & [erg/s] & & & \\
            \hline\hline
            Modified Naima  & $1.1\times10^{44}$ & 2.3 & $6.6\times10^{5}$* & $7.2\times10^{6}$  \\
            \hline  
            Jetset  & $1.1\times10^{44}$ & 2.3 & $6.6\times10^{5}$* & $7.3\times10^{6}$ \\
            \multicolumn{5}{l}{} \\
            \multicolumn{5}{l}{} \\
            \multicolumn{5}{l}{b) Model for \textit{NuSTAR}-2 with a magnetic field of $B'=$0.025\,G} \\
             & $L_{\text{e}}$ & $\alpha_1$ & $\gamma_{\text{br}}'$ &$\gamma_{\text{max}}'$  \\
             & [erg/s] & & & \\
            \hline\hline            
            Modified Naima  & $7.8\times10^{43}$ & 2.5 & $1.9\times10^{5}$* & $1.5\times10^{6}$  \\
            \hline  
            Jetset & $8.2\times10^{43}$ & 2.6 & $1.9\times10^{5}$* & $1.6\times10^{6}$  \\ 
        \end{tabular}
    \end{minipage}%
    \hspace{0.05\textwidth}
    \begin{minipage}{0.45\textwidth}
        \centering
        \begin{tabular}{c c c c c}
            & & & & \\
            & $U_{\text{e}}'$ &  $U_{\text{B}}'$ & $U_{\text{e}}'$/$U_{\text{B}}'$ &  $L_{\text{jet}}$  \\
            &   [erg/cm$^3$] & [erg/cm$^3$] &   & [erg/s]\\
             \hline\hline   
            &$7.1\times10^{-4}$ & $4.0\times10^{-6}$  & 178 & $1.1\times10^{44}$  \\
            \hline
            & $7.1\times10^{-4}$ & $4.0\times10^{-6}$ & 178 & $1.1\times10^{44}$\\
            \multicolumn{5}{l}{} \\
            \multicolumn{5}{l}{} \\
            &  & & & \\
            & $U_{\text{e}}'$ &  $U_{\text{B}}'$ & $U_{\text{e}}'$/$U_{\text{B}}'$ &  $L_{\text{jet}}$  \\
            &   [erg/cm$^3$] & [erg/cm$^3$] &   & [erg/s]\\
             \hline\hline   
            &$5.3\times10^{-4}$ & $2.5\times10^{-5}$ & 21  &$8.9\times10^{43}$ \\
            \hline
            &$5.5\times10^{-4}$ & $2.5\times10^{-5}$ & 22  & $9.3\times10^{43}$  \\
        \end{tabular}
    \end{minipage} 
}
\par\vspace{0.5cm}
*\text{Fixed to the cooling break}\par
\end{table*}

The SED related to the \textit{NuSTAR}-2 observation differs from that of the baseline emission only in the X-ray domain (both soft and hard X-rays). Hence one can describe it with a set of model parameters that are very similar to those from the baseline emission. The magnetic field strength can be kept constant at $0.025$\,G between the two states, and the parameters describing the electron distribution are almost identical, with a slightly higher $\gamma_{\text{max}}'$ for the \textit{NuSTAR}-2 state. We note that the two leptonic frameworks employed agree well with each other, as shown in Fig.~\ref{fig:fits_lep_evolution}b).

On the other hand, the substantially higher emission during the \textit{NuSTAR}-1 observation requires a decrease in the magnetic field strength to $0.01$\,G connected to an increase in $\gamma_{\text{br}}'$, so that the HE peak is not underestimated. Additionally, we need a slightly harder spectral index and higher  $\gamma_{\text{max}}'$ with the higher flux state. The values of the corresponding model parameters are reported in Table~\ref{tab:fits_lep_evolution}, and the model curves are depicted in Fig.~\ref{fig:fits_lep_evolution}a). 

The simple one-zone leptonic approach does explain the evolution of the SEDs reasonably well. However, we still want to test our hypothesis of the low-state of Mrk\,501 being its baseline emission and therefore employ a two-zone scenario building on the hypothesis.

\subsubsection{Two-zone}
\label{subsec:2zone}

\begin{figure*}
\gridline{\fig{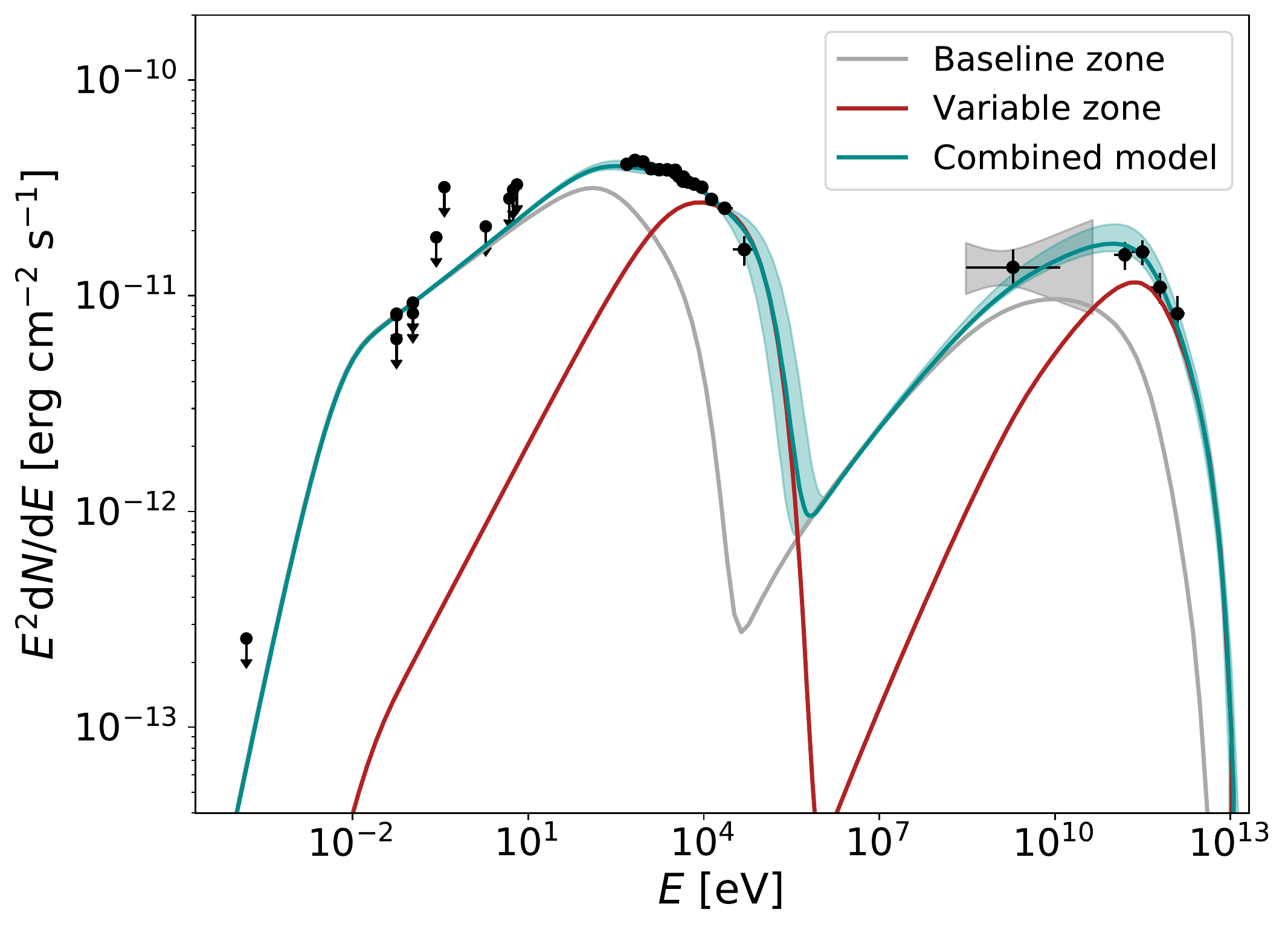}{0.49\textwidth}{(a) SED and two-zone SSC model for the \textit{NuSTAR}-1 observation.}
          \fig{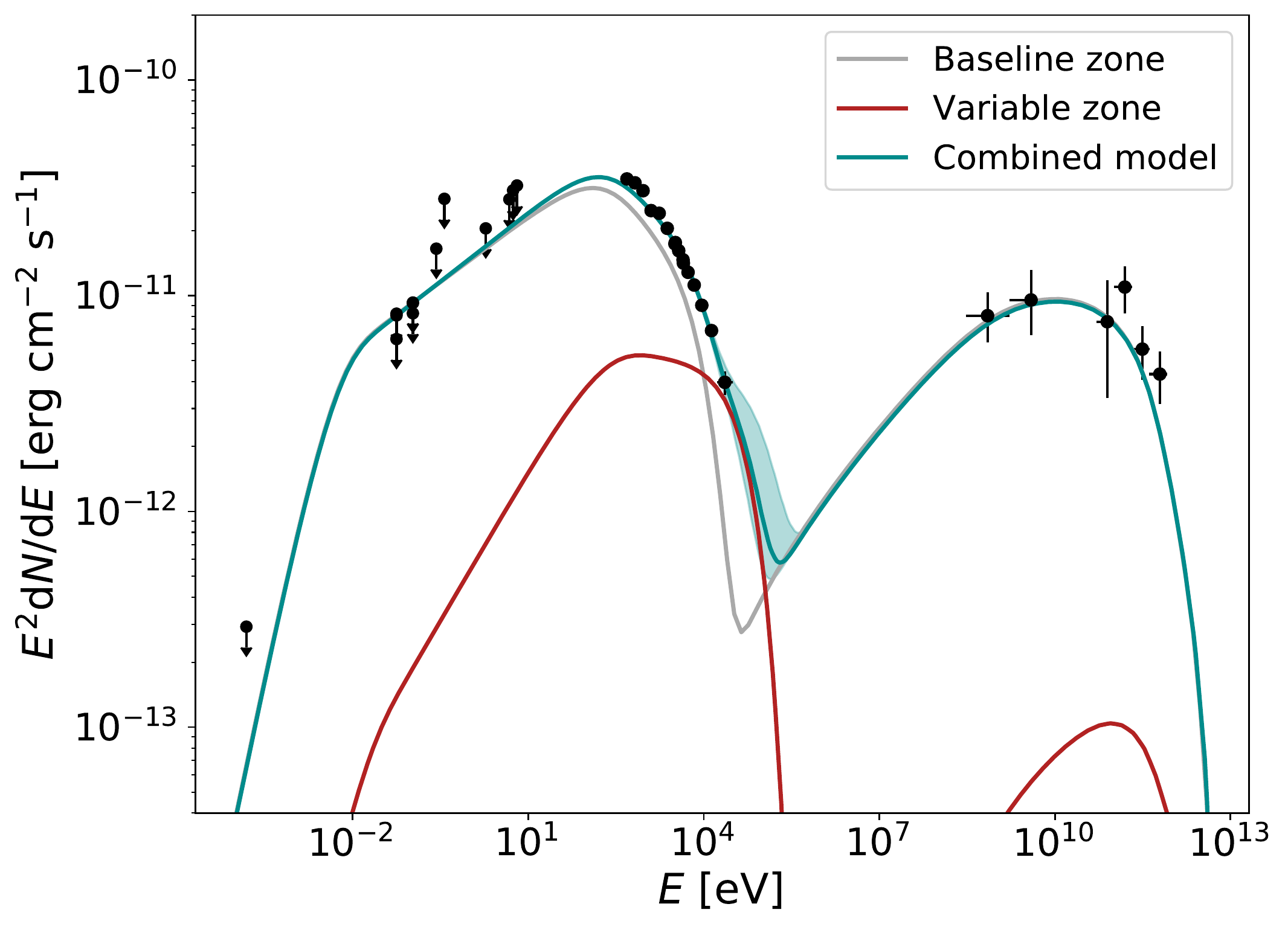}{0.49\textwidth}{(b) SED and two-zone SSC model for the \textit{NuSTAR}-2 observation.}
}
\gridline{\fig{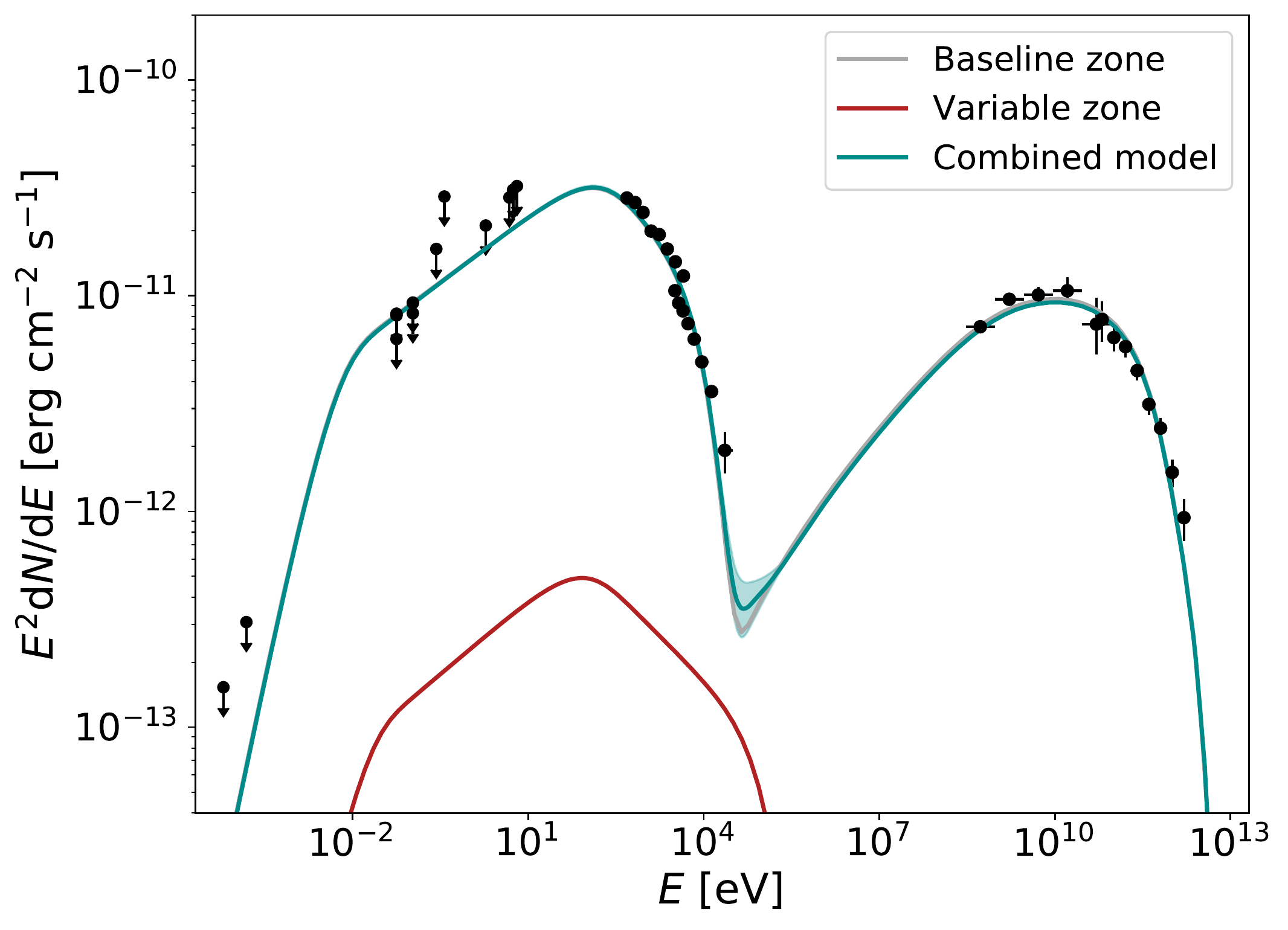}{0.49\textwidth}{(c) SED and two-zone SSC model for the low-state.}
          \fig{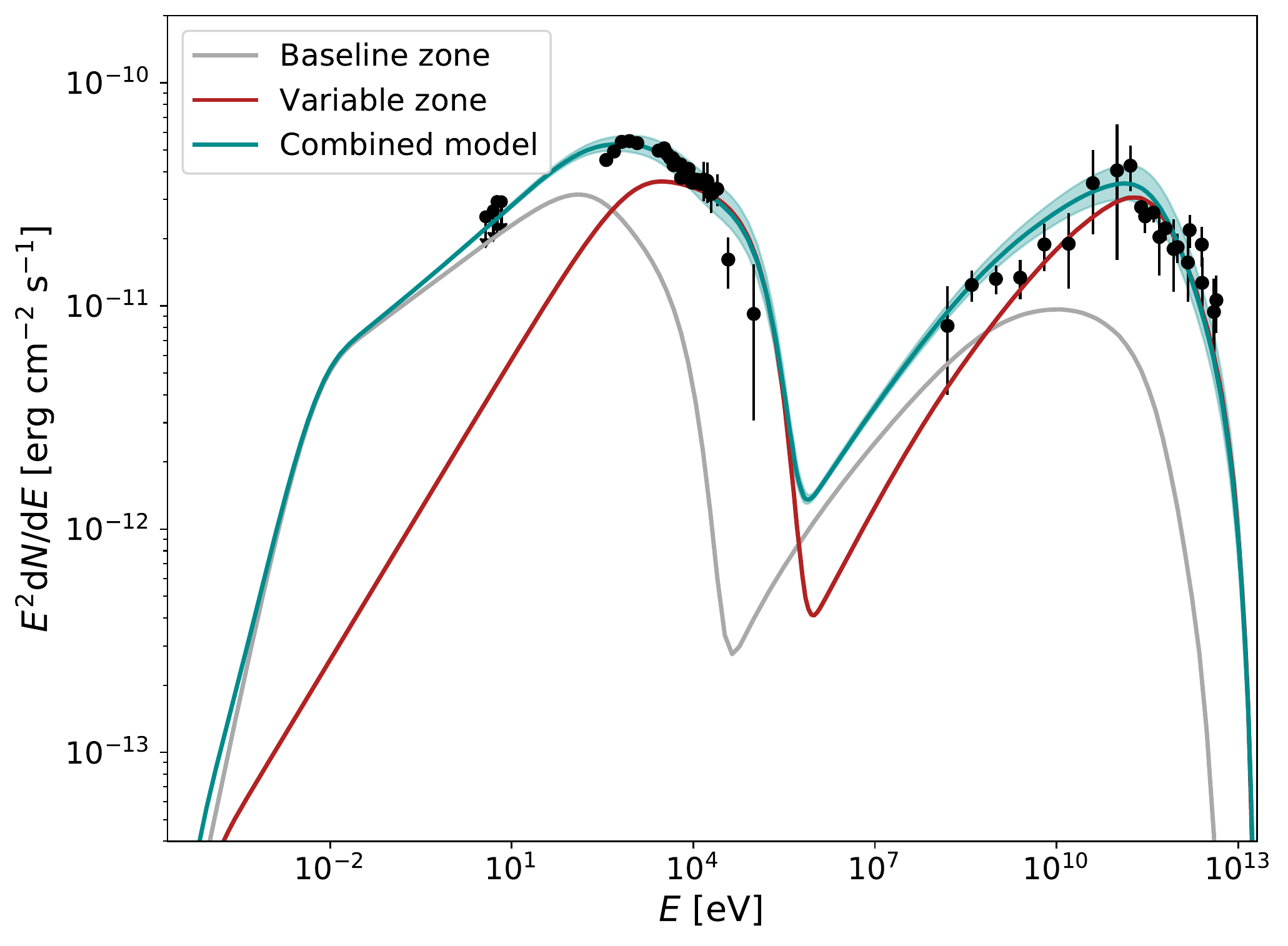}{0.49\textwidth}{(d) SED and related two-zone SSC model for the typical (non-flaring) state, as shown in \citet{Abdo_2011}.}
}

\caption{Leptonic two-zone models that describe the broadband SEDs of Mrk\,501 during the \textit{NuSTAR}-1 and \textit{NuSTAR}-2 observations on 2017-04-28  (MJD 57898) and 2017-05-25 (MJD~57871),  the low-state of Mrk 501 derived with data from 2017-06-17 to 2019-07-23 (MJD~57921 to MJD~58687), and the typical (non-flaring) activity state derived with data from 2009-03-15 to 2009-08-01 (MJD~54905 to MJD~55044), after excluding the 2009-May flare, as reported in \citep{Abdo_2011}. Data with frequencies in the UV or lower are considered as upper limits for the modeling of the blazar emission, and are therefore depicted with arrows.  The baseline zone is set to the one-zone SSC model of the low-activity state described in Table~\ref{tab:fits_lep_bb}, and is depicted in grey. The active (variable) emission zone is described using the jetset framework, and the resulting curves are depicted in red, with the parameters reported in Table~ \ref{tab:fits_lep_2zone}. The curves derived  from the combined broadband emission of these two independent zones are shown in cyan including a confidence interval obtained by the MCMC fit. See Section~\ref{subsec:2zone} for further details.}
    \label{fig:fits_lep_2zone}
\end{figure*}

In our second approach, we investigate the temporal evolution of the SED assuming the existence of two independent emission zones. The first is a stable and always present part of the SED as represented by the model describing the 2-year long low-activity state described in Section~\ref{sec:baseline_lep}. This baseline emission would be often out-shone by other emission regions occurring at different positions in the jet, that show a larger degree of brightness and variability. 
Therefore, we fit the various SEDs assuming two independent zones: one with the properties fixed to the ones given in Table~\ref{tab:fits_lep_bb}, and another zone that is variable.

\begin{table*}
    \centering
    {
    \caption{Parameter values from the active zone in the two-zone leptonic scenario used to describe the broadband SED of Mrk\,501 during the \textit{NuSTAR}-1 observation on 2017-04-28  (MJD 57898) shown in Fig.~\ref{fig:fits_lep_2zone}a), the \textit{NuSTAR}-2 observation on 2017-05-25 (MJD~57871) shown in Fig.~\ref{fig:fits_lep_2zone}b), 
 the low-activity state of Mrk 501 derived with data from 2017-06-17 to 2019-07-23 (MJD~57921 to MJD~58687), shown in Fig.~\ref{fig:fits_lep_2zone}c), 
and the typical (non-flaring) activity state derived with data from 2009-03-15 to 2009-08-01 (MJD~54905 to MJD~55044), as reported in \citet{Abdo_2011}, and shown in Fig.~\ref{fig:fits_lep_2zone}d).
    The baseline broadband emission (first zone) is described with the one-zone SSC model of the low-state reported in Table~\ref{tab:fits_lep_bb}.
    For the active (variable) emission zone, the radiating electron distribution is assumed to be a broken power-law with the spectral indices fulfilling $\alpha_{2}=\alpha_{1} +1$. The Doppler factor $\delta$ is set to 11. The table reports the magnetic field $B'$ and radius $R'$ of the emitting region and, for the assumed broken power-law describing the shape of the electron distribution, the electrons density $n_{\text{e}}'$, the first slope $\alpha_{1}$ and the minimum, maximum and break energies $\gamma_{\text{min}}'$,  $\gamma_{\text{max}}'$, $\gamma_{\text{br}}'$. Additionally, the table also reports the energy densities held by the electron population $U_{\text{e}}'$ and the magnetic field $U_{\text{B}}'$, their ratio, and the total jet luminosity $L_{\text{jet}}$. For the EBL $\gamma$-ray absorption at a redshift $z$=0.034, the model from Franceschini \citep{Franceschini} is used.}
    \label{tab:fits_lep_2zone}
    \begin{minipage}{1.\textwidth}
    \centering
        \begin{tabular}{ c | c c c c c c c}     
              & $R'$ [cm] &  $n_{\text{e}}'$ [cm$^{-3}$] & $B'$ [G] & $\alpha_{1}$  &  $\gamma_{\text{min}}'$  & $\gamma_{\text{br}}'$ &  $\gamma_{\text{max}}'$  \\
            \hline\hline   
            \textit{NuSTAR}-1  & $5.0\times10^{15}$ & 6.5 & $7.4\times10^{-2}$ & 2.0 & $1\times10^{3}$  &  $4.8\times10^{5}$* & $2.8\times10^{6}$ \\
            \textit{NuSTAR}-2  & $2.9\times10^{16 \dagger}$ & $3.4\times10^{-2 \dagger}$ &  $7.4\times10^{-2}$ &  2.1 &  $1\times10^{3}$  & $1.5\times10^{5}$* & $1.8\times10^{6}$   \\
            Low-state  & $5.0\times10^{16 \dagger\dagger}$ & $6.5\times10^{-3 \dagger}$ &  $7.4\times10^{-2}$ & 2.5 &  $1\times10^{3}$  &  $8.6\times10^{4}$*   & $2.6\times10^{6}$  \\
            \hline  
            Typical state  & $1.8\times10^{16}$ & 33.5 & $2.7\times10^{-2}$ & 2.1 & 100  &  $5.1\times10^{5}$* &  $5.1\times10^{6}$ \\
        \end{tabular}
    \end{minipage} 
    \par \vspace{0.75cm}
    \begin{minipage}{1.\textwidth}
    \centering
        \begin{tabular}{c | c c c c}     
             & $U_{\text{e}}'$ & $U_{\text{B}}'$  &  $U_{\text{e}}'$/$U_{\text{B}}'$ & $L_{\text{jet}}$ \\ 
            & $[$erg/cm$^3$] & [erg/cm$^3$]  &  & [erg/s] \\
            \hline\hline   
            \textit{NuSTAR}-1 & $3.6\times10^{-2}$ & $2.6\times10^{-4}$ & 139 & $1.2\times10^{43}$ \\
            \textit{NuSTAR}-2 & $1.5\times10^{-4}$ & $2.2\times10^{-4}$ & 0.7 & $3.7\times10^{42}$    \\
            Low-state & $1.5\times10^{-5}$ & $2.2\times10^{-4}$ & 0.1 & $6.6\times10^{42}$  \\
            \hline  
            Typical state & $1.8\times10^{-2}$ & $2.9\times10^{-5}$ & 621 & $8.7\times10^{43}$  \\
        \end{tabular}    \end{minipage} 
        }

\par\vspace{0.5cm}
*\text{Fixed to the cooling break}\par
$\dagger$\text{ Fixed to expanding $R'$ assuming a spherical blob}\par
$\dagger\dagger$\text{ Fixed to same expanding velocity as obtained from the fits above}\par
\end{table*}

For the additional region that contributes to the two broadband SEDs during the \textit{NuSTAR}-1 \& \textit{NuSTAR}-2 observations, we again assume a minimum electron energy of $\gamma_{min}' = 10^3$. For the \textit{NuSTAR}-1 SED we choose a radius of $R'=5\times 10^{15}$\,cm after trying a grid of $R'$ expanding up to $5\times 10^{16}$\,cm. The radius is limited to be smaller than $5\times 10^{16}$\,cm to not interfere with the baseline region. This is based upon the assumption that the emission region could extend over a large fraction of the cross section of the jet, and if that is the case, the smaller and more active region should be closer to the central engine than the baseline region. Since over time the active region travels along the jet, an upper limit has to be set for the radius, which relates to a minimum distance between the active and baseline regions to ensure our independent treatment of the two zones during the time interval considered in this study. 

Assuming a conical jet model, we follow \citet[][]{Zdziarski_2022} and assume a constant bulk Lorentz factor and therefore $\delta = 11$, as obtained for the low-state model, along the jet.

For the \textit{NuSTAR}-2 SED we allow $R'$ to expand with an upper limit related to the expected maximum change for a jet opening angle of $1/\Gamma_{b} \approx 1/\delta$ and the 27\,days between the two spectra. The electron density $n_{\text{e}}'$ is made dependent on the change in radius, assuming spherical symmetry for the blob and a constant number of electrons for the different models. Again, a broken power-law distribution is used for the radiating electrons, the break energy is fixed to the cooling break, and the two spectral indices, $\alpha_{1}$ and $\alpha_{2}$, are assumed to have a difference of 1. Since the first investigations showed a strong preference of the magnetic field to be around the same value as for the \textit{NuSTAR}-1 state, we fixed it to the same value during the evolution.

Owing to the very good agreement in the results obtained with the two leptonic software packages employed above, for the sake of simplicity, this time we decided to use only the jetset package to conduct the modeling of the data. 
The model parameters that describe well the temporal evolution of the data are reported in Table~\ref{tab:fits_lep_2zone}, and the model curves (for the two zones together and separated) are depicted in Fig.~\ref{fig:fits_lep_2zone}a) and Fig.~\ref{fig:fits_lep_2zone}b). 

We use this scenario to further expand the region for 24 more days until the historically low activity (baseline) starts, and fit the low-state SED with a radius fixed to the same expansion velocity as determined with the previous fits. The same assumptions for the electron density and the Doppler factor as before are used. The result is shown in Fig.~\ref{fig:fits_lep_2zone}c), and agrees with the data. Besides the change in the value of $R'$, the injection mechanism becomes less energetic between the three states: $\gamma_{\text{max}}'$ stays rather constant, but the spectral index softens with decreasing flux.

Additionally, we use the above-mentioned two-zone theoretical scenario to describe the typical broadband SED of Mrk\,501 reported in \citet{Abdo_2011}. For this purpose we use the broadband SED data points depicted in Fig.~8 of \citet{Abdo_2011}, but this time excluding the time interval MJD~54952–54982 when deriving the \textit{Fermi}-LAT spectrum to avoid a spectral hardening above 10 GeV caused by a flaring episode in May 2009 \citep[see Fig.~9 of][and discussion in Section~5.3]{Abdo_2011}. We assume the same connection between $\delta$ and $R'$ as above, but leave all other parameters free to vary. The radius is again limited to be smaller than $5\times 10^{16}$\,cm.  For the electron distribution, we used again a broken power-law. The result is shown in Fig.~\ref{fig:fits_lep_2zone}d). The theoretical model describes the SED data quite well, except for the peaky structure at the lowest end of the X-ray data, whose exact description (within less than 15\% accuracy) would require a complex shape for the electron distribution. We consider that, given the relatively small magnitude of the model-data disagreement, together with systematic uncertainties for the X-ray data \citep[which is already at the level of 10\%-15\%, as reported in ][]{Abdo_2011}, this extra complexity in the theoretical model is not justified.   

\section{Discussion}
\label{sec:discuss}
\subsection{Multi-band variability and correlations}
Section~\ref{sect:MWL_lc} discusses the variability and correlations connected to the low activity of Mrk\,501. Deduced from the obtained fractional variability in Fig.~\ref{fig:Fvar} the main changes between the 4-year period 2017--2020 and the 2-year period with historically low activity take place mainly in the HE X-rays and the VHE $\gamma$-rays.
Usually, an increase of $F_{var}$ with energy is seen for Mrk\,501. However, this behavior is more prominent when flaring episodes are included in the evaluated data set \citep[][]{Mrk501_MAGIC_2009b,Mrk501_MAGIC_2012, Mrk501_MAGIC_2014} than when looking at non-flaring activities \citep[][]{Mrk501_MAGIC_2008,Mrk501_MAGIC_2012}. The $F_{var}$ from the 4-year data set shows a double-peak structure with the HE X-rays and VHE $\gamma$-rays showing the highest variability at a similar level. The same was reported in \citet{Mrk501_MAGIC_2013} for Mrk\,501. $F_{var}$ computed with the 2-year low-state period also increases with energy, but it reaches a plateau at X-rays and $\gamma$-rays. A similar behavior, but limited by the sensitivity at that time, is reported in \citet{Mrk501_MAGIC_2008}.

What should be noted is that, for all mentioned previous results, far shorter time periods are taken into account than our two and four year data sets. Recently, the FACT collaboration \citep{FACT_Mrk501_2021} has reported a study performed with a Mrk\,501 data set that spans over 5.5 years, from December 2012 to April 2018, and therefore overlapping with the 2017--2020 data set featured in this paper. The study performed by the FACT collaboration also reports a double-bump  structure in the fractional variability vs energy, with $F_{var}$ values that are very similar to the ones reported here, except for higher values in the VHE regime. The latter result is not surprising because of the very large VHE activity shown by Mrk\,501 in the year 2014 \citep[][]{Mrk501_HESS_2014,Mrk501_MAGIC_2014}.

Hence, we can conclude that the variability pattern of Mrk\,501 is dominated by variations of the VHE $\gamma$-rays and X-rays, with certain periods showing large and different variability levels, while other time epochs show comparable variability levels. This observation would be consistent with the broadband emission described by a multiple zone scenario where, for the higher flux states, different variability patterns are influenced by different active regions. This variability pattern would be naturally expected from the two-zone model described in Section~\ref{subsec:2zone}, where the active (smaller) region would dominate the emission at X-rays and VHE, and hence the most variable parts of the SED of Mrk\,501.


Our study of possible correlations between the different wavebands, identifies a clear correlation without time lag between the LE/HE X-rays and VHE $\gamma$-rays (See Section~\ref{subsec:var_corr}).  This correlation has been reported multiple times for Mrk\,501 during flaring activities \citep[see e.g.][]{Mrk501_MAGIC_2014,Mrk501_MAGIC_2013,Mrk501_MAGIC_2012}, but is reported here for the first time with a statistical significance above 3$\sigma$ for an extensive period of low activity. It further supports that those scenarios where X-ray and VHE photons are produced by the same particle population, such as in the case for an SSC model, and dominate the broadband emission of Mrk\,501 during all kinds of activity states.

In addition to the correlation with zero time lag, we measure a 3$\sigma$- (DCF value $>$ 3$\sigma$ confidence level) X-ray vs. VHE correlation for a time delay of roughly 30\,days. This correlation exists also when comparing the 0.3--2~keV with the 2--10~keV X-ray fluxes measured with Swift-XRT, with the HE band preceding the LE band. This feature is visible (although with marginal significance) in both the 4-year and the 12-year data set, once the light curves are detrended to remove the long-term behaviour. It is the first time that such a observation is made indicating that a fraction of the X-rays are produced with some delay, particularly those at the lowest energies. This suggests that the acceleration of the high-energy particles may be produced by more than one process, with a potential connection between them. 
However, a word of caution is in order. The time scale of about 30 days coincides with the time scales where an increase in the confidence levels in our periodicity study (see Section~\ref{sec:app_period}) is apparent. Such increase in the confidence levels is probably related to the observing sampling of the source, which is affected by the moon periods that constrain the observations from the MAGIC telescopes (and hence all MWL observations that were coordinated with MAGIC observations). While we tried to include all these effects in our Monte Carlo simulations used to determine the confidence levels, we cannot exclude the existence of some non-accounted for effects that may introduce small artifacts at this time scale. 

For the UV versus R-band, the DCF analysis yields a clear peak centered at zero time lag, but relatively broad, extending over $\pm$1 week. This implies that these two neighbouring energy ranges have the same behavior, with a  variability time scale of the order of about one week, probably due to this emission being produced by relatively low-energy electrons, which have longer radiation timescales. 

The extensive data set spanning from 2008 to 2020 at HE $\gamma$-rays with Fermi-LAT, X-rays  with Swift-XRT and radio with OVRO, allowed us to study potential correlations among these bands with unprecedented precision. While the 4-year data set does not yield conclusive results, the 12-year data set shows 3$\sigma$-correlations between the HE $\gamma$-rays and the X-rays, as well as between HE $\gamma$-rays and radio (See Fig.~\ref{fig:dcfs_1} and Fig.~\ref{fig:dcfs_lt}). 
Such correlated behaviour had not been reported previously at a statistical significance above 3$\sigma$. 

In the case of HE $\gamma$-rays vs. X-rays, the DCF analysis shows a correlation that is highest at about zero time lag (for both X-ray energy bands, 0.3--2~keV and 2--10~keV), but with a very broad peak that extends $\pm$1 year. This positive correlation, extending over a large range of time lags, is ascribed to the long-term flux increase for multiple years, together with a decrease around the year 2017, which is observed in the LCs from all these instruments (see Fig.~\ref{fig:long_LC_comb}). 
A time shift of several months in these LCs would still keep the overall long-term behaviour, and hence the positive correlation obtained by our study. When the long-term flux variations in our light curves are removed, as described in Section~\ref{sec:correlations}, the correlation plots yield a single bump centered at a time lag zero, and with a width of about $\pm$10 days (see Fig.~\ref{fig:dcfs_lt}c) and Fig.~\ref{fig:dcfs_lt}d)). The DCF value for a time lag zero is clearly above the 3$\sigma$ confidence level, and hence statistically significant. Overall, the correlation plots from Fig.~\ref{fig:dcfs_lt} show that the keV and the GeV emissions are clearly correlated on both long (months, years) and short (weeks) timescales. This indicates that the radiation at these two energy bands is produced, at least partially, by the same population of particles, which further supports the SSC scenarios for the variable emission of this source. 
Further support for the SSC scenario is given by the correlation between VHE and HE $\gamma$-rays revealed after removing the long term trend from the 4-year data set. 

On the other hand, in the correlation between OVRO and \textit{Fermi}-LAT, the radio lags behind the $\gamma$-rays by more than 100 days, with slightly higher DCF values at $\sim$ 126\,days and 238\,days.  Using a 5.5 year data set from Mrk\,501, from 2012 to 2018, the FACT collaboration \citep[][]{FACT_Mrk501_2021} has recently shown a positive correlation  between the $\textit{Fermi}$-LAT fluxes and those from OVRO, with a relatively flat behaviour and with time lags extending from -300\,days to +300\,days; the significance of such correlated behaviour was not computed \citep[see Fig.~9 of ][]{FACT_Mrk501_2021}. In our study, that employs a 12-year data set, we show that the correlation is statistically significant ($>$\,3$\sigma$) only for time lags larger than -100 days, implying the radio emission is connected to the $\gamma$-rays with a delay larger than 3 months. This correlation, however, disappears when the light curves are detrended (see Fig.~\ref{fig:dcfs_1}b)), hence indicating that the radio and $\gamma$-ray emission are related only when considering the flux variations with time scales of a few months. A delay between radio and $\gamma$-ray fluxes with time delays of several tens or even hundred days has been reported with a statistical significant larger than 3$\sigma$ for other blazars  \citep[][]{Corr_gamma_radio,2021MNRAS.504.1427A}.
Such delays are usually explained by moving disturbances that travel along the jet. The time delay of the radio emission can then be converted to the distance between the locations dominating the radio and $\gamma$-ray emission using Eq. 1 in \citet{Corr_gamma_radio}. With a maximum jet speed of $\beta$=0.9 and a maximum Doppler factor in the radio regime of $\delta$=2 \citep{Lister_2021} we obtain the bulk Lorentz factor $\Gamma=1.45$ using Eq. 4 in \citet{Hovatta_2009}. Hence, for the time delay of 126\,days (238\,days) the emission regions are at maximum 0.27\,pc (0.51\,pc) apart. Following the recipe for the calculations of the distance from the central engine of the radio core emission $d_{core}$ for Mrk 421 in \citet{Corr_gamma_radio} replacing the radio core size with 0.13\,mas for Mrk\,501 \citep[][]{Weaver_2022} we obtain $d_{core}=2.05$\,pc. Hence, the $\gamma$-ray emission region is at least 1.78\,pc (1.54\,pc) away from the central engine, and located closer to the radio emission than the turbulent inner regions of the blazar, as already found for Mrk421 \citep{Corr_gamma_radio,2021MNRAS.504.1427A}. However, the Doppler factor of 11 obtained for the $\gamma$-ray emission site in Section~\ref{sec:low-activity-SED} is in contradiction to assuming a constant $\delta$=2 for the whole region between the two emission sites. Assuming $\delta$=11 for the whole region would give a distance between the sites of 5.66\,pc (10.69\,pc) which is bigger than $d_{core}$ and can therefore be excluded. If we consider a linear decrease from $\delta$=11 to $\delta$=2 between the two sites, we obtain a separation of 2.59\,pc (4.88\,pc). Already this simplified assumption relaxes the overshooting of $d_{core}$. More developed models such as decelerating jets have been proposed before \citep[][]{Meyer_2011} to explain the tension between Doppler factors measured in the radio regime and the ones needed to explain MWL SEDs.

The DCF analysis derived with the HE $\gamma$-ray data from \textit{Fermi}-LAT and the optical R-band yields a hint of correlation, although not significant. In the SSC models, one would expect a direct correlation between the eV and the GeV emission, since these two energy bands are produced by the same particle population, and hence are good tracers of the dynamics of these particles. The lack of a clear correlation could be explained by the existence of additional contributions to these energy bands, perhaps coming from different regions which are not physically connected. This would yield an uncorrelated behaviour that would worsen the correlated behaviour expected from the most simple one-zone SSC scenarios. Additionally, it could point to a hadronic nature of the emission which would not show a correlation between the two wavebands.

The absence of correlation observed between the polarization degree and the multi-band fluxes could be a sign of different mechanisms than shock acceleration being at work in the jet \citep[][]{Jorstad_2007}.  Additionally, the missing correlation between optical and radio in polarization degree and angle is hinting towards different emission zones producing the radiation. Nonetheless, since both optical and radio emission are known to have additional components, apart from the main blazar emission, the correlation could be partially washed out. Together with the rather sparse radio coverage, this does not allow us to draw significant conclusions. For the polarization angle, the preferred angles in the optical R-band and the 43\,GHz radio measurements coincide with the jet direction previously determined for Mrk\,501 \citep[][]{Weaver_2022}. The agreement between the measured polarization angles and the obtained jet direction could point towards a magnetic field perpendicular to the jet direction contributing to the collimation of the jet. However, the situation might be more complex taking into account relativistic effects \citep[][]{Lyutikov_2005}.

\subsection{Physics insights from the theoretical modeling of the broadband SEDs}

The 2017--2020 data set includes a 2-year time interval with the historical low activity of Mrk\,501 from mid-2017 to mid-2019. This provides us with a remarkable opportunity to study the baseline emission without disturbances from other contributions that are more variable and normally dominate the broadband emission of an active Mrk\,501. From previous correlation studies, as well as the ones included in this work, the preferred scenario for explaining the variable part of the blazar emission is of leptonic origin. However, for the baseline and thus stable part of the emission, this does not necessarily apply. Therefore, we consider both electrons and protons to be possible emitters for the observed low-state emission.

From the modeling results in Section~\ref{sec:spect} we can see that both relativistic electrons as well as protons can explain reasonably well the observed low-activity SED. Indeed, even the hadronic model can be constructed with a low magnetic field of 3\,G preserving the required low variability. The consideration of including hadronic components in the blazar emission origin is of particular importance since it allows to estimate the possible contribution from blazars to the flux of neutrinos and ultra-high-energy cosmic rays (UHECRs, E$>$10$^{18}$ eV). Taking into account the multi-messenger picture that is nowadays provided by neutrino telescopes like IceCube, we can use the upper limits in \citet{IceCube_UL2020} to check our predicted neutrino emission. The obtained neutrino rates of $10^{-5}$\,per\,year in our hadronic models are well below the upper limit of 1 neutrino per year (10.3 best fit neutrino in the first 10 years of IceCube data). For the lepto-hadronic models, higher neutrino rates are expected. However, our lepto-hadronic models prediction of $10^{-3}$ neutrinos per year is perfectly consistent with the non-detection of Mrk\,501 by IceCube. The higher neutrino flux is accompanied with higher estimated jet powers compared to the hadronic case. This can be accounted for by fine-tuning the low-energy part of the proton distribution. Further, the hadronic scenario provides very high proton energies up to $\gamma_{max}'\sim10^{10}$ potentially indicating BL Lacs as UHECR accelerators. For the available data set, leptonic, hadronic and lepto-hadronic models are valid scenarios to explain the observed low-activity state of Mrk\,501. As one of the currently missing pieces of the multi-messenger blazar picture, data in the MeV energy range would be extremely efficient in reducing the degeneracy among the various emission models. Furthermore, future X-ray/$\gamma$-ray polarization measurements would help in distinguishing between theoretical models due to the different predictions in polarization properties between the models. 

To explain such a long-lasting stable emission, a standing shock scenario \citep[see e.g.][]{Marscher_2008,Marscher_2014} provides a reasonable scenario as previously discussed for the quiescent behavior of Mrk421 in \citet{Abdo_2011_Mrk421}. Shock acceleration has been shown before to be a plausible scenario explaining the behavior of Mrk\,501 \citep[][]{Baring_2016} and would be in line with the obtained spectral indices in all our applied models. It has been further strengthened for Mrk\,501 by the first X-ray polarization measurements reported in \cite{IXPE_Mrk501} whose higher polarization degree compared to the optical measurements supports  a shock acceleration scenario. In this scenario, the particles are accelerated when the jet flow crosses the standing shock, and subsequently radiation is emitted. As long as the particle flow and shock properties remain stable, constant acceleration and emission is taking place. 

In what concerns the time evolution of the SED in the months before the historically low activity that starts in mid-2017, only leptonic scenarios (one-zone and two-zone models) are considered to explain the variations in the broadband emission, because of the tight correlations between X-rays and $\gamma$-rays (both HE and VHE).

In earlier studies, transitions between different states were commonly attributed to changes in the break energy $\gamma_{br}'$ for Mrk\,501, and therefore to the injection of electrons \citep[][]{Mrk501_MAGIC_2014, Mrk501_MAGIC_2012, Mrk501_MAGIC_2006}. Within the one-zone leptonic scenario described in Section~\ref{subsec:1zone}, the most important parameter change occurs in the magnetic field $B'$, that increases from 0.01 to 0.025\,G as Mrk\,501 transitions towards the historical low activity in mid-2017. The observed changes in $B'$ go hand in hand with changes in $\gamma_{\text{br}}'$ since they are linked through the cooling break in our scenario. Furthermore, the electron distribution requires small adjustments with the power-law indices $\alpha_1$ (which is linked to $\alpha_2$) becoming softer, and the maximum energy $\gamma_{\text{max}}'$ decreasing with time, as Mrk\,501 reaches the low activity.  We propose that a small increase in the ambient parameter $B'$ explains the observed broadband SED time evolution when the injected electron distribution flowing through the shock adjusts to the surrounding.

For all three emission states in the 4-year data set, we determined the synchrotron peak frequency $\nu_s$. First we used the phenomenological description in \citet{blazar_seq} applied to all data points to determine $\nu_s$. Additionally, we exploited the one-zone leptonic and hadronic modeling results, which are described further below, for further estimates of $\nu_s$. Table~\ref{tab:nu_s_states} summarized the peak frequencies, which for the low-state are of the order of $5\times10^{15}$ to $5\times10^{16}$\,Hz. The differences between the estimates have their root in the treatment of the low-energy points as upper limits for the theoretical models while for the phenomenological fits this assumption is not made so as to be able to compare it to previously published results more easily. However, the synchrotron peak is not very well covered by our data set and can therefore not be identified more specifically. While for the \textit{NuSTAR}-2 the peak frequencies hardly shift, the \textit{NuSTAR}-1 state shows a clear shift to higher frequencies between $2\times10^{16}$ to $3\times10^{17}$\,Hz. The phenomenological evaluation clearly places Mrk\,501 in the HSP regime.

In order to test our assumption of the low-state being the baseline emission of Mrk\,501, our second scenario assumes two emission zones as described in Section~\ref{subsec:2zone}. We consider that there is one region that is responsible for the historical low-activity broadband SED, and a distinct and independent region that is responsible for the main variations in the SED. These variations can occur on timescales of days or weeks, and are particularly dramatic in the X-ray and $\gamma$-ray energy range. The low-activity broadband emission could be produced by high-energy electrons or high-energy protons, as demonstrated in Section~\ref{sec:low-activity-SED}. Since it is assumed to be steady (in reality there could be small long-term flux variations, which are not considered here), it is irrelevant whether we use a leptonic or hadronic framework in our two-zone scenario. For simplicity, we used the one-zone SSC scenario described in Section~\ref{sec:baseline_lep}. Because of the above-mentioned flux variability and correlations, as we do with the one-zone scenario, we consider that the broadband emission produced in the second region is dominated by high-energy electrons.
We assume a region with a size $R'$ smaller than that of the baseline one, and with a higher $B'$ field, potentially indicating a position closer to the central engine. Over time, the active region expands, the density of radiating particles becomes smaller, and the spectral index softens. If we assume a dependence of the jet magnetic field on the size of the blob using the relation in  \citet[][]{Tramacere_2022}, we can conclude that the variable region stays at a stable position in the jet when expanding. Additionally, it is only valid if the radius of the variable region makes up only a fraction of the low-state radius which suggests that at least the variable region does not cover the whole jet radius. However, the magnetic field could also be dominated by local contribution traveling together with the shock region inside the jet.

The radio emission is not reproduced by the models since it is assumed to originate from a more extended region in the outer part of the jet than considered here. Therefore, at least a third if not further multiple or more complex zones would be required to reconstruct the full SEDs as for example demonstrated by \citet{2019MNRAS.482.4798L}. Other alternative models could potentially also reproduce the low Doppler factor observed in the radio regime as shown by \citet{2005A&A...432..401G} using a structured jet or with a decelerating jet by \citet{2003ApJ...594L..27G}.
It is, however, remarkable that the flux density predicted by the baseline model matches reasonably well with what is observed with millimetre-wavelength VLBI.  Using global mm-VLBI array observations, \citet{Giroletti2008} measured a flux of $S_\mathrm{86\,GHz}\sim45$\,mJy for the central component seen at the jet base, corresponding to $\nu F_\nu \sim 4.0\times10^{-14}$ erg\,cm$^{-2}$\,s$^{-1}$. The deconvolved size of such a emission region is smaller than $\sim 5\times 10^{16}$\,cm (in the observer's frame). It is therefore possible that at least the low-state baseline VHE emission has a counterpart that is directly accessible with mm-VLBI (while the more compact components responsible for the variable emission remain self-absorbed).

This  two-zone scenario can also be used to describe the typical (non-flaring) SED of Mrk\,501 derived with data from 2009 and reported in \citet{Abdo_2011}. To account for the higher radio flux (at 15 GHz) in 2009 with respect to the low-activity state after mid-2017 (see Fig.~\ref{fig:long_LC_comb}), while considering the correlations between the radio and the $\gamma$-rays reported in Section~\ref{sec:correlations}, we had to decrease the minimum energy of the high-energy electrons, down to $\gamma_{\text{min}}'=100$. The active region produces a radio emission that is comparable to that of the baseline region (see Fig.~\ref{fig:fits_lep_2zone}d)), and hence account for variations in radio, X-rays and $\gamma$-rays. 
In this realization of the model, the active region is quite far from equipartition ($U_{\text{e}}'/U_{\text{B}}' > 5 \times 10^{2}$), but well within the values that are considered possible for HSPs such as Mrk\,501  \citep[][]{Mrk501_MAGIC_2009b,Mrk501_MAGIC_2012, MAGIC_EHBL}. Such high deviation from equilibrium has been, for example, consistently reproduced using  relativistic, oblique, magnetohydrodynamic shocks by \citet{Baring_2016}.


\section{Summary and Conclusions}
\label{sec:summary}

This paper reports a detailed characterization of the time evolution of the broadband emission of Mrk\,501 during an extended period of very low activity, that spans from 2017 to 2020. The coordinated observations involve a large number of instruments, including MAGIC, \textit{Fermi}-LAT, \textit{Swift}, GASP-WEBT and OVRO. Additionally, three 10-hour long observations  with \textit{NuSTAR} yielded a precise measurement of the falling segment of the low-energy bump, which is expected to be dominated by the highest energy electrons at the source.

During this extended period of low activity, we identify clear variability throughout the electromagnetic spectrum, with the highest flux variations occurring at the X-ray and VHE $\gamma$-ray energies, which are found to be positively correlated. The correlated variations in the X-rays and VHE $\gamma$-ray fluxes from Mrk\,501 have been reported many times for flaring activities, but it has been elusive during low activity. This observation indicates that the mechanisms that dominate the X-ray/VHE variations during very low activity are not substantially different from those that are responsible for the emission during flaring activity.


Additionally, we use a 12-year data set, from 2008 to 2020, to evaluate the correlations among the radio, X-rays and HE $\gamma$-ray emission. The extension and precision of this data set allows us to see, for the first time for Mrk\,501, statistically significant ($>$\,3$\sigma$) correlations between the X-ray and the HE $\gamma$-ray emission, as well as between the HE $\gamma$-ray and the radio fluxes, with the radio emission lagging the HE $\gamma$-ray emission by more than 100\,days. 
The X-ray/HE $\gamma$-ray correlation, which occurs on both short (weeks) and long (months and years) time scales,
unambiguously indicates a common origin between (at least a fraction of) the emission in these two bands. The radio/HE $\gamma$-ray correlation, which only happens on long (months) timescales, may also indicate a relation between the origin of these two emissions, but the large time lag between them introduces an additional complexity. Correlations between radio and HE $\gamma$-rays, with delays of several tens or hundreds of days, have been observed in a number of blazars, and are often considered as a signature of the radio emission being downstream of the $\gamma$-ray emission. For Mrk\,501 this would place the $\gamma$-ray region either close to the radio region or very close to the central engine depending on the assumptions made for the corresponding Doppler factor.

Triggered by the various claims of periodicity in the emission of Mrk\,501 \citep[][]{Osone_2006,Kranich_1999,Bhatta_2019}, and by the regular enhancements in the DCF values with a periodicity of about 30 days when using both the 4-year and the 12-year data sets (see Fig.~\ref{fig:dcf_xrt-le_xrt-he} and Fig.~\ref{fig:dcfs_detrend_xrt}), we exploited the 12-year data set from Fig.~\ref{fig:long_LC_comb} to search for periodicity in the radio, X-ray and HE $\gamma$-rays. We do not see any significant ($>3\sigma$) periodicity for any of the energy bands considered (see Appendix \ref{sec:app_period} for details). There are some indications for periodic behaviour in the data at certain timescales (e.g. $\sim$30 days for X-rays). They are, however, mostly related to the binning and sampling of the light curves.

One of the most interesting results from this study is the identification of a  2-year long epoch, from mid-2017 to mid-2019, when the X-ray and VHE $\gamma$-ray fluxes of Mrk\,501 are the lowest detected to date, and may be considered as the baseline emission of this archetypal TeV blazar. The broadband SED of this historically low activity could be accurately characterized, and modeled reasonably well, within various theoretical frameworks that consider distinct origins for the high-energy emission of the source, namely a purely leptonic, a purely hadronic, and a lepto-hadronic scenario. The size of the emitting region responsible for this baseline emission coincides with the scales probed by mm-VLBI observations, which also match the flux density expected at the low energy tail of the SED model.

Owing to the results derived from the correlation studies reported in this paper, as well as those previously published that relate to typical and/or flaring activities of Mrk\,501, leptonic scenarios are preferred to describe the variable components in the broadband emission. However, for the bulk of the stable (baseline) emission of Mrk\,501, these arguments based on variability and correlations do not necessarily apply, and hence considering both high-energy electrons and high-energy protons as contributors for the high-energy emission of this blazar seems viable. Our study shows that, even with a well measured broadband SED, the degeneracy among these very distinct theoretical models is large, and the current data do not offer the necessary knobs to distinguish between them. This means that on the one hand we should continue the MWL monitoring with our currently available instruments to generate more long term blazar data sets covering different emission states for which time-dependent models could reduce the degeneracy. On the other hand, we should push towards collecting even more MWL and multi-messenger information on this source, for example, precise polarization measurements of the X-ray emission, or measuring the $\gamma$-ray emission in the MeV enery range, where the different models differ substantially. In this context, the 
recently launched IXPE satellite\footnote{\url{https://www.nasa.gov/mission_pages/ixpe/index.html}}, or the satellite missions e-ASTROGAM \citep[][]{eAstrogram}, COSI \citep{COSI} and AMEGO \citep{amego}, which are being constructed or considered for construction in the next years, could play a crucial role in unraveling the different emission mechanisms at work in Mrk\,501, and blazars in general. Last, but not least, an accurate measurement of the high-energy neutrino flux from Mrk\,501 (and its potential flux variation) would clearly break many model degeneracies. Such a measurement is unlikely to be recorded by the current IceCube detector (even if one integrates over ten more years), but may be provided by the future generations of neutrino telescopes, such as IceCube-Gen2 \citep[][]{IceCube_Gen2} and KM3NET \citep{KM3net_2016}. 

\section*{Author contribution}
A. Arbet Engels: MAGIC analysis cross-check; M. Cerruti: theoretical modeling and interpretation, paper drafting; S. Gasparyan: theoretical modeling and interpretation, paper drafting; L. Heckmann: project leadership, coordination of MWL data analysis, MAGIC and Fermi data analysis, variability and correlation analysis, theoretical modeling and interpretation, paper drafting; D. Paneque: organization of the MWL observations and coordination of the MWL data reduction, theoretical interpretation, paper drafting; N. Sahakyan: theoretical interpretation; The rest of the authors have contributed in one or several of the following ways: design, construction, maintenance and operation of the instrument(s) used to acquire the data; preparation and/or evaluation of the observation proposals; data acquisition, processing, calibration and/or reduction; production of analysis tools and/or related Monte Carlo simulations; overall discussions about the contents of the draft, as well as related refinements in the descriptions.

\begin{acknowledgments}
We would like to thank the Instituto de Astrof\'{\i}sica de Canarias for the excellent working conditions at the Observatorio del Roque de los Muchachos in La Palma. The financial support of the German BMBF, MPG and HGF; the Italian INFN and INAF; the Swiss National Fund SNF; the grants PID2019-104114RB-C31, PID2019-104114RB-C32, PID2019-104114RB-C33, PID2019-105510GB-C31, PID2019-107847RB-C41, PID2019-107847RB-C42, PID2019-107847RB-C44, PID2019-107988GB-C22 funded by MCIN/AEI/ 10.13039/501100011033; the Indian Department of Atomic Energy; the Japanese ICRR, the University of Tokyo, JSPS, and MEXT; the Bulgarian Ministry of Education and Science, National RI Roadmap Project DO1-400/18.12.2020 and the Academy of Finland grant nr. 320045 is gratefully acknowledged. This work was also been supported by Centros de Excelencia ``Severo Ochoa'' y Unidades ``Mar\'{\i}a de Maeztu'' program of the MCIN/AEI/ 10.13039/501100011033 (SEV-2016-0588, SEV-2017-0709, CEX2019-000920-S, CEX2019-000918-M, MDM-2015-0509-18-2) and by the CERCA institution of the Generalitat de Catalunya; by the Croatian Science Foundation (HrZZ) Project IP-2016-06-9782 and the University of Rijeka Project uniri-prirod-18-48; by the Deutsche Forschungsgemeinschaft (SFB1491 and SFB876); the Polish Ministry Of Education and Science grant No. 2021/WK/08; and by the Brazilian MCTIC, CNPq and FAPERJ.
The \textit{Fermi} LAT Collaboration acknowledges generous ongoing support from a number of agencies and institutes that have supported both the development and the operation of the LAT as well as scientific data analysis. These include the National Aeronautics and Space Administration and the Department of Energy in the United States, the Commissariat \`a l'Energie Atomique and the Centre National de la Recherche Scientifique / Institut National de Physique Nucl\'eaire et de Physique des Particules in France, the Agenzia Spaziale Italiana and the Istituto Nazionale di Fisica Nucleare in Italy, the Ministry of Education, Culture, Sports, Science and Technology (MEXT), High Energy Accelerator Research Organization (KEK) and Japan Aerospace Exploration Agency (JAXA) in Japan, and the K.~A.~Wallenberg Foundation, the Swedish Research Council and the Swedish National Space Board in Sweden. Additional support for science analysis during the operations phase is gratefully acknowledged from the Instituto Nazionale di Astrofisica in Italy and the Centre National d'\'Etudes Spatiales in France. This work performed in part under DOE Contract DE-AC02-76SF00515.
\end{acknowledgments}
\begin{acknowledgments}
L.~H. acknowledges the support from the 2019 Fermi Summer School hosted by the University of Delaware, and the Fermi Science Support Center and the lead organizers and instructors Joe Asercion, Erik Blaufuss, Regina Caputo, Mattia Di Mauro, Joe Eggen, Manel Errando, Henrike Fleischhack, Adam Goldstein, Colby Haggerty, Liz Hays, Jamie Holder, Julie McEnery, Jeremy Perkins, Judy Racusin, Alex Reustle, Jacob Smith, and Leo Singer.
M.C. has received financial support through the Postdoctoral Junior Leader Fellowship Programme from la Caixa Banking Foundation, grant No. LCF/BQ/LI18/11630012
A.A.E and D.P acknowledge support from the Deutsche Forschungs Gemeinschaft (DFG, German Research Foundation) under Germany’s Excellence Strategy EXC-2094 – 390783311.
M.\,B. acknowledges support from the YCAA Prize Postdoctoral Fellowship and from the Black Hole Initiative at Harvard University, which is funded in part by the Gordon and Betty Moore Foundation (grant GBMF8273) and in part by the John Templeton Foundation.
The Abastumani team acknowledges financial support by the Shota Rustaveli
NSF of Georgia under contract FR-19-6174.
The R-band photometric data from the University of Athens Observatory (UOAO) were obtained after utilizing the robotic and remotely controlled instruments at the facilities.
This research was partially supported by the Bulgarian National Science Fund of the Ministry of Education and Science under grants DN 18-10/2017,  KP-06-H28/3 (2018), KP-06-H38/4 (2019) and KP-06-KITAJ/2 (2020) and the National RI Roadmap Project D01-383/18.12.2020.
The Skinakas Observatory is a collaborative project of the University of Crete, the Foundation for Research and Technology -- Hellas, and the Max-Planck-Institut f\"ur Extraterrestrische Physik.
Part of this work is based upon observations carried out at the Observatorio Astron\'omico Nacional on the Sierra San Pedro M\'artir (OAN-SPM), Baja California, M\'exico.
This article is partly based on observations made with the IAC-80 operated on the island of Tenerife by the Instituto de Astrofisica de Canarias in the Spanish Observatorio del Teide. Many thanks are due to the IAC support astronomers and telescope operators for supporting the observations at the IAC-80 telescope.  This article is also based partly on data obtained with the STELLA robotic telescopes in Tenerife, an AIP facility jointly operated by AIP and IAC.  
G.\,D. acknowledges observing grant support from the Institute of Astronomy and Rozhen NAO BAS through the bilateral joint research project “Gaia Celestial Reference Frame (CRF) and fast variable astronomical objects” (2020–2022, head – G. Damljanovic). This research was supported by the Ministry of Education, Science and Technological Development of the Republic of Serbia (contract No 451-03-68/2022-14/200002). 
\end{acknowledgments}
\begin{acknowledgments}
M.\,D.\,J thanks the Brigham Young University Department of Physics and Astronomy for continued support of the ongoing extragalactic monitoring program at the West Mountain Observatory.
S.K. acknowledges support from the European Research Council (ERC) under the European Unions Horizon 2020 research and innovation programme under grant agreement No.~771282.
This research has made use of data from the OVRO 40-m monitoring program (Richards, J. L. et al. 2011, ApJS, 194, 29), supported by private funding from the California Insitute of Technology and the Max Planck Institute for Radio Astronomy, and by NASA grants NNX08AW31G, NNX11A043G, and NNX14AQ89G and NSF grants AST-0808050 and AST- 1109911.
This publication makes use of data obtained at Mets\"ahovi Radio Observatory, operated by Aalto University in Finland.
The Medicina radio telescope is funded by the Italian Ministry of University and Research (MUR) and is operated as a National Facility by the Italian National Institute for Astrophysics (INAF).
The Submillimeter Array is a joint project between the Smithsonian Astrophysical Observatory and the Academia Sinica Institute of Astronomy and Astrophysics and is funded by the Smithsonian Institution and the Academia Sinica.
The RATAN-600 observations were supported in the framework of the national project "Science" by the Ministry of Science and Higher Education of the Russian Federation under the contract 075-15-2020-778. 
\end{acknowledgments}
\begin{acknowledgments}
The research at Boston University was supported in part by NASA Fermi GI grants  80NSSC20K1567 and 80NSSC22K1571,  National Science Foundation grant AST-2108622, and the NRAO Student Observing Support Program.  
This study was based (in part) on observations conducted using the 1.8 m Perkins Telescope Observatory (PTO) in Arizona (USA), which is owned and operated by Boston University.
The VLBA is an instrument of the National Radio Astronomy Observatory, USA. The National Radio Astronomy Observatory is a facility of the National Science Foundation operated under cooperative agreement by Associated Universities, Inc.
The IAA-CSIC co-authors acknowledge financial support from the Spanish "Ministerio de Ciencia e Innovaci\'o" (MCINN) through the "Center of Excellence Severo Ochoa" award for the Instituto de Astrof\'isica de Andaluc\'ia-CSIC (SEV-2017-0709). Acquisition and reduction of the POLAMI data was supported in part by MICINN through grants AYA2016-80889-P and PID2019-107847RB-C44. The POLAMI observations were carried out at the IRAM 30m Telescope. IRAM is supported by INSU/CNRS (France), MPG (Germany) and IGN (Spain).
This publication makes use of data products from the Near-Earth Object Wide-field Infrared Survey Explorer (NEOWISE), which is a joint project of the Jet Propulsion Laboratory/California Institute of Technology and the University of Arizona. NEOWISE is funded by the National Aeronautics and Space Administration.
\end{acknowledgments}

%



\bibliography{bibliography_paper}{}
\bibliographystyle{aasjournal}

\appendix

\section{Additional information - Long term light curve}
\begin{figure*}
   \centering
   \includegraphics[width=0.9\textwidth]{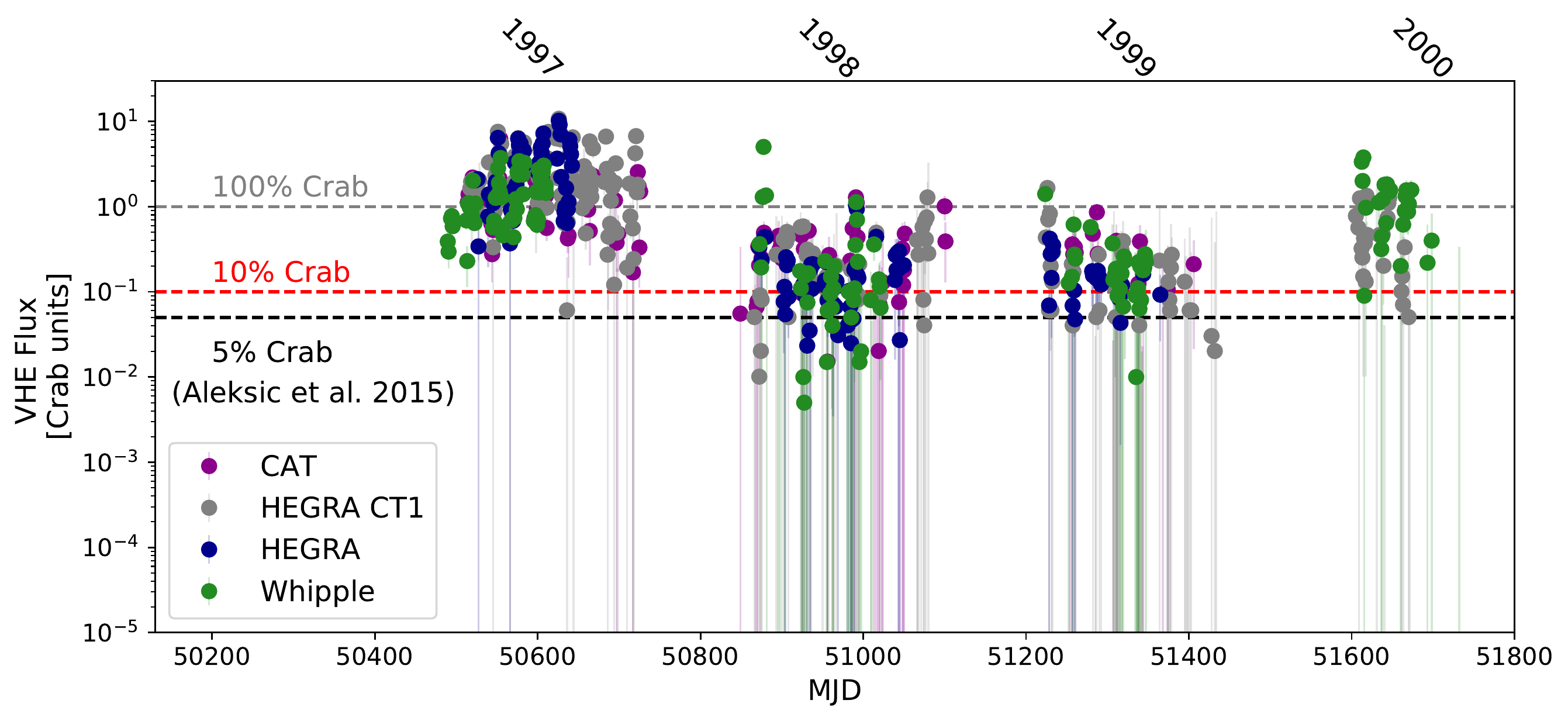}
   \caption{VHE data for Mrk\,501 collected between 1997 and 2000 as summarized in \citet{Mrk501_2005}  scaled to the Crab Nebula flux (Crab unit). Flux values of 100\% (grey), 10\% (red) and 5\% (black) that of the Crab Nebula are indicated by the dashed lines for reference.}
   \label{fig:long_LC_old}
\end{figure*}

Fig.~\ref{fig:long_LC_old} shows data from the first generation IACTs for Mrk\,501 as described in Section~\ref{sec:intro}.
\clearpage
\newpage

\section{Additional information - Instruments and analysis}
In this section additional information for the descriptions in Section~\ref{sec:instruments} is provided. Table~\ref{tab:spec_para_MAGIC} and Table~\ref{tab:spec_para_Fermi} summarize the spectral parameters for the MAGIC and \textit{Fermi}-LAT analyses for the spectra described in Section~\ref{sec:spect}. Table~\ref{tab:Offset_optical} summarizes the offsets applied to the optical R-band data using the KVA data as a reference as described in Section~\ref{subsec:inst_opt}.

\begin{table}
\centering
\begin{tabular}{ c | c  c c}     
 & N$_0$& $\alpha$ & E$_0$  \\
  & [$10^{-11}$\,cm$^{-2}$\,s$^{-1}$\,TeV$^{-1}$] &  & [TeV]  \\
\hline
Low-state & 2.56 $\pm$ 0.09 & -2.67 $\pm$ 0.04 & 0.3 \\
\textit{NuSTAR}-1 & 9.96 $\pm$ 0.64 & -2.35 $\pm$ 0.09 & 0.3  \\
\textit{NuSTAR}-2 & 4.34 $\pm$ 0.45 & -2.48 $\pm$ 0.16 & 0.3 \\
\textit{NuSTAR}-3 & 1.69 $\pm$ 0.46 & -2.65 $\pm$ 0.45 & 0.3  \\
\end{tabular}
\caption{Spectral parameters for the MAGIC analysis described in Section~\ref{sec:instruments} for the spectra described in Section~\ref{sec:spect}. Displayed are the spectral parameters of the applied power-law: the prefactor N$_0$, the spectral index $\alpha$, the energy scaling parameter E$_0$.} 
\label{tab:spec_para_MAGIC}
\end{table}

\begin{table}
\centering
\begin{tabular}{ c | c  c c}     
 & N$_0$& $\alpha$ & E$_0$  \\
  & [$10^{-12}$\,cm$^{-2}$\,s$^{-1}$\,MeV$^{-1}$] &  & [MeV]  \\
\hline
Low-state & 5.19 $\pm$ 0.22 & -1.92 $\pm$ 0.03 & 1000 \\
\textit{NuSTAR}-1 & 8.39 $\pm$ 1.72 & -2.00 $\pm$ 0.09 & 1000  \\
\textit{NuSTAR}-2 & 5.09 $\pm$ 1.20 & -1.83 $\pm$ 0.16 & 1000 \\
\textit{NuSTAR}-3 & 3.43 $\pm$ 1.62 & -1.59 $\pm$ 0.23 & 1000  \\
\end{tabular}
\caption{Spectral parameters for the \textit{Fermi}-LAT analysis described in Section~\ref{sec:instruments} for the spectra described in Section~\ref{sec:spect}. Displayed are the spectral parameters of the applied power-law: the prefactor N$_0$, the spectral index $\alpha$, the energy scaling parameter E$_0$.} 
\label{tab:spec_para_Fermi}
\end{table}

\begin{table}
\centering
\begin{tabular}{ c | c}     
Instrument & Offset [mJy] \\
\hline
West Mountain (91 cm) & -1.20 \\
Vidojevica (140 cm) & -1.88 \\
Vidojevica (60 cm) & -2.66 \\
University of Athens Observatory (UOAO)  & -3.99 \\
Tijarafe (40 cm) & -0.93 \\
Teide (STELLA-I) & -0.48 \\
Teide (IAC80) & -0.14 \\
St. Petersburg & -0.38 \\
Skinakas & -1.40 \\
San Pedro Martir  (84 cm) & -0.13 \\
Rozhen (200 cm) & -0.65 \\
Rozhen (50/70 cm) & -1.07 \\
Perkins & -0.80 \\
New Mexico Skies (T21) & -3.61 \\
New Mexico Skies (T11) & -1.41 \\
Lulin (SLT) & -0.09 \\
Hans Haffner & -1.35 \\
Crimean (70cm; ST-7; pol) & -0.26 \\
Crimean (70cm; ST-7) & -0.47 \\
Crimean (70 cm; AP7) & -0.49 \\
Connecticut (51 cm) & -1.09 \\
Burke-Gaffney & -3.76 \\
Belogradchik & -2.30 \\
AstroCamp (T7) & -3.37 \\
Abastumani (70 cm) & -5.00 \\
AAVSO  & -4.77 \\
\end{tabular}
\caption{Offsets applied to the optical R-band data using the KVA data as a reference, as described in Section~\ref{subsec:inst_opt}.} 
\label{tab:Offset_optical}
\end{table}

\section{Additional information - MWL data}
\begin{figure*}
   \centering
   \includegraphics[width=0.9\textwidth]{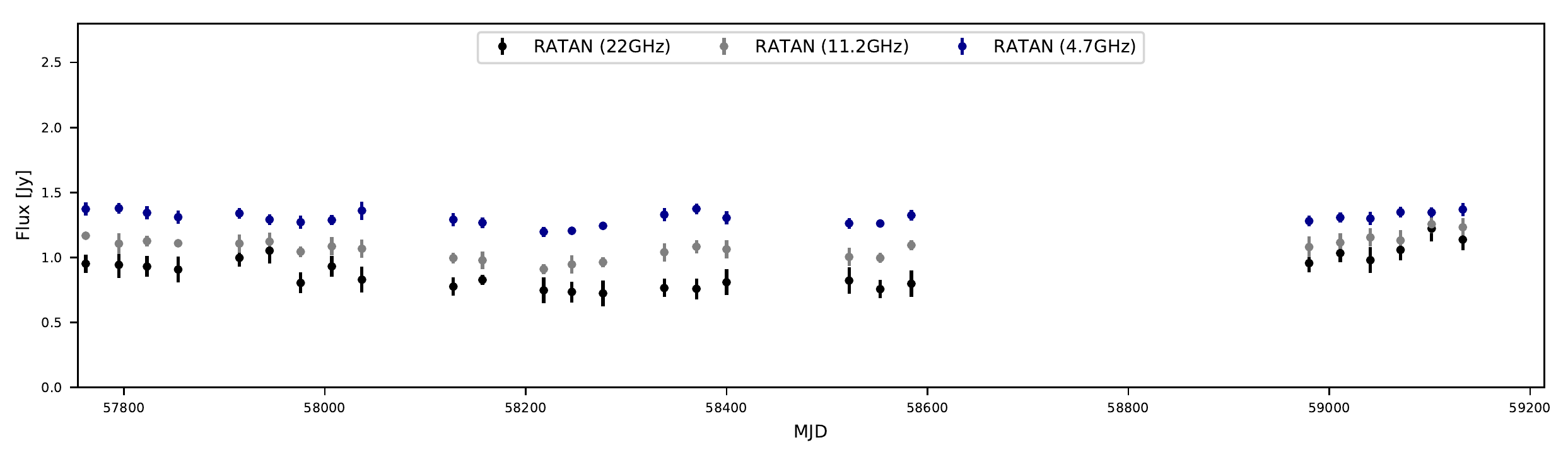}
   \caption{Radio light curve between MJD 57754 to MJD 59214 for Mrk\,501 obtained using the RATAN-600 instrument at different frequencies.}
   \label{fig:lc_ratan}
\end{figure*}

\begin{figure*}
\gridline{\fig{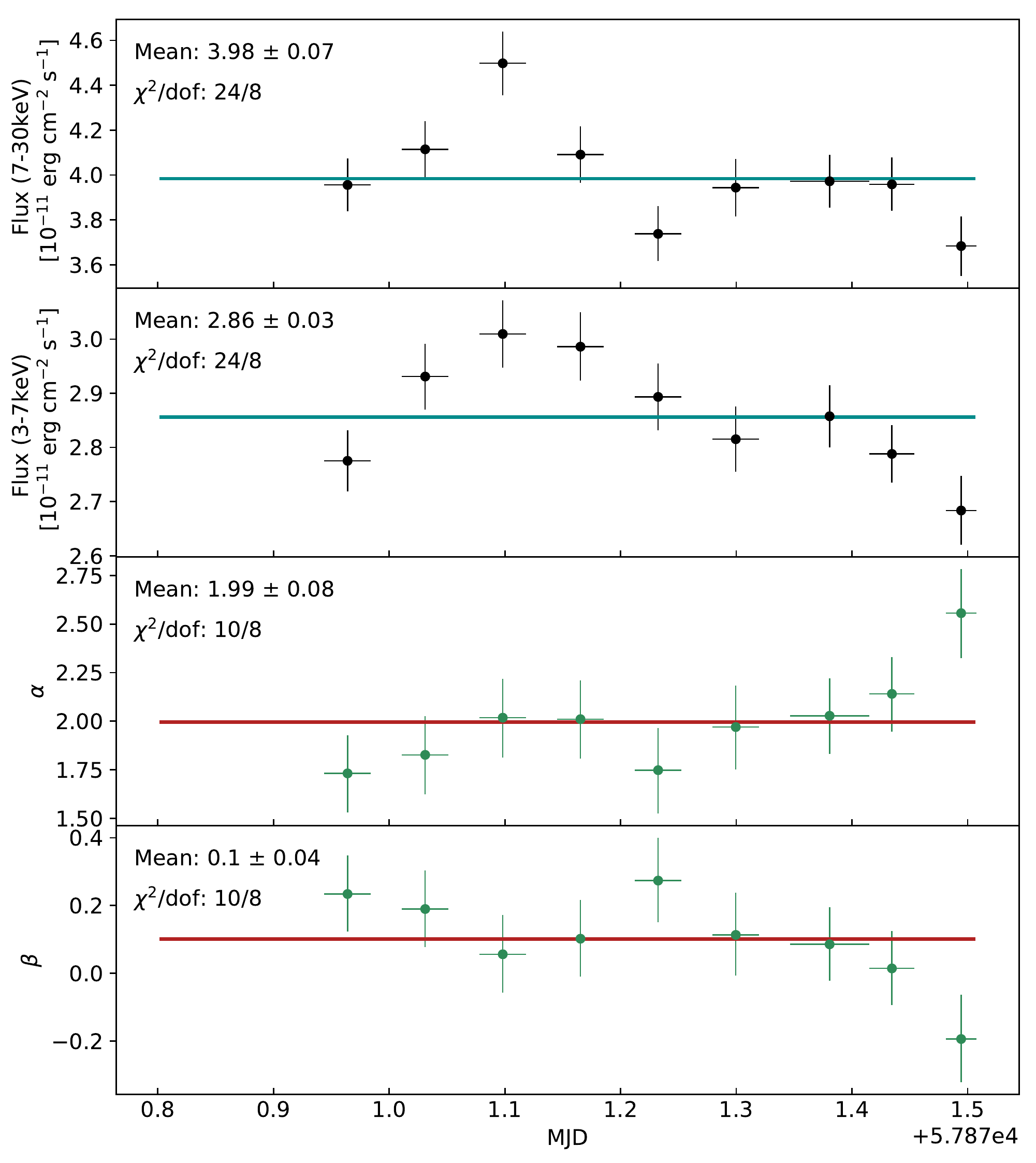}{0.49\textwidth}{(a) First \textit{NuSTAR} observation on 2017-04-28.}
          \fig{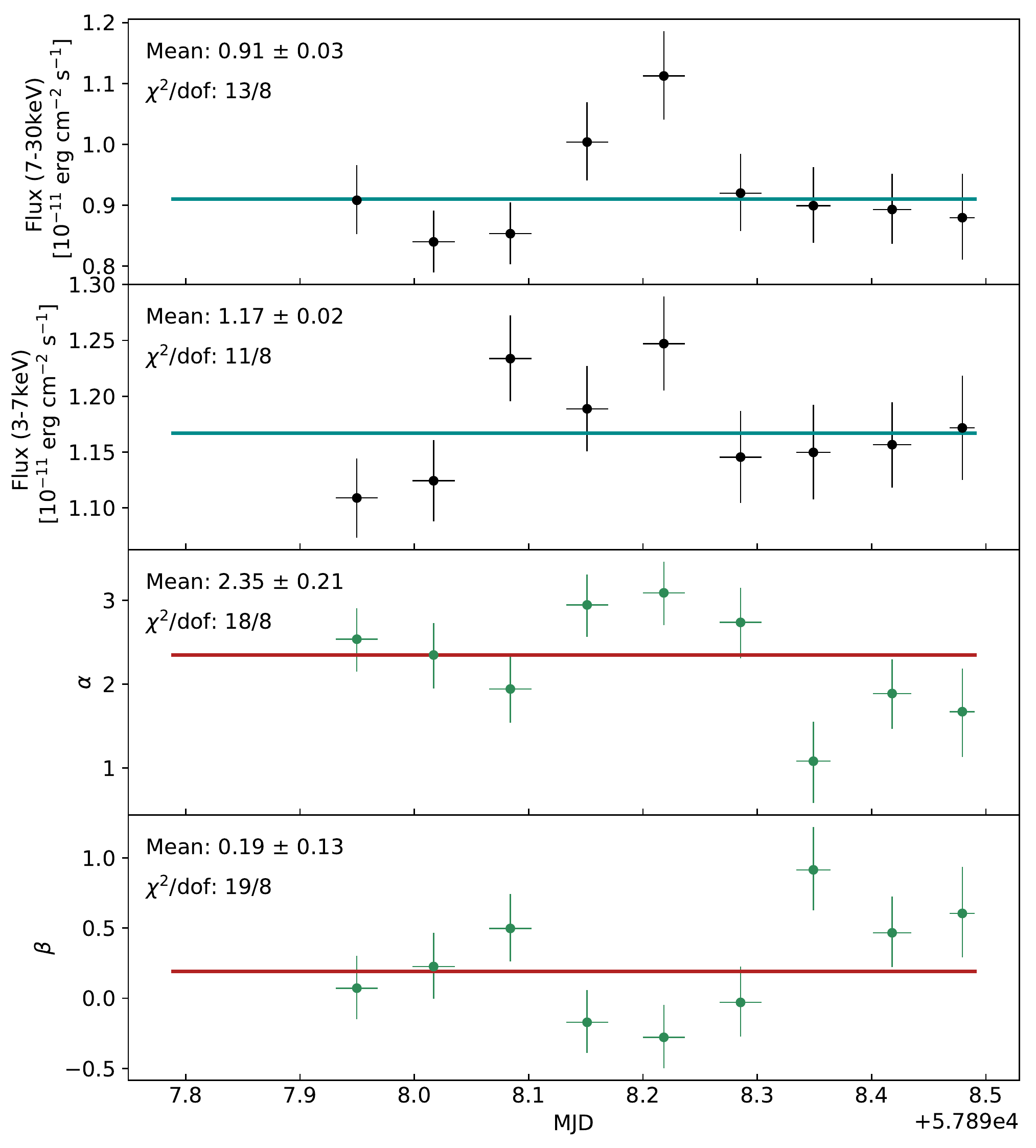}{0.49\textwidth}{(b) Second \textit{NuSTAR} observation on 2017-05-25.}
}
\gridline{\fig{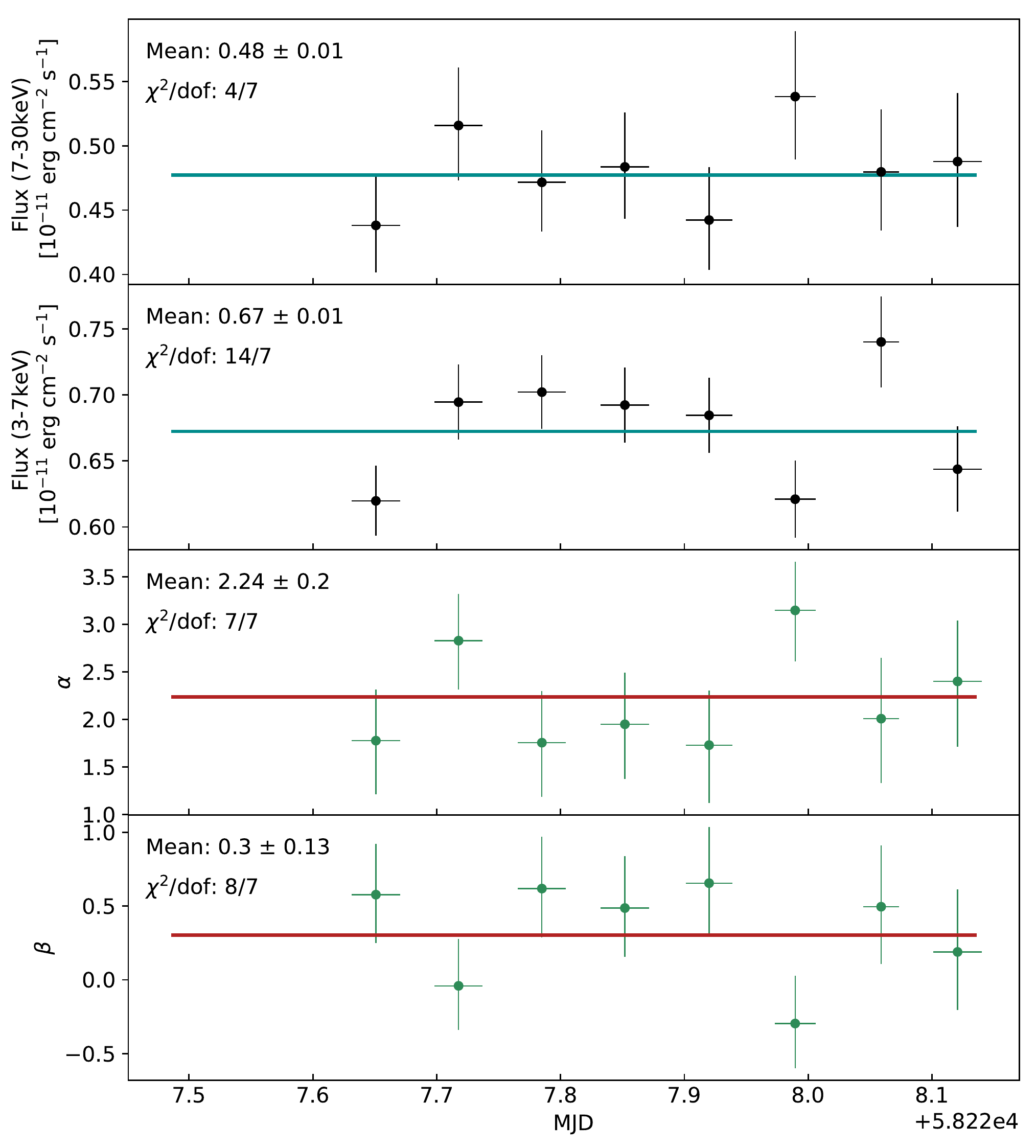}{0.49\textwidth}{(c) Third \textit{NuSTAR} observation on 2018-04-20.}
}

\caption{Mrk\,501 intra-night light curves for the \textit{NuSTAR} observations during the 2017 to 2020 campaign. Shown are the flux in the (7-30\,keV) and (3-7\,keV) energy bands as well as the parameters of the spectral fit using a log parabola: $\alpha$ (power-law index) and $\beta$ (curvature index). Additionally, the results from constant fits to the data are also displayed.}
    \label{fig:nustar_intranight}
\end{figure*}

\begin{figure*}
\gridline{\fig{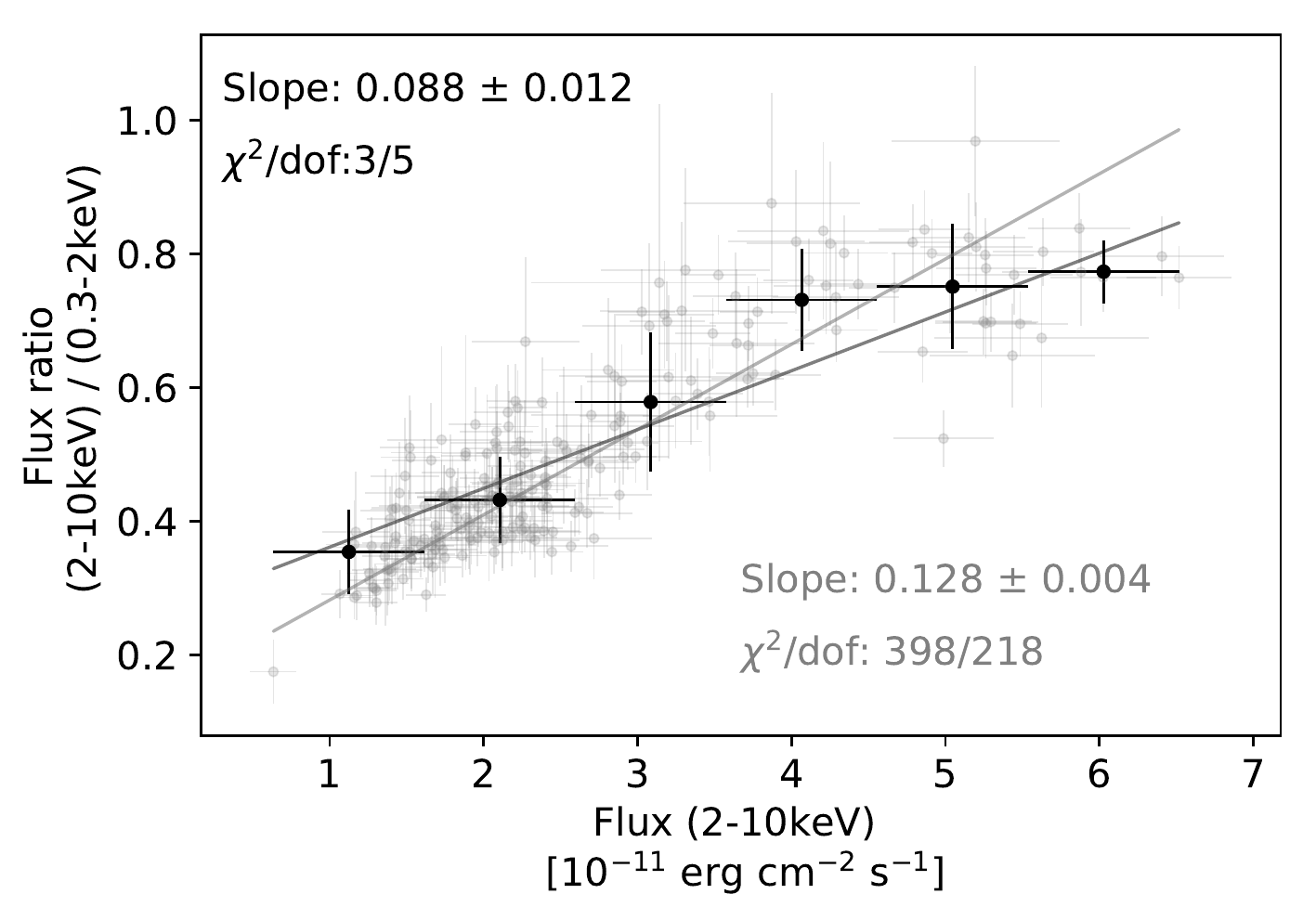}{0.49\textwidth}{(a) \textit{Swift}-XRT hardness ratio (flux ratio of the higher energy range (2-10\,keV) over the lower energy band (0.3-2\,keV)) compared to the flux at (2-10\,keV).}
          \fig{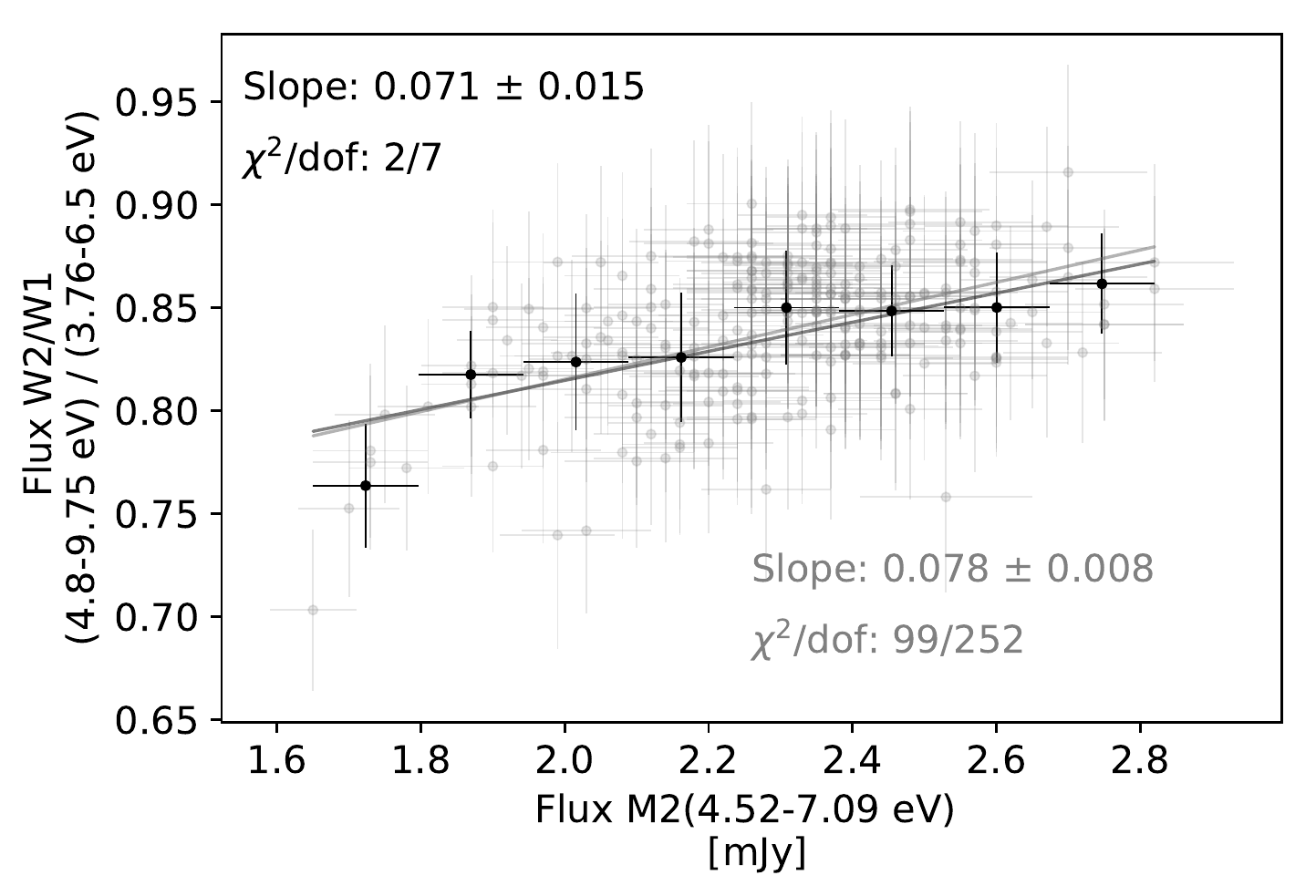}{0.49\textwidth}{(b) \textit{Swift}-UVOT hardness ratio (flux ratio of the highest energy filter W2 (4.8-9.75\,eV) over the lowest energy filter W1 (3.76-6.5\,eV)) compared to the flux from the filter in the middle M2 (4.52-7.09\,eV).}
}

\caption{Hardness ratios over flux values computed for different instruments. In grey, the single data points are shown together with the results of a gradient fit applied to them. In black, the single data points are binned in flux and the mean and standard deviation computed per flux bin. A gradient fit is also applied to the binned data, and the results depicted in black.}
    \label{fig:hardness}
\end{figure*}

In this section additional figures for the MWL data description in Section~\ref{sect:MWL_lc} are presented. Fig.~\ref{fig:lc_ratan} shows the radio light curve between MJD 57754 to MJD 59214 for Mrk\,501 obtained using the RATAN-600 instrument at different frequencies. Table~\ref{tab:NuSTAR_data} summarizes the three \textit{NuSTAR} pointing results and Fig.~\ref{fig:nustar_intranight} shows the intra-night behavior for the three observations. In Fig.~\ref{fig:hardness} the "harder when brighter" behavior of the \textit{Swift} measurements is presented for the XRT and UVOT instruments.

\begin{table*}
\centering
\begin{tabular}{ c | c c c c c c c}   
ObsID & date & Live time & Count rate (3-30\,keV) & simple photon  & Flux (3-7\,keV) \\
& & [s] & [cts s$^{-1}$] & power-law index & [erg cm$^{-2}$s$^{-1}$] \\
\hline
60202049002  & 2017 April 27 & 17000 & 0.876 $\pm$ 0.008 & 2.26$\pm$ 0.05 & $2.9\times10^{-11}$ \\
60202049004 & 2017 May 24 & 19000 & 0.309 $\pm$  0.004 & 2.78$\pm$  0.05 & $1.2\times10^{-11}$ \\
60466006002 & 2018 April 19 & 18800 &  0.174$\pm$  0.003 & 2.81$\pm$  0.05 & $0.7\times10^{-11}$ \\
\end{tabular}
\caption{\textit{NuSTAR} spectral results for the three Mrk\,501 observations performed during the 2017--2020 MWL campaign.}
\label{tab:NuSTAR_data}
\end{table*}


\clearpage
\newpage

\begin{figure*}
\gridline{\fig{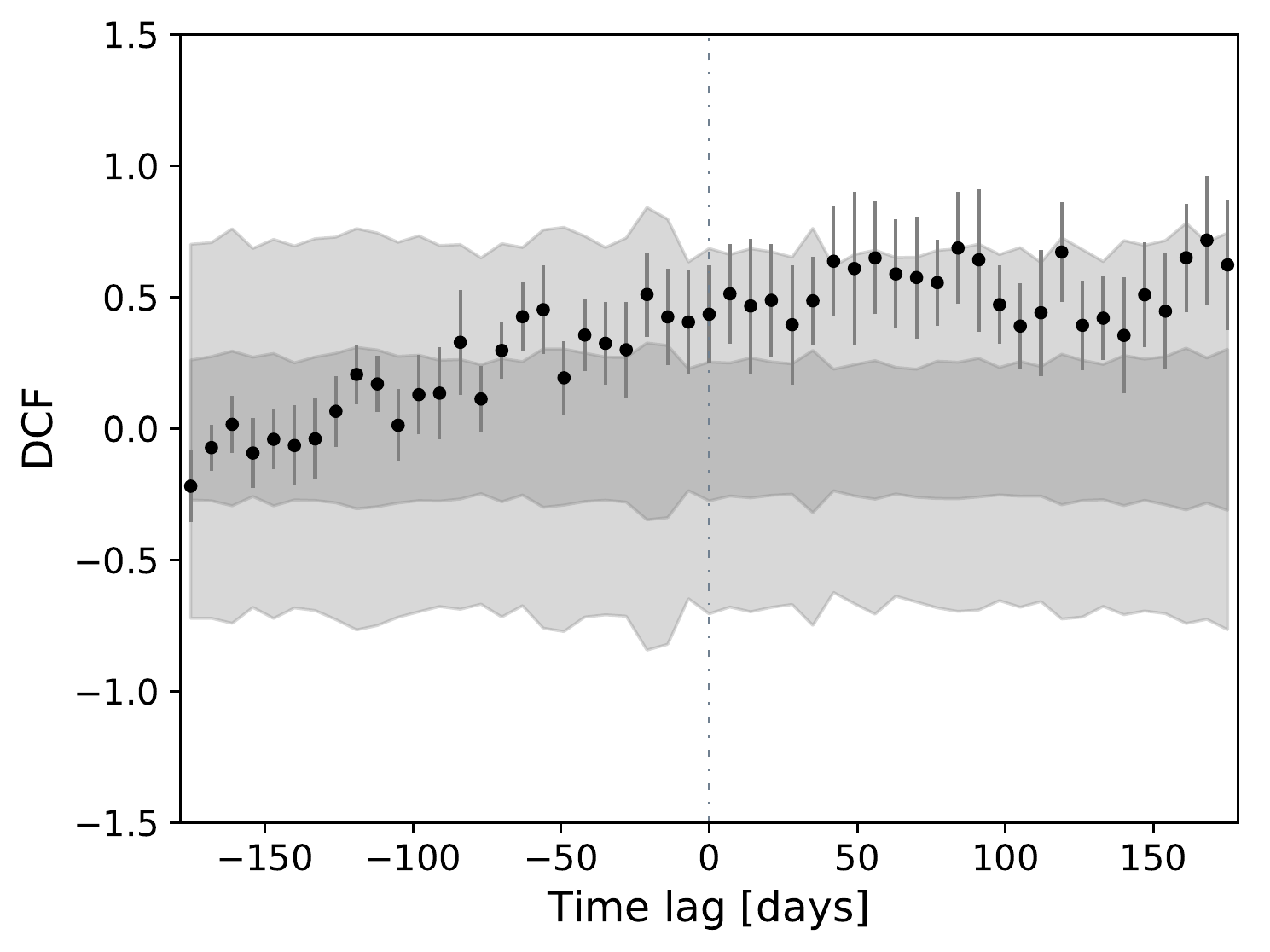}{0.45\textwidth}{(a) \textit{Swift}-XRT (2-10\,keV) vs. OVRO (15\,GHz).}
          \fig{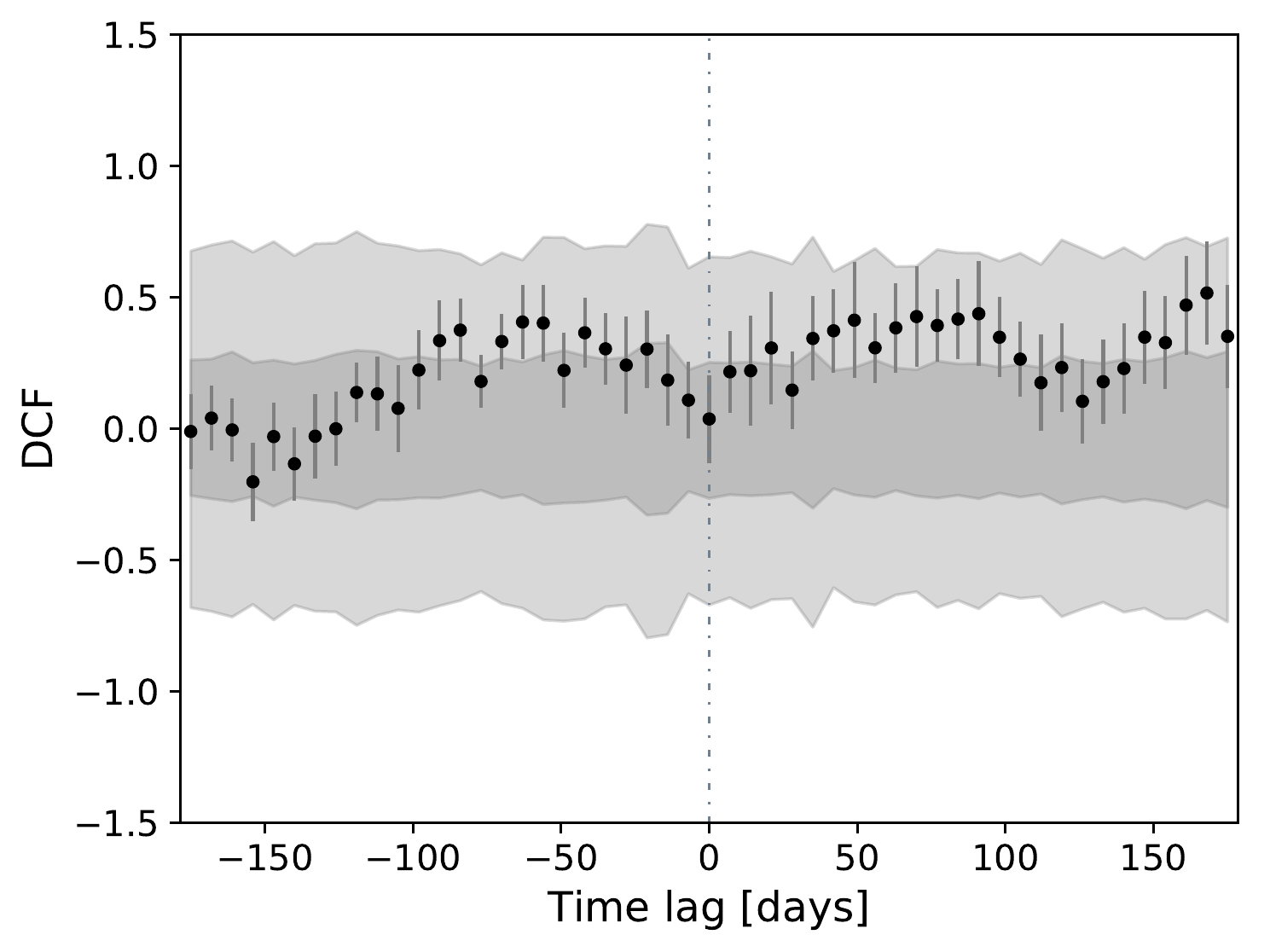}{0.45\textwidth}{(b) \textit{Swift}-XRT (0.3-2\,keV) vs. OVRO (15\,GHz).}
}
\gridline{\fig{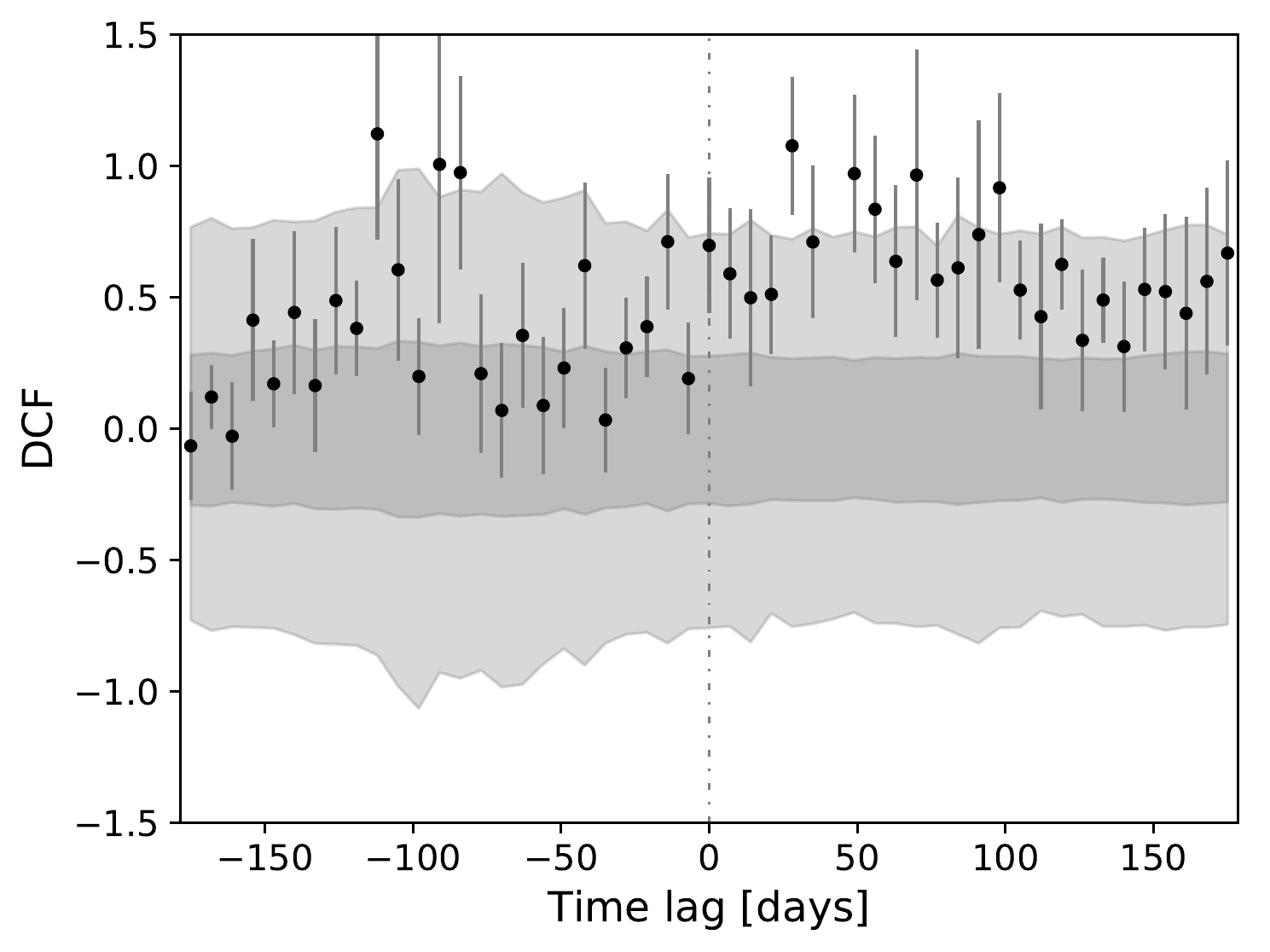}{0.45\textwidth}{(c) MAGIC ($>$0.2\,TeV) vs. OVRO (15\,GHz).}
}

\caption{Discrete correlation function DCF computed for different pairs of the light curves shown in Fig.~\ref{fig:MWL_LC}  using a binning of 7\,days. It is computed for different time shifts, time lags, applied to the LCs. The 1$\sigma$ and 3$\sigma$ confidence levels obtained by simulations  as described in Section~\ref{subsec:var_corr} are shown by the dark and light grey bands, respectively, respectively.}
    \label{fig:dcfs_2}
\end{figure*}

\begin{figure*}
\gridline{\fig{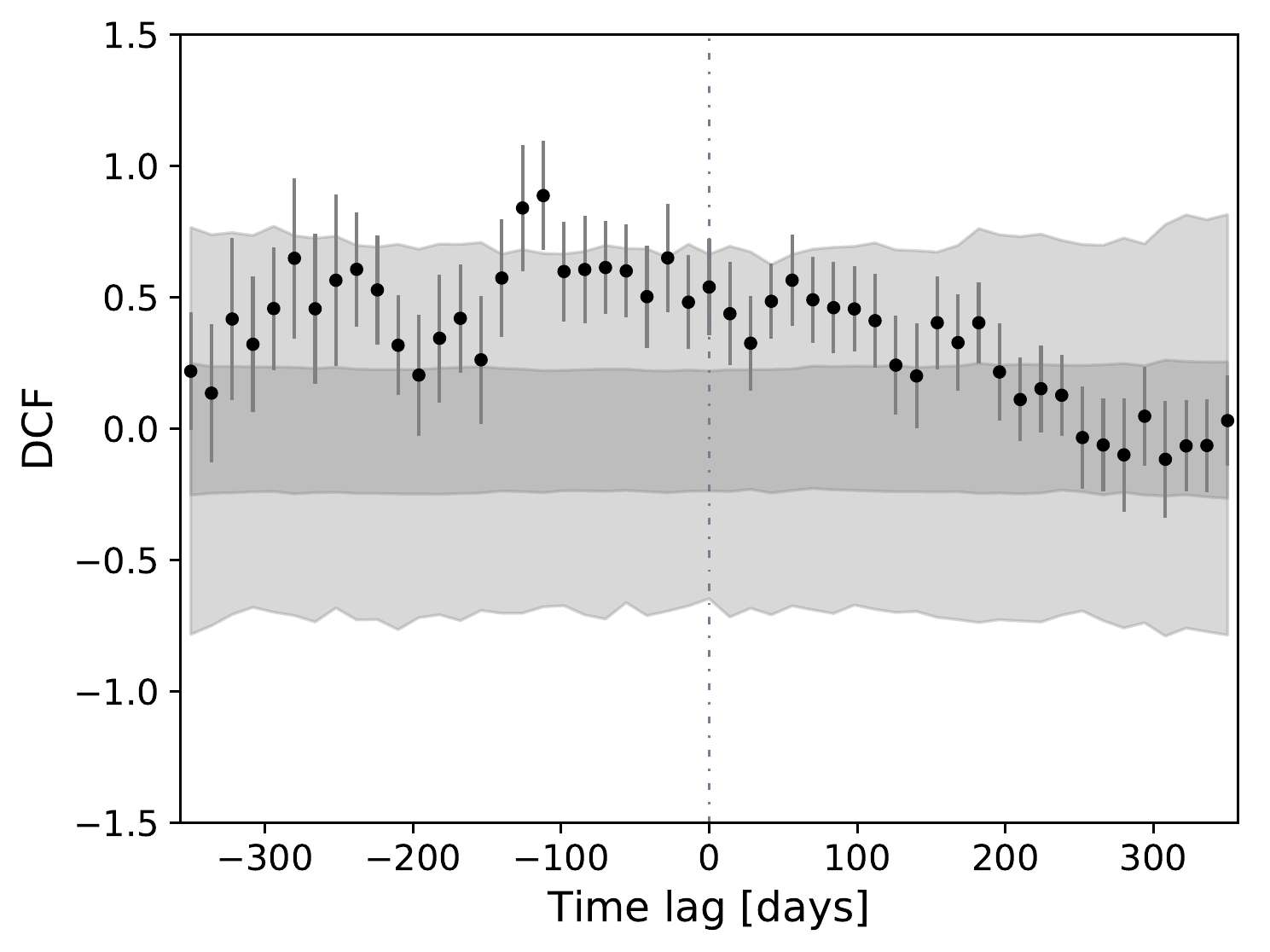}{0.45\textwidth}{(a) \textit{Fermi}-LAT (0.3-500\,GeV) vs. OVRO (15\,GHz) using a binning of 14\,days.}
          \fig{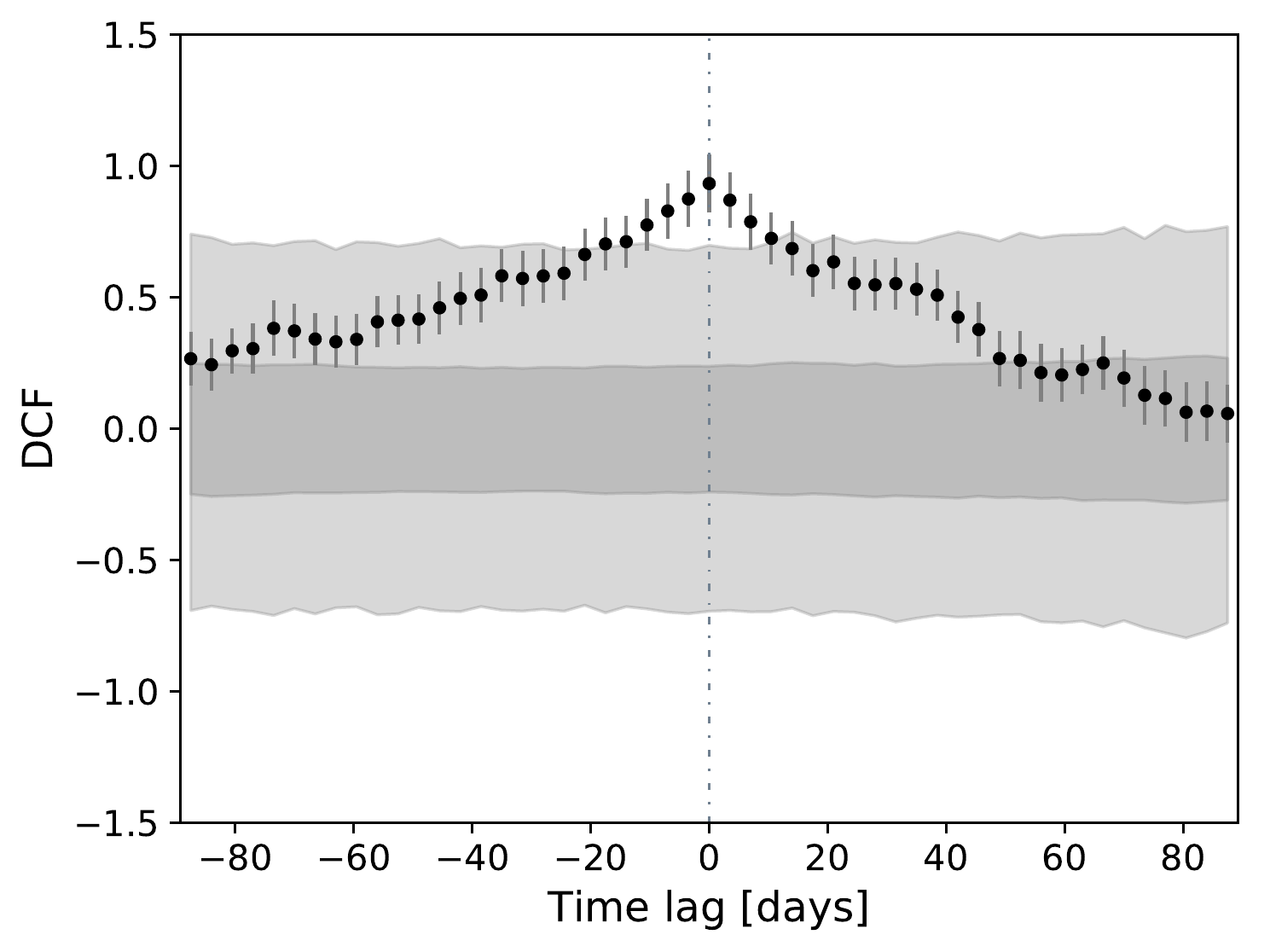}{0.45\textwidth}{(b) UVOT W1 (3.76-6.5\,eV) vs. Optical (R-band) using a binning of 3.5\,days.}
}
\gridline{\fig{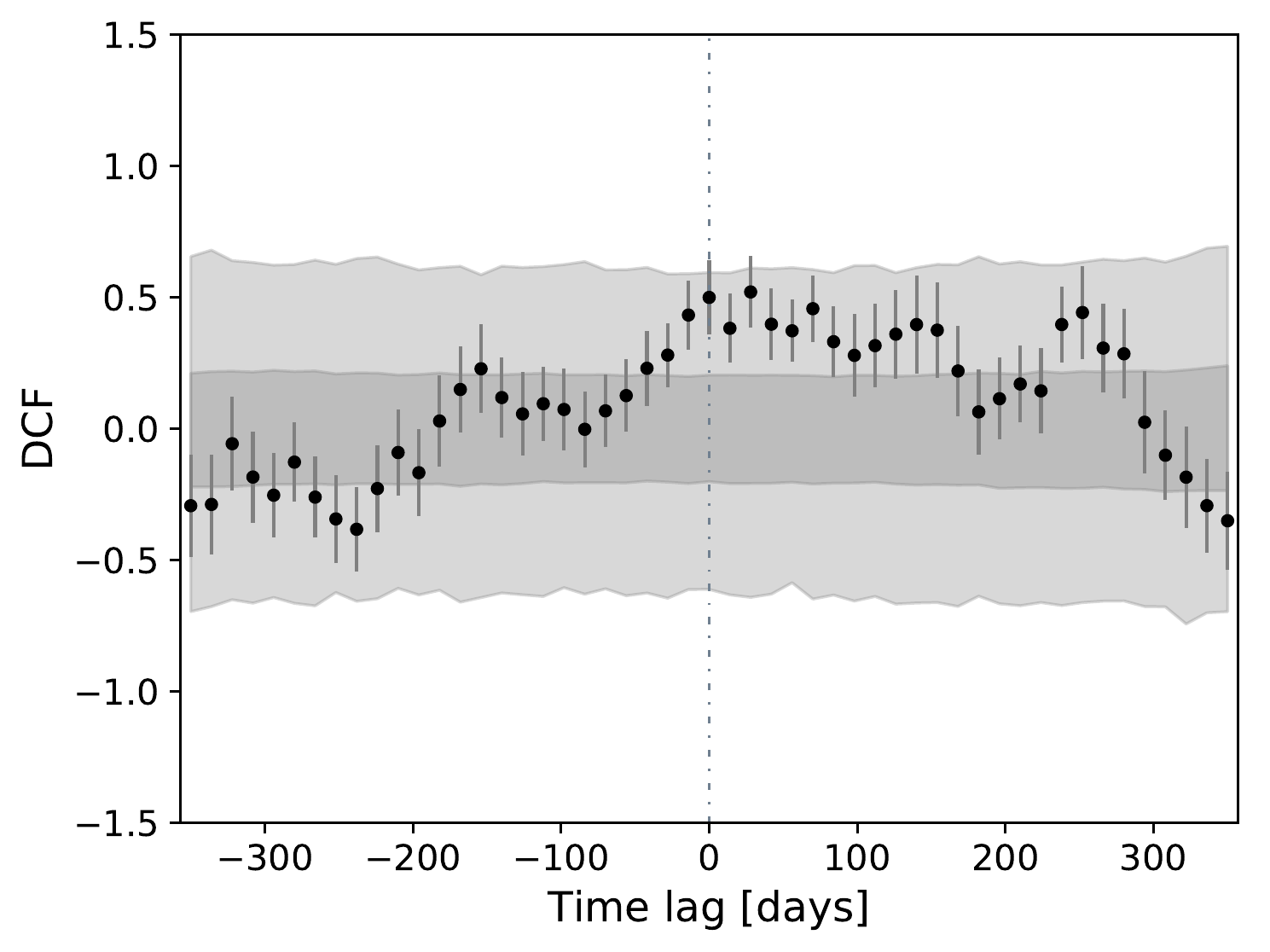}{0.45\textwidth}{(c) \textit{Fermi}-LAT (0.3-500\,GeV) vs. Optical (R-band) using a binning of 14\,days.}
    \fig{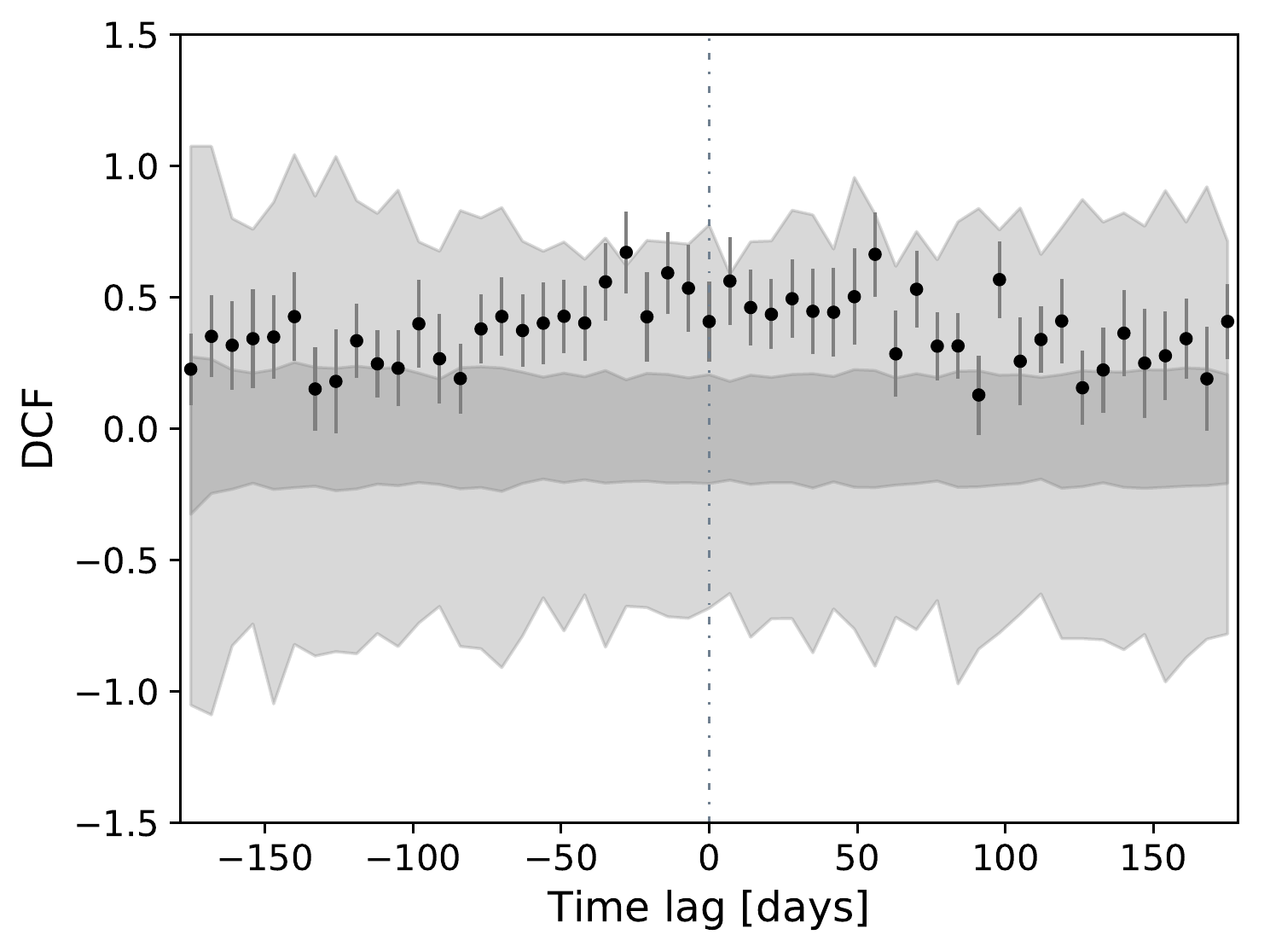}{0.45\textwidth}{(d) OVRO (15\,GHz) vs. Mets\"ahovi (37\,GHz) using a binning of 7\,days.}
}

\caption{Discrete correlation function DCF computed for different pairs of the light curves shown in Fig.~\ref{fig:MWL_LC}. It is computed for different time shifts, time lags, applied to the LCs. The 1$\sigma$ and 3$\sigma$ confidence levels obtained by simulations  as described in Section~\ref{subsec:var_corr} are shown by the dark and light grey bands, respectively.}
    \label{fig:dcfs_3}
\end{figure*}

\begin{figure}
    \centering
    \includegraphics[width=0.47\textwidth]{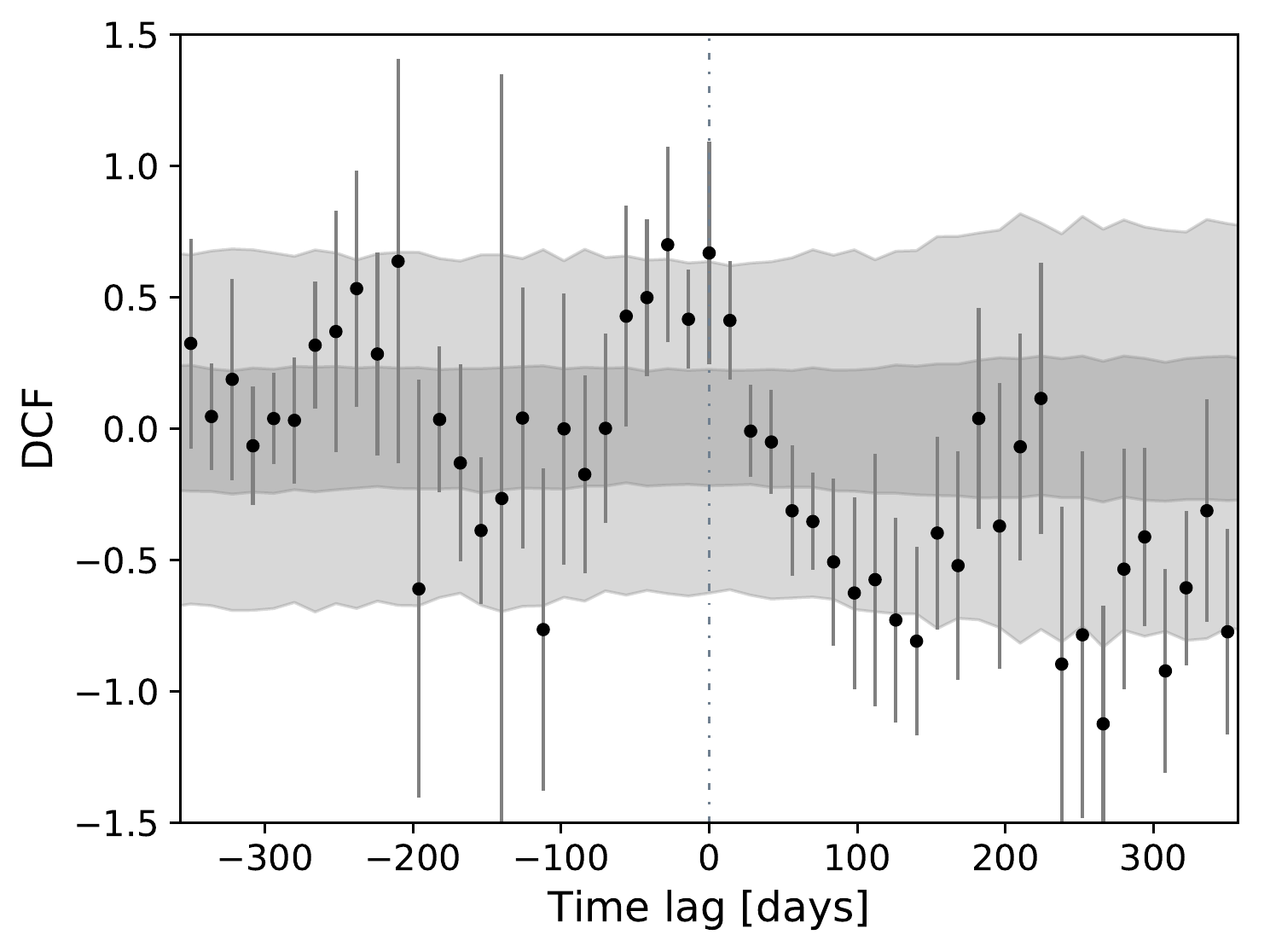}
    \caption{Discrete correlation function DCF computed for the MAGIC ($>$2 TeV) and \textit{Fermi}-LAT (0.3-500 GeV) light curves shown in Fig.~\ref{fig:MWL_LC} using a binning of 14\,days with the light curves detrended, as described in Section~\ref{subsec:var_corr}, before computing the correlations . The 1$\sigma$ and 3$\sigma$ confidence levels obtained by simulations as described are shown by the dark and light grey bands, respectively.}  
    \label{fig:dcf_magic_fermi_detrend}
\end{figure}


\begin{figure*}
\gridline{\fig{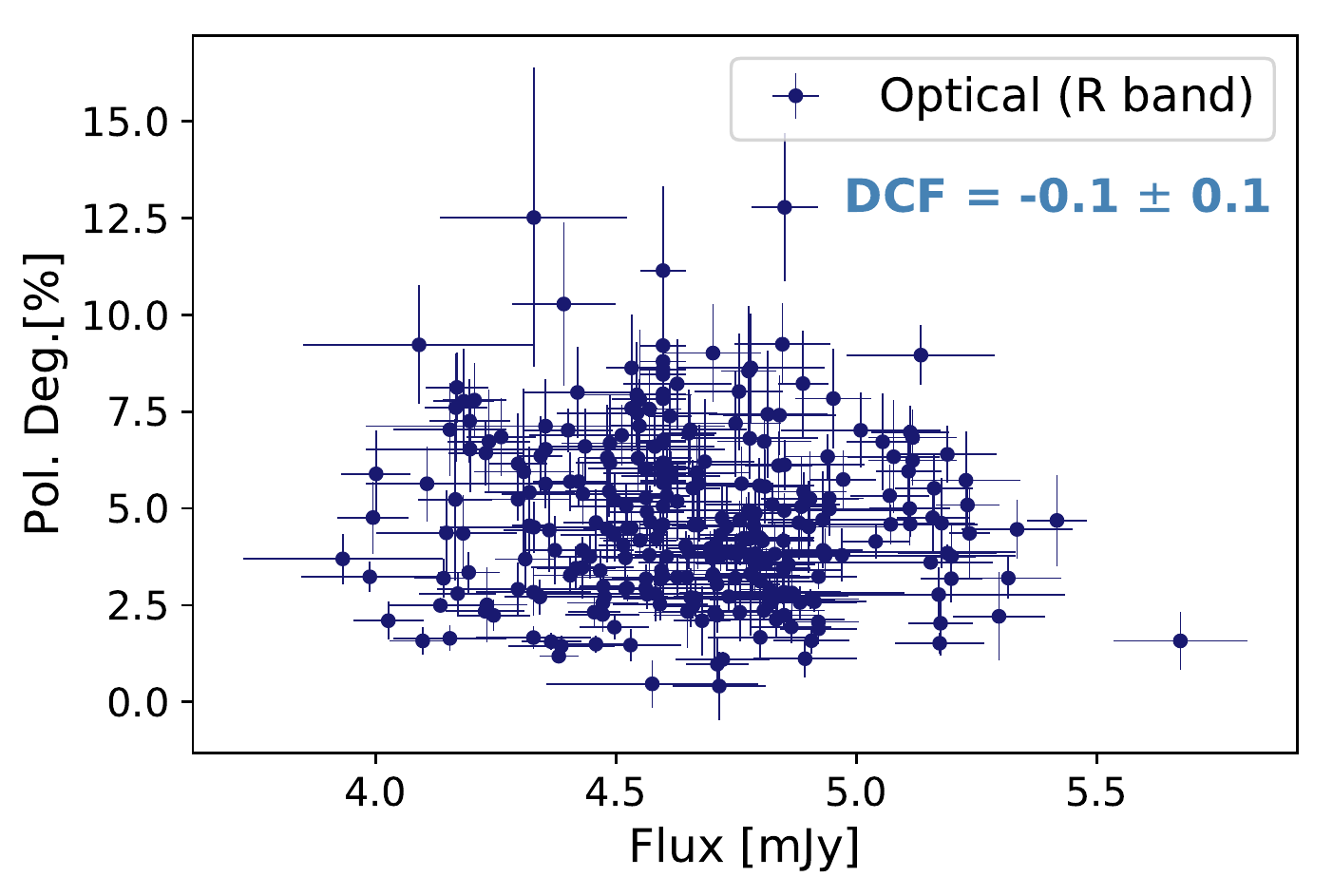}{0.49\textwidth}{(a) Optical (R-band), no correlation is found with DCF $\sim$ 0.}
          \fig{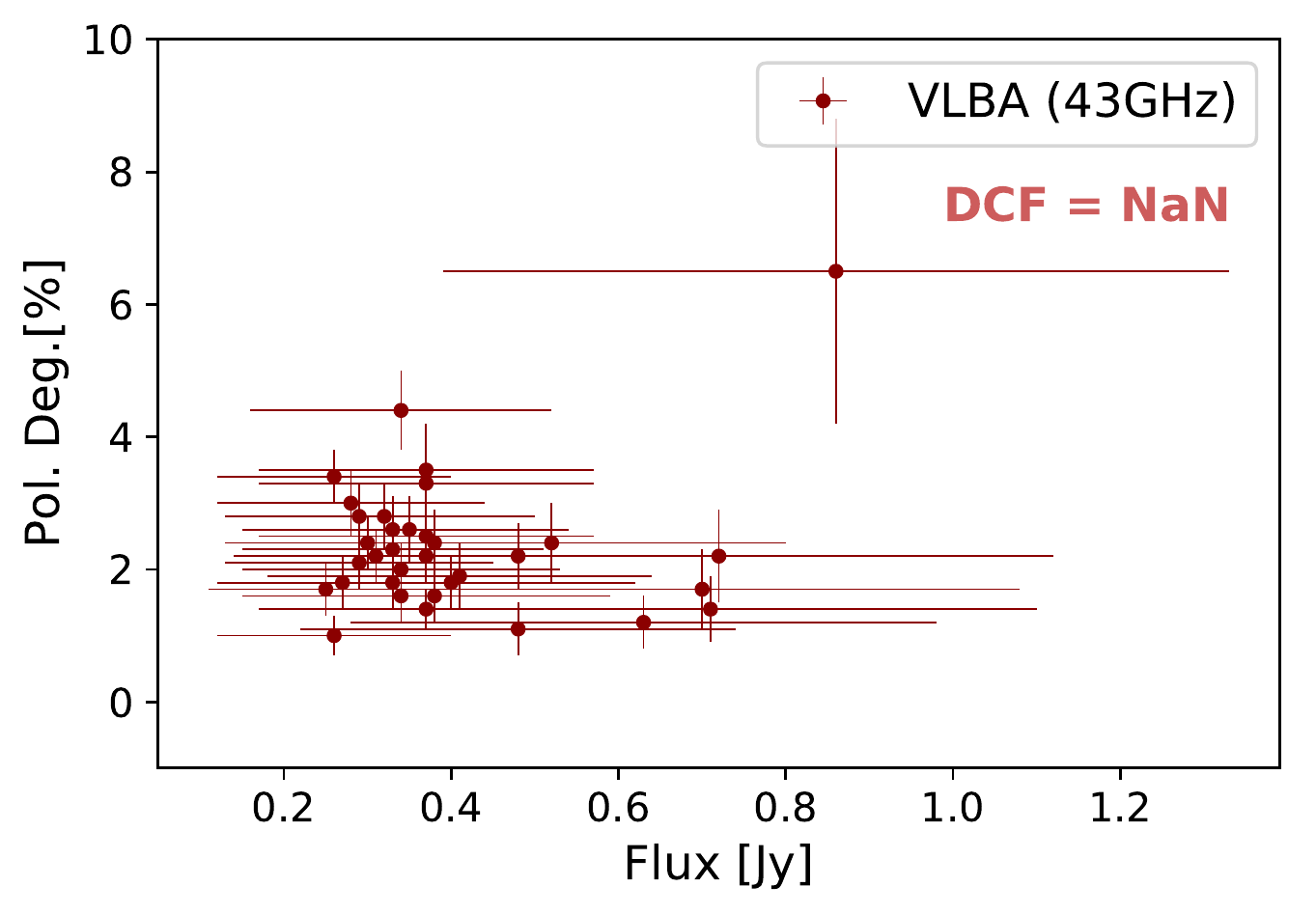}{0.48\textwidth}{(b) VLBA (43\,GHz), no DCF can be computed since the errors on the flux values are bigger than the variations in flux.}
}
\caption{Distribution of the polarization degree among flux measurements for different energy bands for Mrk\,501 data taken from 2017 to 2020. The correlation is quantified using the Discrete Correlation Function (DCF).}
    \label{fig:pol_deg_corr}
\end{figure*}


\begin{figure*}
\gridline{\fig{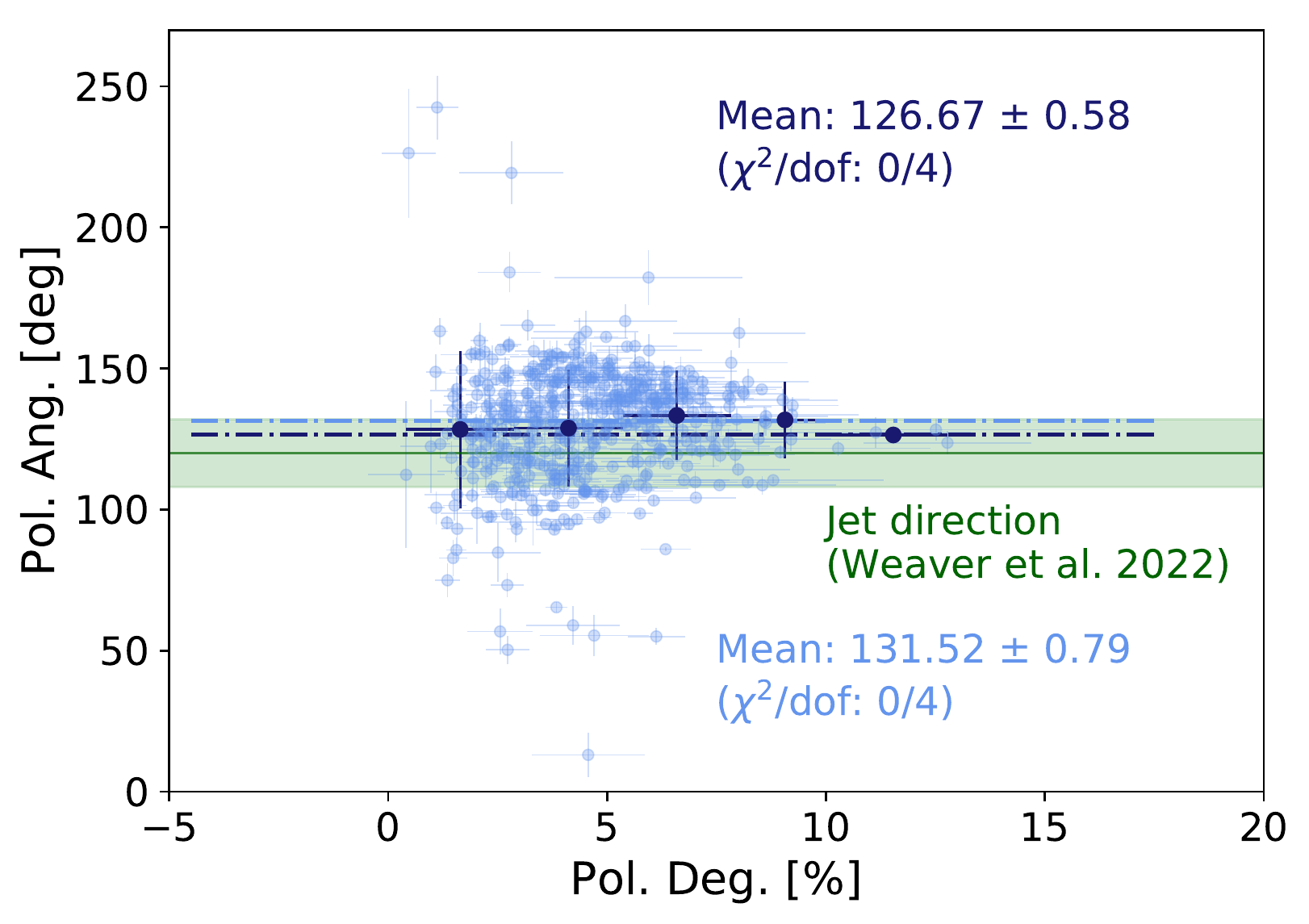}{0.49\textwidth}{(a) Optical R-band data: In light blue the single data points are shown together with the results of a constant fit applied to them with the mean values shown by the dashed line. In dark blue the single data points are binned in polarization degrees and the mean and standard deviation computed per flux bin. Again a constant fit is applied and shown in dark blue.}
          \fig{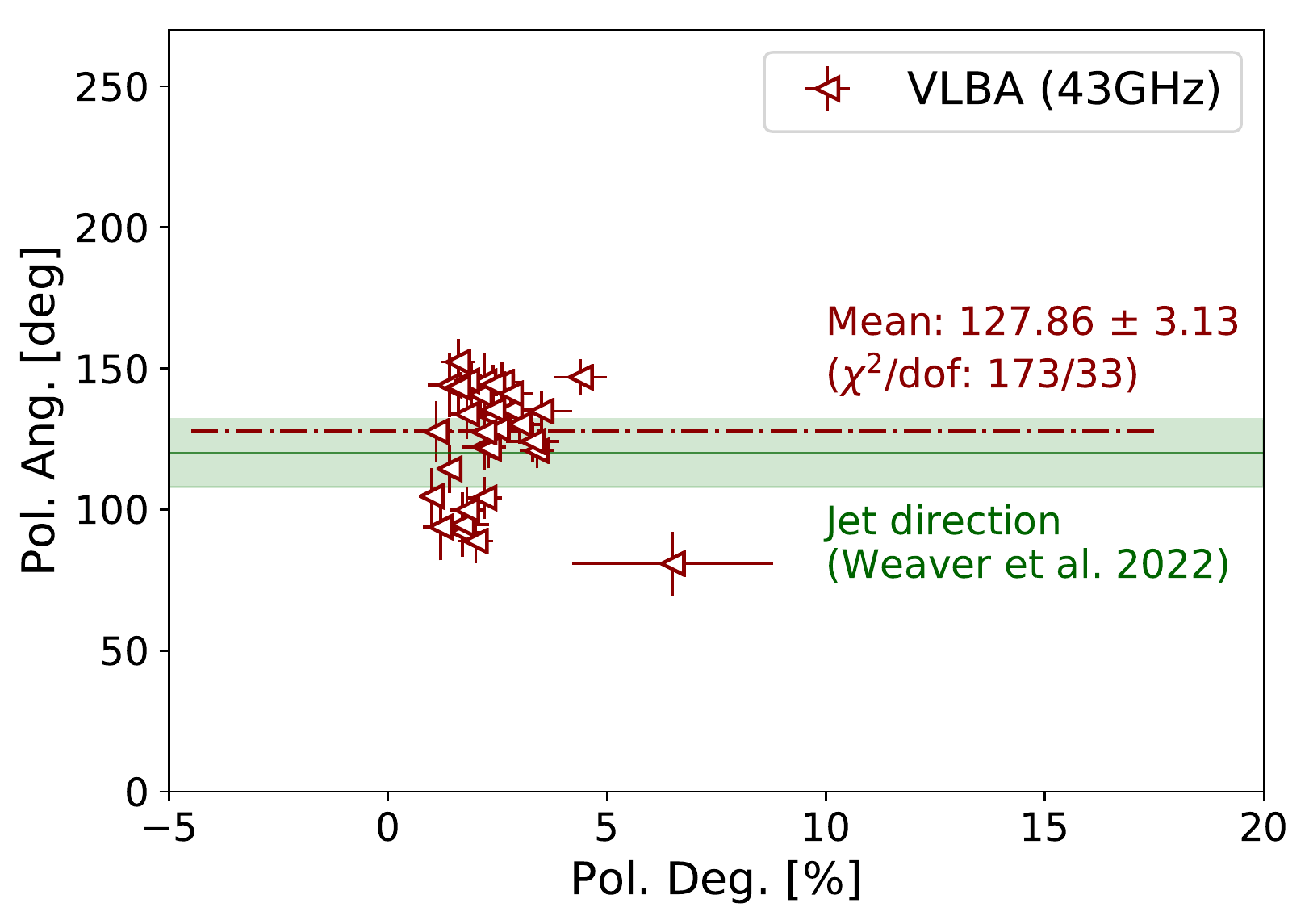}{0.49\textwidth}{(b) Radio data taken with different instruments at different frequencies. A constant fit is applied to each data set and the mean values shown by the dashed lines.}
}
\caption{Polarization angle over polarization degree for Mrk\,501 data taken from 2017 to 2020. The  jet direction for Mrk\,501 determined in \citet[][]{Weaver_2022} is shown in green with 1$\sigma$ uncertainties.}
    \label{fig:pol_ang_optical}
\end{figure*}


\begin{figure*}
\gridline{\fig{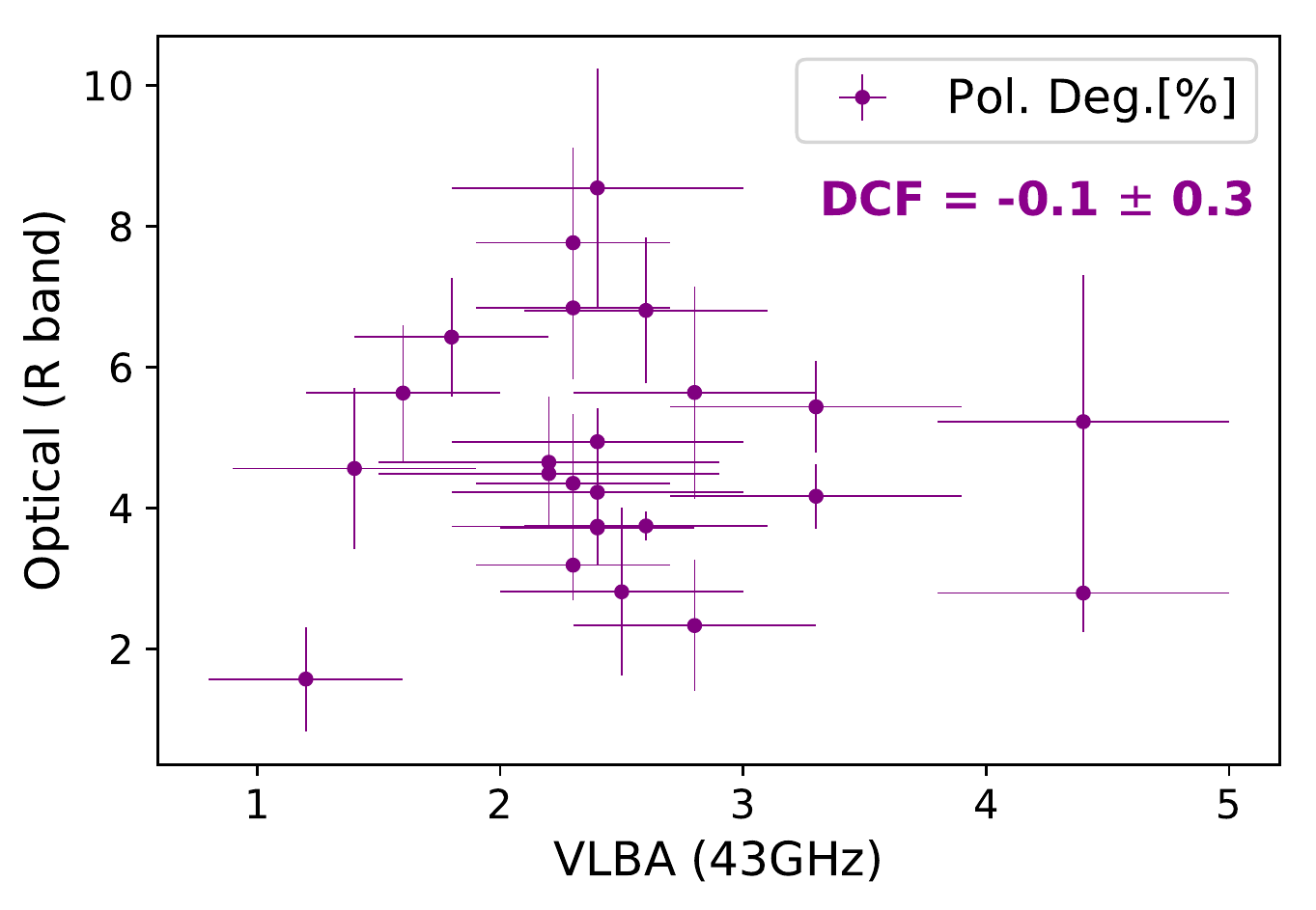}{0.48\textwidth}{(a) Polarization degree.}
          \fig{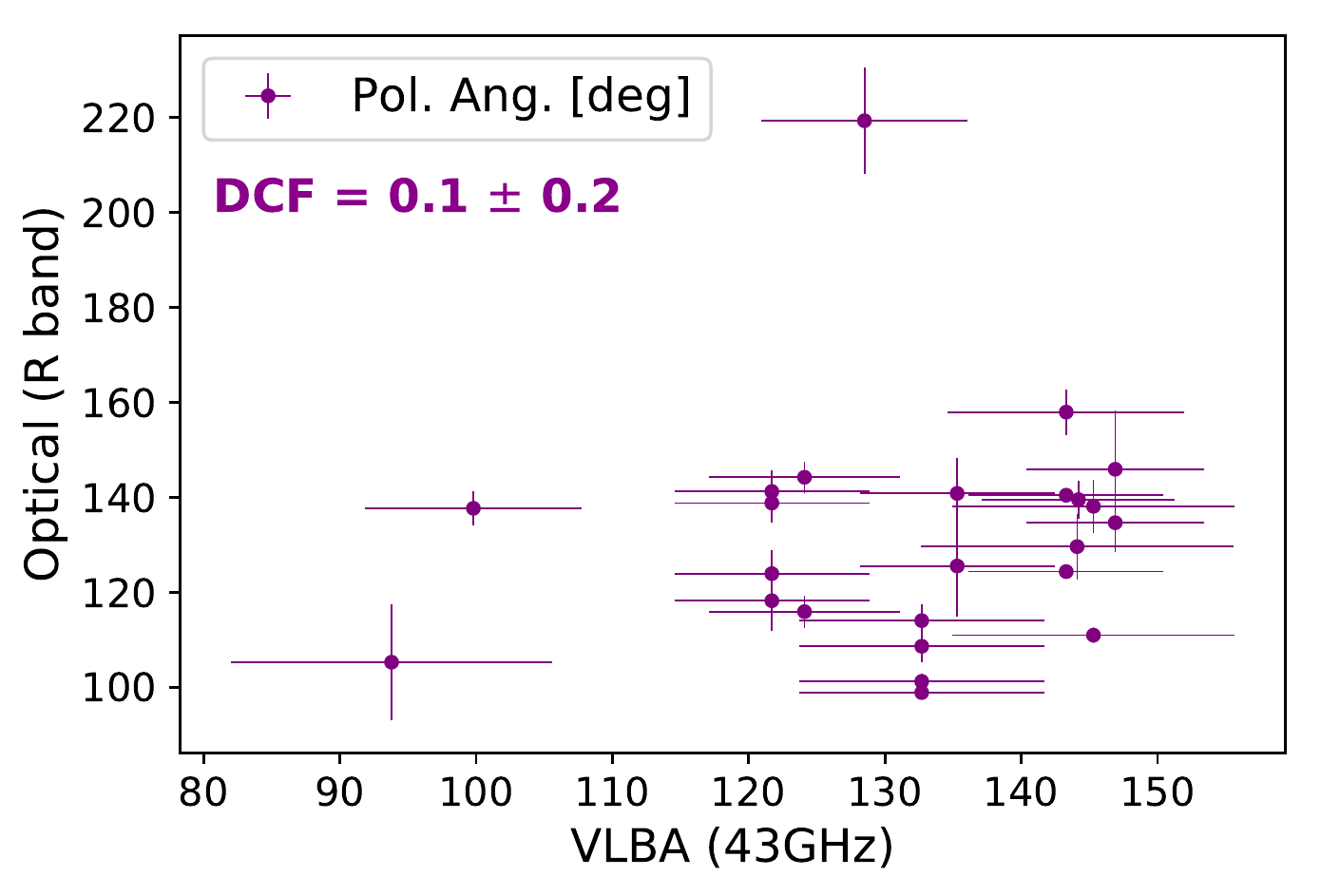}{0.49\textwidth}{(b) Polarization angle.}
}
\caption{Comparison of the polarization parameters measured in the optical R band and the radio regime for Mrk\,501 data simultaneous within 2 days taken from 2017 to 2020. The correlation is quantified using the Discrete Correlation Function (DCF).}
    \label{fig:pol_corr_opt_rad}
\end{figure*}

\clearpage
\newpage

\section{Additional information - Periodicity}
\label{sec:app_period}
For the periodicity analysis described in Section~\ref{sec:periodicity}  we use the LombScargle timeseries package provided by astropy v5.0.1 \citep[][]{astropy} and apply it to our light curves shown in Fig.~\ref{fig:long_LC_comb}. Since we do not expect to be sensitive to a periodicity below 1\,day due to the time sampling, we bin our not yet equally spaced light curves (\textit{Swift}-XRT and OVRO) in 1\,day bins. Following the recommendations in \citet[][]{VanderPlas_2018} we carefully evaluate the frequency grid to choose for each waveband. As a minimum frequency we use our time window of 12\,years $f_{min}=1/(12\times365)$\,days$^{-1}$. Due to the applied even sampling our maximum frequency is given by the bin size with $f_{max} \sim 1/2t_{\text{bin}}$\,days$^{-1}$. The number of frequencies to be evaluated in between is thereafter chosen according to Eq. 44 in \citet{VanderPlas_2018}. To estimate the 1$\sigma$ and 3$\sigma$ detection significance we apply the same procedure to the simulated light curves described in Section~\ref{sec:correlations}.

\begin{figure*}
\gridline{\fig{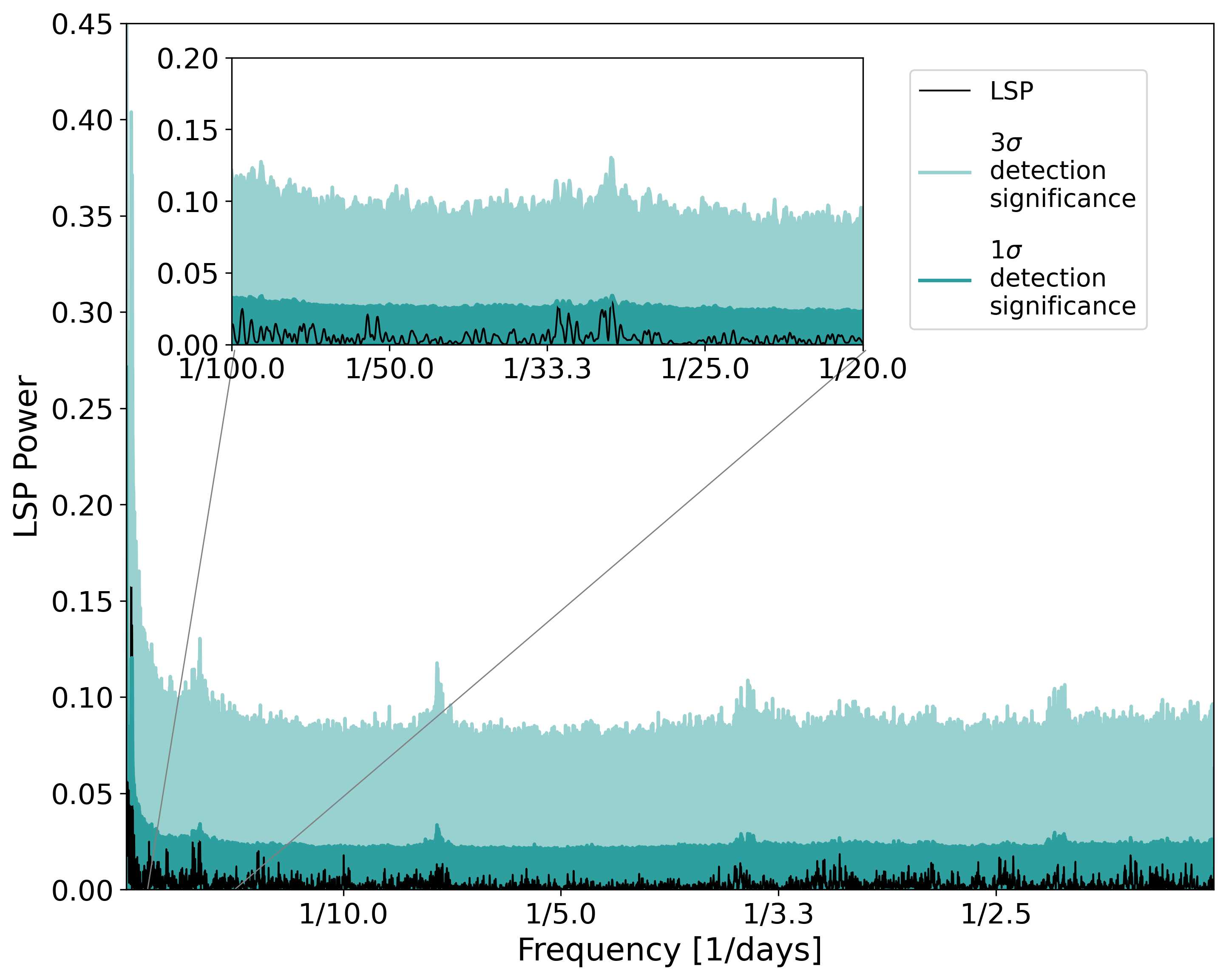}{0.49\textwidth}{(a) LSP for the \textit{Swift}-XRT (2-10\,keV) light curve including a more detailed view of the frequency range corresponding to a periodicity between 10 to 100\,days.}
          \fig{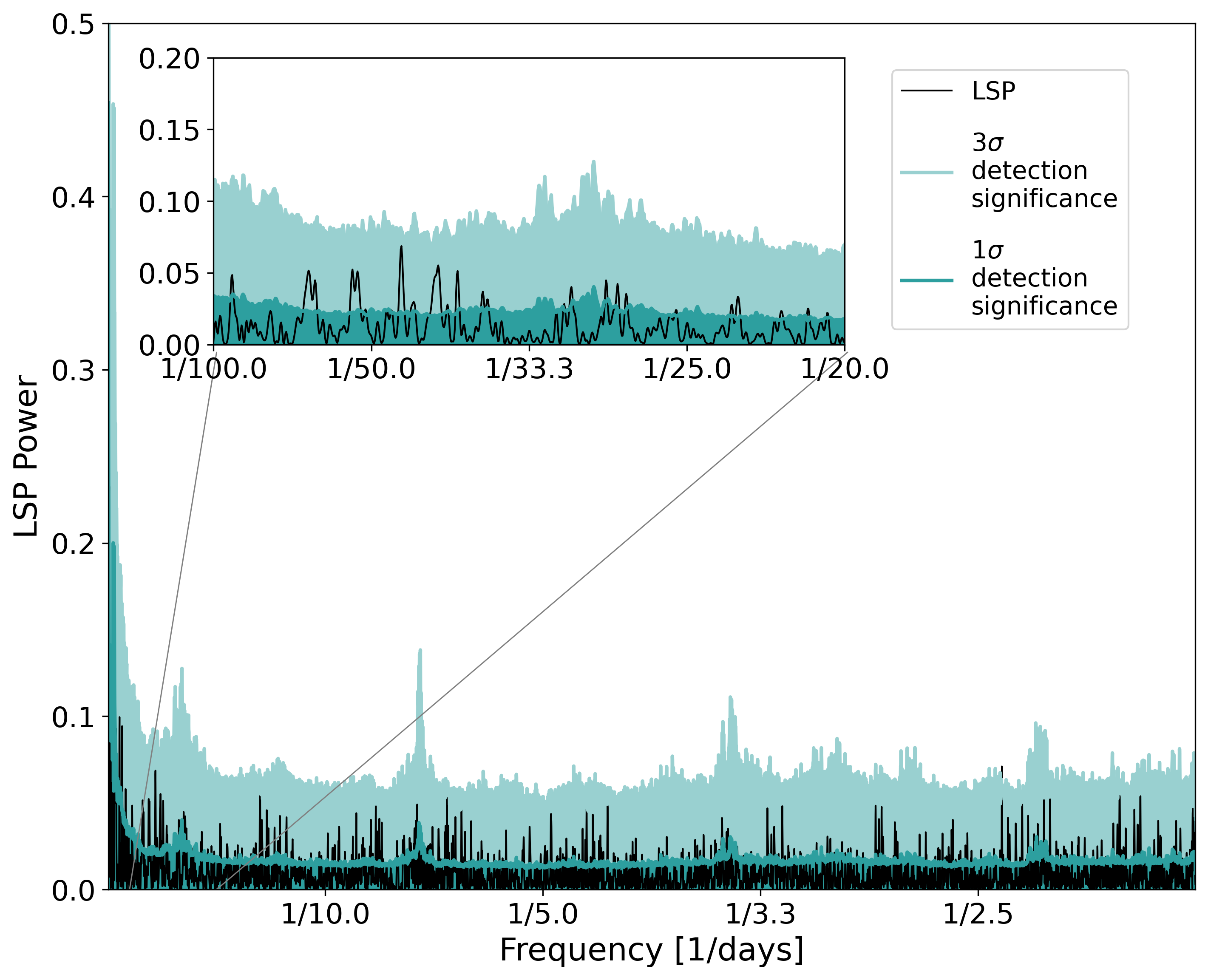}{0.49\textwidth}{(b) LSP for the \textit{Swift}-XRT (0.3-2\,keV) light curve including a more detailed view of the frequency range corresponding to a periodicity between 10 to 100\,days.}
}
\caption{Lomb-Scargle Periodigram (LSP) for different light curves of our 12-year data set shown in Fig.~\ref{fig:long_LC_comb}. The 1$\sigma$ and 3$\sigma$ detection significance levels obtained by simulations are shown as well as described in Section \ref{sec:app_period}.}
    \label{fig:LSP_xrt}
\end{figure*}

\begin{figure}
    \centering
   \includegraphics[width=0.49\textwidth]{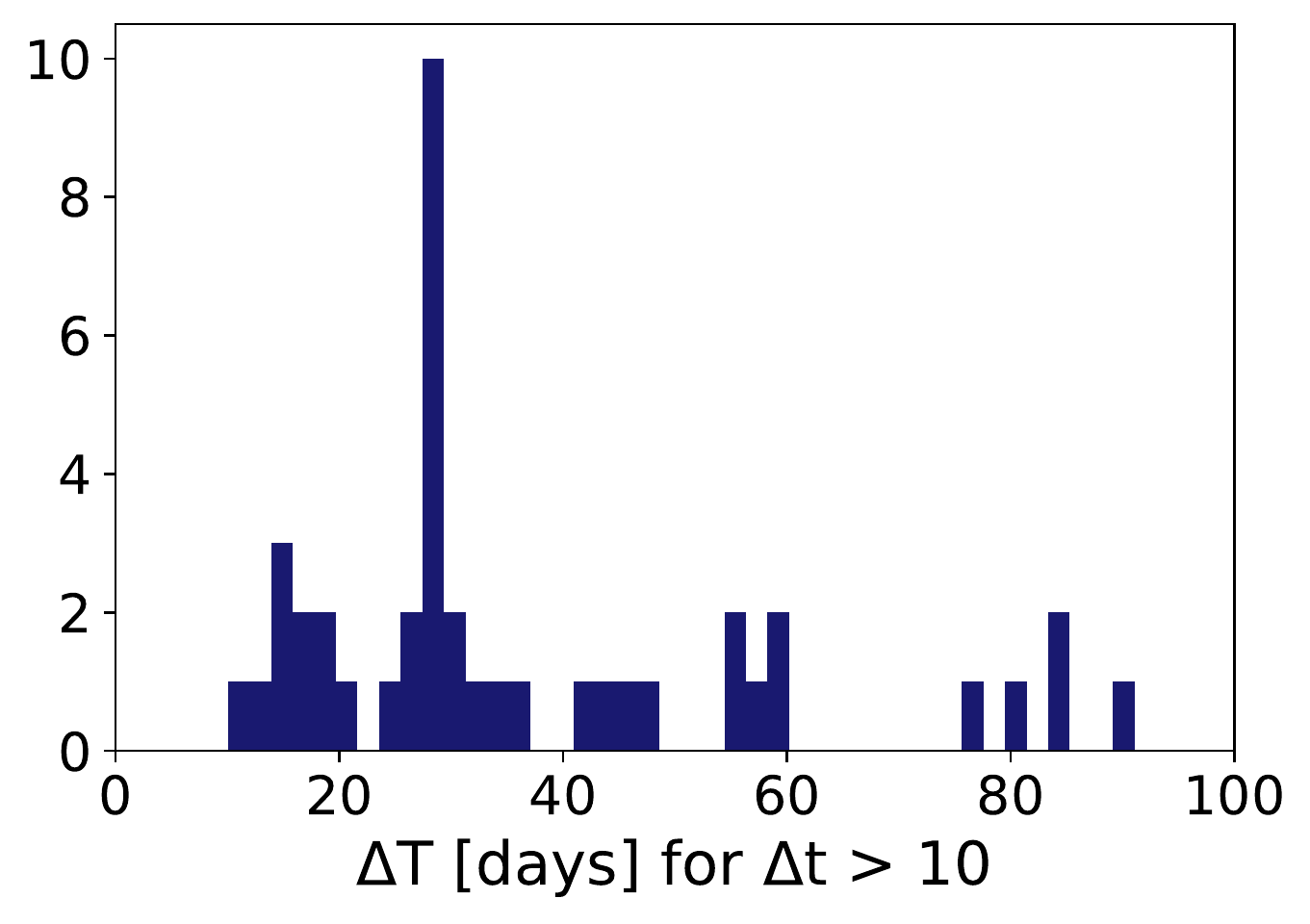}   \caption{Distribution of time gaps between the starting dates of observational periods for the \textit{Swift}-XRT in 12-year data set shown in Fig.~\ref{fig:long_LC_comb} displayed for the most dominant time scale (0-100 days). Observational periods are defined by the time gaps of adjacent measurement bigger than 10\,days.}
    \label{fig:mjd_break_XRT_lt}
\end{figure}

Considering our 12 years of X-ray data, we can see the LSP powers and significant levels rising towards the edges of the chosen frequency range as shown in Fig~\ref{fig:LSP_xrt}. This is expected and explained by the 1\,day binning. At $f=1/30$\,days$^{-1}$ we can see slightly raised LSP powers in our signal as well as in our detection levels. It can be attributed to the time sampling of the X-ray light curve which is coordinated together with the Earth based telescopes. Fig.~\ref{fig:mjd_break_XRT_lt} depicts the time gaps between the start of different observational periods. Whenever we see a time gap of more than 10\,days between adjacent measurements, we define it as the start of an observational period, which is often at a cadence of $\sim$30\,days. This is caused by bright moon light limiting the Earth based telescopes explaining the '30-day' - or better '28-day' periodicity.


For the VHE data we are limited by the fact that the various data sets in Fig.~\ref{fig:long_LC_comb} are computed using different energy thresholds. They can therefore only be used as a combined light curve by carefully evaluating and adding systematic errors accounting for the differences. However, we can use the confidence levels obtained for the X-ray data sets for a first estimation of the periodicity significance in the VHE band. The X-ray confidence bands are certainly lower than the VHE ones would be. On the one hand because both the measurement uncertainties as well as the time sampling in the X-ray is better than in the VHE. On the other hand because the mentioned systematics would further increase the confidence levels. When we apply the LSP to the VHE data, we obtain a very similar pattern as for the X-ray data. Around the region of $f=1/30$\,days$^{-1}$ the maximum height of the peaks is around 0.05 with none of them standing out from the random fluctuations. Connecting this to the 3$\sigma$ confidence bands in Fig~\ref{fig:LSP_xrt}, we can conclude that no significant periodicity can be detected in the VHE band. 

\begin{figure*}
\gridline{\fig{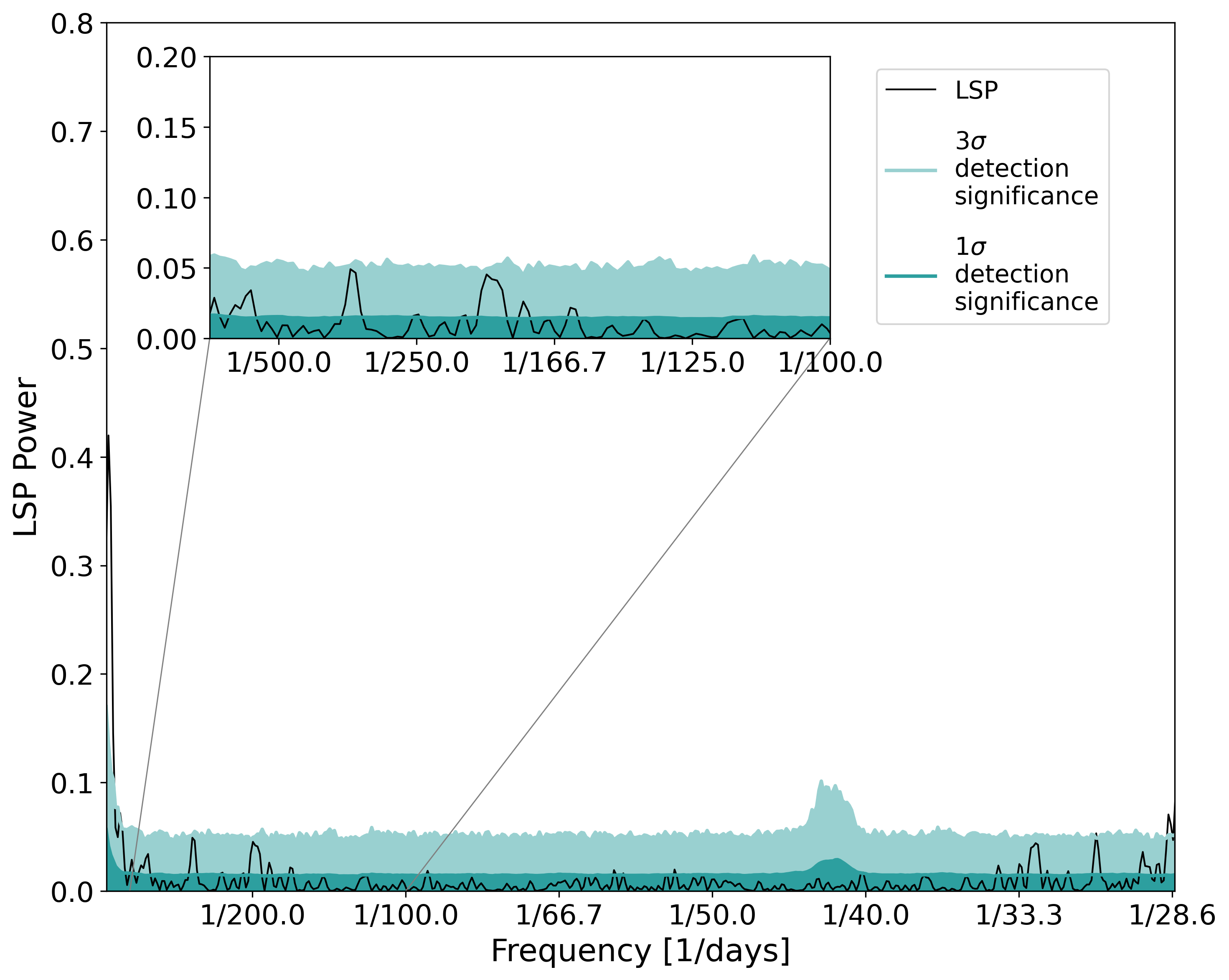}{0.49\textwidth}{(a) LSP for the \textit{Fermi}-LAT (0.3-500\,GeV) light curve including a more detailed view of the frequency range corresponding to a periodicity between 100 to 1000\,days.}
          \fig{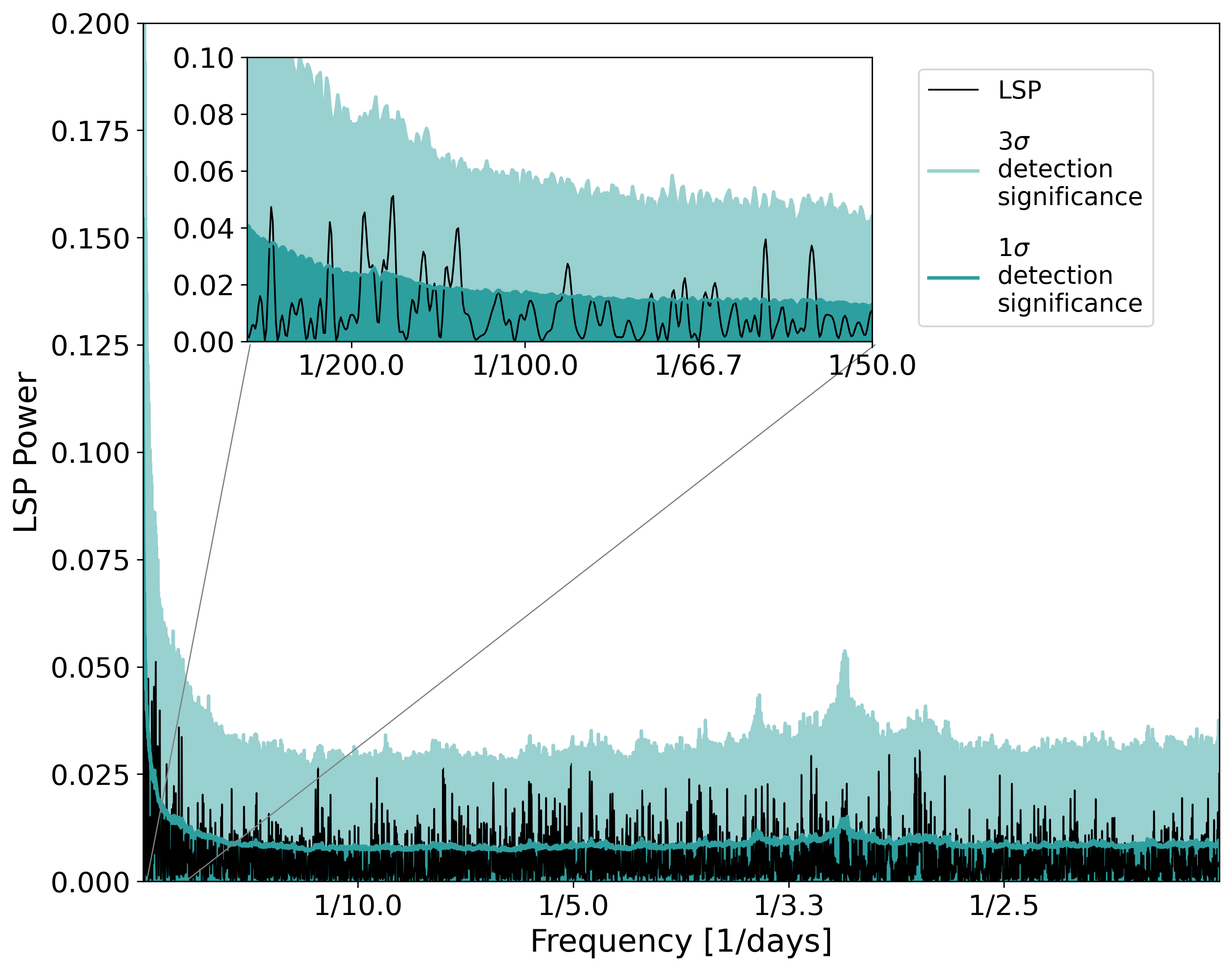}{0.49\textwidth}{(b) LSP for the OVRO (15\,GHz) light curve including a more detailed view of the frequency range corresponding to a periodicity between 40 to 400\,days.}
}
\caption{Lomb-Scargle Periodigram (LSP) for different light curves of our 12-year data set shown in Fig.~\ref{fig:long_LC_comb}. The 1$\sigma$ and 3$\sigma$ detection significance levels obtained by simulations are shown as well as described in Section \ref{sec:app_period}.}
    \label{fig:LSP2}
\end{figure*}

Moreover, no significant peaks are seen for the \textit{Fermi}-LAT light curve (Fig.~\ref{fig:LSP2}a)), if we have a closer look at the previously claimed '330-day' periodicity. We can confirm the slightly raised LSP power around this frequency with a significance of 2.3\,$\sigma$ at $f=1/325$\,days$^{-1}$. Similarly we can confirm another weak hint with a significance of 2.2\,$\sigma$ at $f=1/200$\,days$^{-1}$. This together with the time scale close to the duration of one year, and therefore possible yearly background fluctuations, refrain us from making any scientific claims for periodicity in the $\gamma$-ray range.
However, a significant peak can be seen at the lowest energies at a $f=1/3300$\,days$^{-1}$. Since the time span is too long to be covered by our data set more than once, no periodicity can be claimed. Nonetheless, it indicates that a sinusoidal distribution with a period of $1/3300$\,days$^{-1}$ might describe the data better than a constant or just random flux distribution over time.

For our last long term light curve, OVRO, no significant periodicity can be found (Fig~\ref{fig:LSP2}b)). 

As mentioned in Section~\ref{sec:correlations}, the 'look elsewhere effect' would need to be taken into when evaluating different frequencies. Therefore, all significance values stated here are local significances. However, since none of the values can be considered significant ($>$3$\sigma$) and the global significance values would be even lower, there is no need to apply the correction.


\clearpage
\newpage

\section{Additional information - Low-state SED}
\begin{figure}
    \centering
   \includegraphics[width=0.49\textwidth]{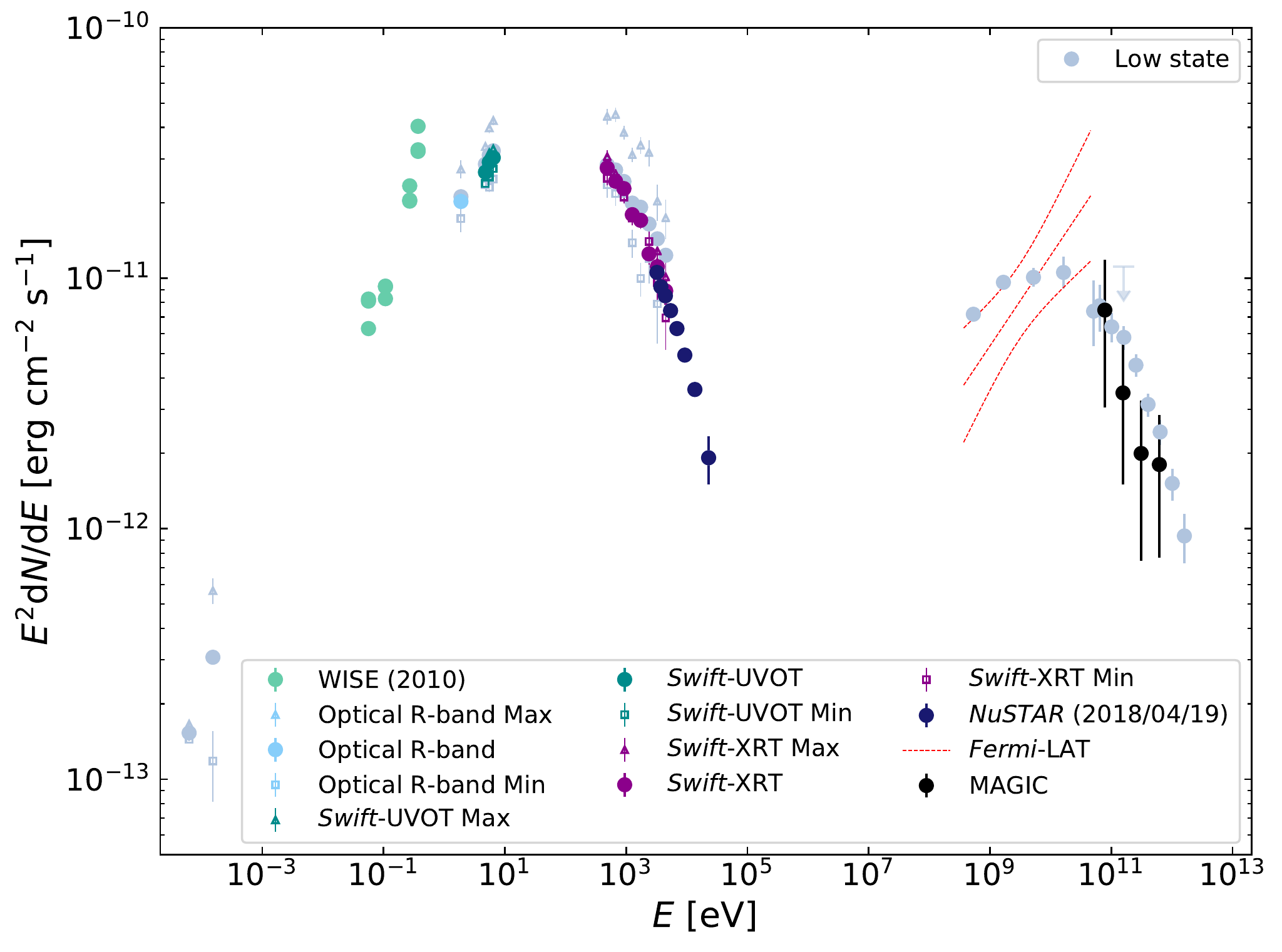}
   \caption{Broadband SED around the third \textit{NuSTAR} observation on 2018-04-20 (MJD~58228).
For \textit{Fermi}-LAT, the $\gamma$-ray spectrum is derived using data from a 2-week interval centered at the \textit{NuSTAR} observation. Archival WISE data from 2010 are also shown. For comparison purposes, the low-activity SED from Fig.~\ref{fig:seds}a) is shown with light-grey markers.}
   \label{fig:sed_nustar3}
\end{figure}

In this section a comparison of the SED around the third \textit{NuSTAR} observation and the low-state is presented in Fig.~\ref{fig:sed_nustar3} as well as the synchroton peak frequencies $\nu_s$ for the different flux states of Mrk\,501 in Table~\ref{tab:nu_s_states}.

\begin{table}
\centering
\begin{tabular}{ l | c c c}  
& $\nu_s^{\text{low-state}}$ & $\nu_s^{\text{\textit{NuSTAR}-1}}$ & $\nu_s^{\text{\textit{NuSTAR}-2}}$ \\
\hline
Phenomenological &  $5.3\times10^{15}$ & $2.0\times10^{16}$  & $5.1\times10^{15}$\\
Leptonic & $3.1\times10^{16}$ & $2.9\times10^{17}$ &$3.1\times10^{16}$ \\
Hadronic (LeHa) & $2.7\times10^{16}$ & - & - \\
Hadronic (SOPRANO) &   $4.6\times10^{15}$ &  - & - \\
Lepto-hadronic (LeHa) &  $2.7\times10^{16}$ & - & - \\
Lepto-hadronic (SOPRANO) &  $3.3\times10^{16}$ & - & - \\

\end{tabular}
\caption{Synchroton peak frequencies $\nu_s$ for the different SEDs shown in Fig.~\ref{fig:seds} using the phenomenological description of \citet{blazar_seq} and the one-zone leptonic and hadronic modelling results described in Section~\ref{subsec:1zone}.}
\label{tab:nu_s_states}
\end{table}

\clearpage
\newpage

\section{Additional information - Theoretical models}
\begin{figure}
    \centering
   \includegraphics[width=0.49\textwidth]{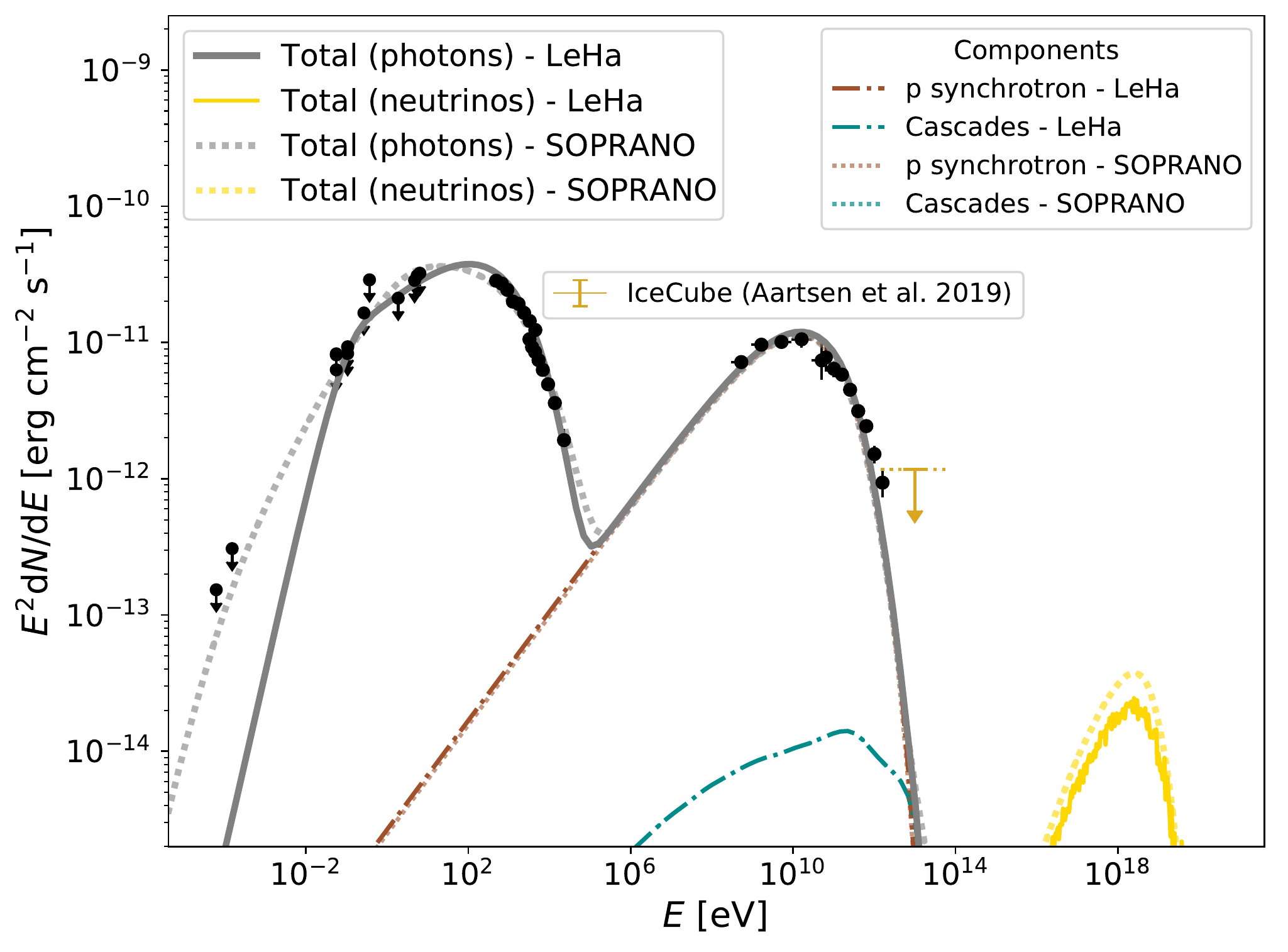}
   \caption{Hadronic one-zone models that describe the broadband SED of the low-state of Mrk\,501 derived with data from 2017-06-17 to 2019-07-23 (MJD~57921 to MJD~58687), as described in Section~\ref{sec:baseline_lep}. The solid lines describes the results of the LeHa code while the model obtained with the SOPRANO code is shown by the dashed lines. The proton synchrotron and cascade components are shown in brown and turquoise. The corresponding model parameters are reported in Table~\ref{tab:fits_had} and Table~\ref{tab:fits_had_Sargis}. Data with frequencies in the UV or lower are considered as upper limits for the modeling of the blazar emission,  and therefore depicted with arrows. Additionally, the neutrino flux estimate is shown by the yellow curve (solid for the LeHA results, dashed for the SOPRANO result), together with the upper limit from Icecube \citep{IceCube_UL2020} depicted by the golden upper limit.}
   \label{fig:fit_hadronic_comp}
\end{figure}

\begin{figure} 
    \centering
   \includegraphics[width=0.49\textwidth]{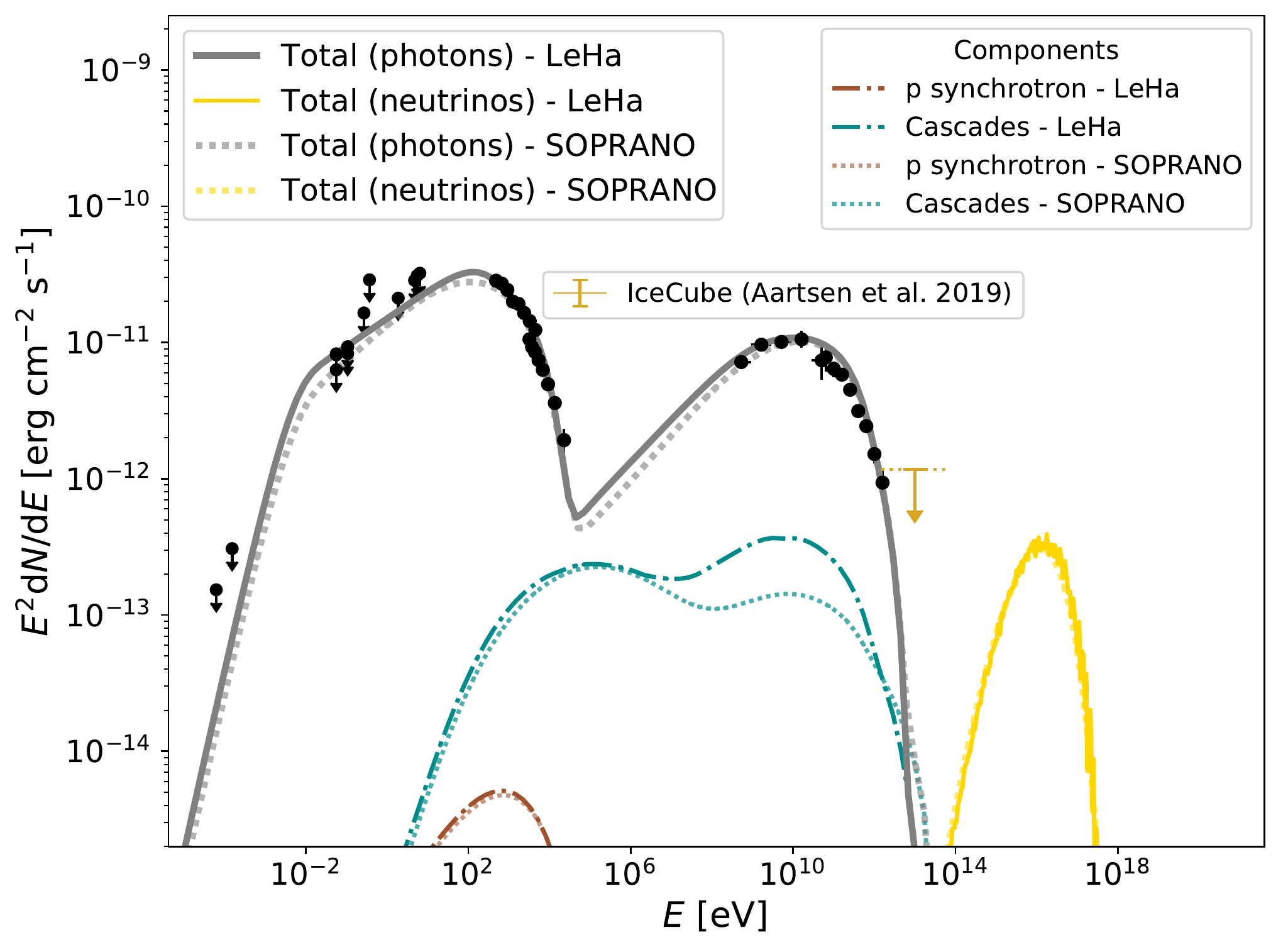}
   \caption{Lepto-hadronic one-zone models that describe the broadband SED of the low-state of Mrk\,501 derived with data from 2017-06-17 to 2019-07-23 (MJD~57921 to MJD~58687), as described in Section~\ref{sec:baseline_lep}. The solid lines describes the results of the LeHa code while the model obtained with the SOPRANO code is shown by the dashed lines. The proton synchrotron and cascade components are shown in brown and turquoise. The corresponding model parameters are reported in Table~\ref{tab:fits_had} and Table~\ref{tab:fits_had_Sargis}. Data with frequencies in the UV or lower are considered as upper limits for the modeling of the blazar emission,  and therefore depicted with arrows. Additionally, the neutrino flux estimate is shown by the yellow curve (solid for the LeHA results, dashed for the SOPRANO result), together with the upper limit from Icecube \citep{IceCube_UL2020} depicted by the golden upper limit.}
   \label{fig:fit_leptohadronic_comp}
\end{figure}

In this section, a more detailed view on the hadronic and lepto-hadronic models obtained by the two frameworks (LeHa and SOPRANO) described in Section~\ref{subsec:hadronic} and Section~\ref{subsec:lepto-hadronic} is given. Fig.~\ref{fig:fit_hadronic_comp} shows the comparison for the hadronic one-zone models displaying the distinguishing components between our hadronic and lepto-hadronic solutions, the proton synchrotron and the cascade emissions. What should be noted is that for the SOPRANO code not all cascade emission can be displayed since the electron-positron pairs are not tagged, but evolve self consistently without distinguishing primary and secondary particles. Therefore, only an estimation of the component can be given and leads to differences between the models. Fig~\ref{fig:fit_leptohadronic_comp} shows the same comparison for the lepto-hadronic one-zone model and Table~\ref{tab:fits_had_Sargis} shows the parameters for the models obtained with the SOPRANO code. Here it should be again stated that when comparing these parameters with the ones in Table~\ref{tab:fits_had}, the ones obtained the SOPRANO code describe the initially injected particle distribution while the ones obtained with the LeHa code describe the radiating distributions.

\begin{table}
    \centering
    {\caption{Parameter values obtained by the SOPRANO code for the hadronic and lepto-hadronic one-zone models used to describe the low-state SED of Mrk\,501 derived with data from 2017-06-17 to 2019-07-23 (MJD~57921 to MJD~58687), as described in Section~\ref{subsec:hadronic} and shown in Fig.~\ref{fig:fit_hadronic}, Fig.~\ref{fig:fit_hadronic_comp} (hadronic), and in Section~\ref{subsec:lepto-hadronic} and Fig.~\ref{fig:fit_leptohadronic}, Fig.~\ref{fig:fit_leptohadronic_comp} (lepto-hadronic).
    In these model realizations, the radius of the emission region $R'$ is fixed to $1.14\times10^{17}$\,cm, and the Doppler factor to $\delta=$11. Since the models described the injected particle distributions simple power-laws are used. The table reports the magnetic field $B'$, the radius $R'$ of the emitting region and, for the assumed initial power-law distribution for the electrons and protons, the densities N$_{\text{0,e}}'$ and N$_{\text{0,p}}'$, the slopes $\alpha_{\text{e}}$, $\alpha_{\text{p}}$ and the minimum and maximum energies $\gamma_{\text{min,e}}'$,  $\gamma_{\text{min,p}}'$,  $\gamma_{\text{max,e}}'$,  $\gamma_{\text{max,p}}'$. We also show the  energy densities held by the electron population $U_{\text{e}}'$, the proton population $U_{\text{p}}'$ and the magnetic field $U_{\text{B}}'$, their ratios and the total jet luminosity $L_{\text{jet}}$. For the EBL $\gamma$-ray absorption at a redshift $z$=0.034, the model from Franceschini \citep{Franceschini} is used.}
    \label{tab:fits_had_Sargis}
        \begin{tabular}{ c | c  | c}     
            & Hadronic & Lepto-hadronic \\
            \hline \hline
            $B'$ [G] & 3 & 0.025 \\
             $R'$ [cm] &  $1.14 \times 10^{17}$ & $1.14 \times 10^{17}$ \\
             N$_{\text{0,e}}'$ [1/cm$^{3}$] & 0.9  & $6.2\times 10^{3}$\\
             N$_{\text{0,p}}'$ [1/cm$^{3}$] & 7.1 & $2.4\times 10^{2}$ \\
             $\alpha_{\text{e}}$ & 1.7  & 2.45\\
             $\gamma_{\text{min,e}}'$ & 2000  & 1000 \\
             $\gamma_{\text{max,e}}'$  & $6.3 \times 10^{4}$  & $1.3 \times 10^{6}$ \\
             $ \alpha_{\text{p}}$ & 2.2 & 2.0\\
             $\gamma_{\text{min, p}}'$ & 1 & 1\\
             $\gamma_{\text{max, p}}'$ & $1.2 \times 10^{10}$ & $2 \times 10^{7}$ \\
            &&\\
            $U_{\text{e}}'$ [erg/cm$^{3}$]  & $3.6\times 10^{-5}$ & $4.9 \times 10^{-4}$ \\
            $U_{\text{B}}'$ [erg/cm$^{3}$] & 0.36 & $2.5 \times 10^{-5}$  \\
            $U_{\text{p}}'$ [erg/cm$^{3}$] & $5.2\times 10^{-2}$ & 5.9 \\
            $U_{\text{e}}'$ / $U_{\text{B}}'$ & $1.0 \times 10^{-4}$ & 19.6 \\
            $U_{\text{p}}'$ / $U_{\text{B}}'$ & 0.15 & $2.4\times 10^{5}$ \\
            $L_{\text{jet}}$ [erg/s] & $3.0 \times 10^{46}$ & $4.4 \times 10^{47}$\\
        \end{tabular}
        }
\end{table}





\end{document}